# Revealing the missing heritability via cross-validated genome-wide association studies

Xia Shen

Division of Computational Genetics, Department of Clinical Sciences, Swedish University of Agricultural Sciences, Box 7078, SE-750 07 Uppsala, Sweden.

Correspondence to: xia.shen@slu.se

**Presented here is a simple method for cross-validated genome-wide association studies (cvGWAS). Focusing on phenotype prediction, the method is able to reveal a significant amount of missing heritability by properly selecting a small number of loci with implicit predictive ability. The results provide new insights into the missing heritability problem and the underlying genetic architecture of complex traits.**

Recently, the case of missing heritability has drawn a lot of attention in genetics of complex traits[1,2,3,4,5,6,7]. It has been widely noticed that for many complex traits, the loci uncovered by means of *e.g.* genome-wide association studies (GWAS) could only explain a minor proportion of the phenotypic variance, even though the observed heritability of the trait is much higher. Strategies have been proposed to search for the sources of such missing heritability[4], *e.g.* capturing additive genetic variance using polygenic effects across the genome[5,8] and mapping quantitative trait loci (QTL) using a powerful design[7]. However, even the use of all the genomic variants could not fully explain the missing heritability. Here, I propose a simple method to perform association mapping based on genomic variants' predictive ability, explain the reason why the estimation of narrow sense heritability using all the markers across the genome is not reliable, and show that the underlying heritability can be much higher than the conventional estimate and can even be well captured by a rather small number of QTL.

Forty-nine traits in *Arabidopsis thaliana*[9] (9 flowering, 21 developmental, 12 defense and 7 ionomics) were analyzed, with the sample size varied from 84 to 194 inbred lines. The *Arabidopsis* accessions were genotyped using a 250K SNP (single nucleotide polymorphism) array, where 216,130 SNPs were available in the analysis. Instead of screening the genome using ordinary GWAS *p*-values, each SNP was assessed for their individual predictive ability by a 5-fold cross validation, *i.e.* the samples were split into a training (80%) and a test (20%) set for five replicates, without overlap among the five test sets. A linear regression of the phenotype on the SNP genotype was fitted in the training set for each marker, and the estimated model was used to perform out-of-sample prediction in the test set. The predictive ability of an individual SNP was evaluated via an $R^2$, which is the squared correlation coefficient between the true phenotypic measurements and their predicted values in each of the five test sets. The mean of the five $R^2$ values, denoted as $R^2_{SNP}$, provided an estimate of the proportion of phenotypic variance captured by the SNP. It should be noted that such a predictive ability measurement is not a function of the *p*-value in ordinary GWAS (*e.g.* **Fig. 1**). Namely, the *p*-values obtained in GWAS tend to under-estimate the predictive performance of the SNPs. Comparison of the association results based on *p*-values and $R^2_{SNP}$ for all the analyzed traits are given in **Supplementary Figure 1-49**.

For each trait, among the top 0.05% of the SNPs (*n* = 108) that had the highest $R^2_{SNP}$, the best subset with no more than 5 SNPs was selected by a forward stepwise selection procedure, based on a cross-validated assessment of their joint predictive power[10]. In order to compare the narrow sense heritability explained by the selected subset of SNPs ($h^2_{QTL}$) with that explained by the entire genome ($h^2_G$), another 5-fold cross validation was conducted. Both a random effects model using only the selected SNPs (the QTL) as explanatory variables and a whole-genome ridge regression (SNP-BLUP[11,12]) were fitted in the training sets and used for predicting the phenotypic values in the test sets. $h^2_{QTL}$ and $h^2_G$ were estimated as the mean of the corresponding five squared correlation coefficients between the true and the predicted values[5]. For most traits, as shown in **Figure 2**, the small number of QTL had substantial advantage over the whole genome in terms of captured narrow sense heritability. The results indicated similar genetic architecture for the traits that belong to the same type. For instance, the defense and ionomics traits showed rather sparse architecture, whereas the flowering traits tended to be more polygenic. Interestingly, two gene expression traits of *FRI* and *FLC*, although regarded as flowering-related, appeared to have sparse architectures. For all the analyzed traits, details about the selected QTL and the heritability estimates are provided in **Supplementary Table 2**.

As a proof of concept, the results clearly showed that assessing the total narrow sense heritability using a large number of markers across the genome[5,7,8] is not a valid approach. The main reason is that one has to substantially sacrifice the precision of the estimated QTL effects when incorporating too many markers as explanatory variables. When the QTL effects or the effects of the SNPs tagging the causal loci are properly estimated, the heritability inherited by the causal loci can be much better revealed than the entire genome. The results indicated that most of the missing heritability was missed by improper analytical methods. Beyond statistical significance in GWAS, more functional loci of complex traits can actually be revealed via assessments based on predictive performance.

## Methods

***Software & URLs***. The *Arabidopsis thaliana* GWAS data set is available at: https://cynin.gmi.oeaw.ac.at/home/resources/atpolydb. All the analysis was conducted in R[13]: http://www.R-project.org/. The association mapping based on cross-validated $R^2_{SNP}$ has been implemented in the "cvGWAS" package: https://r-forge.r-project.org/projects/cvgwas/. The forward stepwise selection procedure was executed by the "FWDselect" package[14]: http://cran.r-project.org/web/packages/FWDselect/. The random QTL effects model was fitted by the "hglm" package[15]: http://cran.r-project.org/web/packages/hglm/. The whole-genome ridge regression (SNP-BLUP) was fitted by the "bigRR" package[12]: http://cran.r-project.org/web/packages/bigRR/.

**Figure Legends**

**Figure 1: Comparison of the SNPs predictive ability and *p*-values for *FRI* gene expression**. The predictive ability is assessed by $R^2_{SNP}$ ("Proportion of phenotypic variance explained via CV"), where CV stands for "cross validation". The *p*-values were obtained using Wilcoxon test. The horizontal line indicates the Bonferroni-corrected genome-wide significant threshold, and the vertical line shows the cut-off that determines which SNPs are to be passed onto the forward selection procedure.

**Figure 2: Comparison of the narrow sense heritability captured by the selected QTL ($h^2_{QTL}$) and the whole genome ($h^2_G$)**. Each colored point represents an analyzed trait. The color of each point shows the type of the trait, where blue, red, green and pink refer to flowering, developmental, defense and ionomics traits, respectively. The size of each point is proportional to the number of QTL selected (from 2 to 5). The cross on each point shows the standard error estimates based on the cross validation. The dashed line indicates equality of $h^2_{QTL}$ and $h^2_{QTL}$ as a visual guide.

**Supplementary Table 1-2** and **Supplementary Figure 1-49** are available as Supplementary Information from the journal website.

**Acknowledgement**

X.S. is funded by a Future Research Leaders grant from Swedish Foundation for Strategic Research (SSF) to Örjan Carlborg.

Figure 1

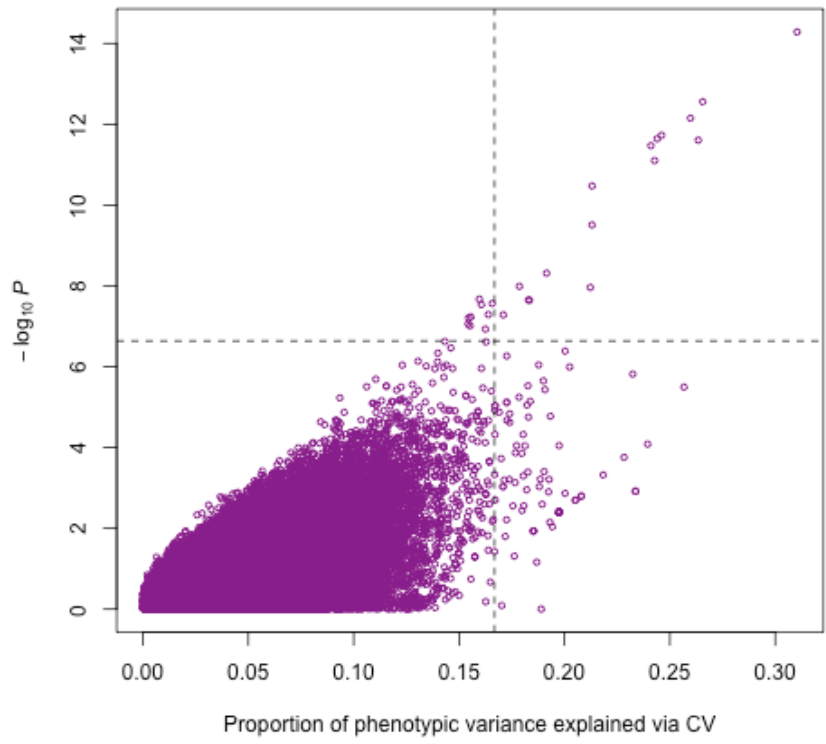

Figure 2

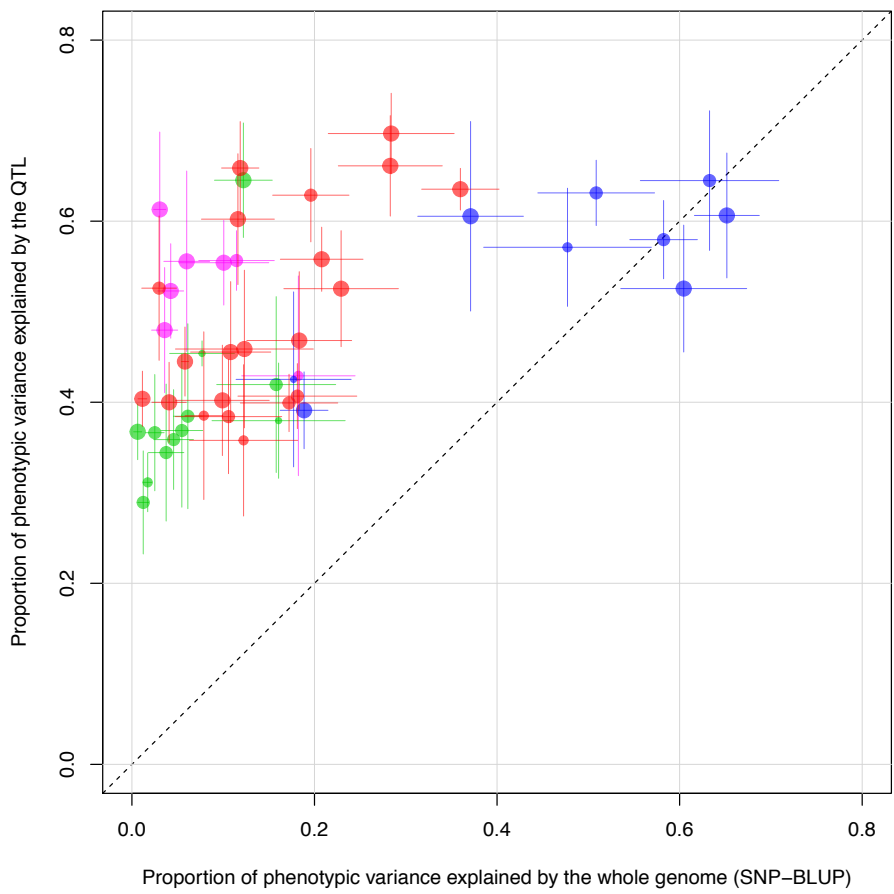

This document contains the Supplementary Information for

*Revealing the missing heritability via cross-validated genome-wide association studies*

by Shen, X. (2013)

Remarks:

- The Excel table format of **Supplementary Table 2** can be downloaded at: https://docs.google.com/file/d/0B2ixEvB0Gwt6SWhZcW1KQmx2Wmc/edit?usp=sharing

- In **Supplementary Figure 1-49**, the horizontal dashed line in each top panel is the Bonferroni-corrected genome-wide significance threshold, and the vertical is the cut-off for selecting candidate SNPs to be passed onto the forward selection procedure. The phenotypic variance explained in the bottom panels was estimated by $R^2_{SNP}$.

- The link provided in **Methods** for the package "cvGWAS" is the project home page. For package download: https://r-forge.r-project.org/R/?group_id=1694. If the package is being built on R-Forge, refer to https://r-forge.r-project.org/scm/viewvc.php/pkg/R/cvscore.R?view=markup&revision=2&root=cvgwas for the source code of the main function cvscore(), which is a directly usable add-on function for the GWA analysis package **GenABEL** (Aulchenko 2007 Bioinformatics).

**Supplementary Table 1: Phenotypes analyzed.**
Refer to Atwell *et al*. (2010) for further details about phenotype description and scoring.

| Phenotype | ID | Type | Sample size | Description | Growth Conditions |
|---|---|---|---|---|---|
| SD | 3 | Flowering | 162 | Days to flowering time (FT) under Long Day (LD) and Short Day(SD) +/- vernalization. | 18°C, 8 hrs daylight. |
| FT10 | 5 | Flowering | 194 | Flowering time (FT) | 10°C, 16 hrs daylight. |
| FT16 | 6 | Flowering | 193 | | 16°C, 16 hrs daylight. |
| FT22 | 7 | Flowering | 193 | | 22°C, 16 hrs daylight. |
| Emco5 | 9 | Defense | 86 | Disease presence or absence following inoculation with each isolate. | 20-22°C, 10 hrs daylight, 70% humidity. |
| Emwa1 | 10 | Defense | 85 | | |
| Hiks1 | 12 | Defense | 84 | | |
| Lithium (Li7) | 14 | Ionomics | 93 | *In planta* ion concentration. | 20°C, 16 hrs daylight. |
| Sulfur (S34) | 19 | Ionomics | 93 | | |
| Potassium (K39) | 20 | Ionomics | 93 | | |
| Manganese (Mn55) | 22 | Ionomics | 93 | | |
| Iron (Fe56) | 23 | Ionomics | 93 | | |
| Cobalt (Co59) | 24 | Ionomics | 93 | | |
| Zinc (Zn66) | 27 | Ionomics | 93 | | |
| AvrRpm1 | 33 | Defense | 84 | Hypersensitive response. | 20°C, 12 hrs daylight. |
| FLC | 43 | Flowering | 167 | FLC and FRI gene expression. | Growth in greenhouse, ~20-22°C, 16 hrs daylight. |
| FRI | 44 | Flowering | 164 | | |
| 8W GH LN | 46 | Flowering | 163 | LN at FT. | 20-22°C, natural light from the middle of October 2002 till March 2003, vernalized (8 wks, 4°C, 8 hrs daylight). |
| 0W GH LN | 48 | Flowering | 135 | | 20-22°C, natural light from the middle of October 2002 till March 2003. |
| FT Diameter Field | 58 | Flowering | 180 | Plant diameter at flowering (field). | Growth in field or greenhouse (20°C, 16 hrs daylight), started in October. |

| Phenotype | ID | Type | Sample size | Description | Growth Conditions |
|---|---|---|---|---|---|
| At1 | 65 | | 175 | | |
| At1 CFU2 | 66 | | 175 | | |
| As CFU2 | 68 | | 175 | | |
| Bs | 69 | | 175 | | |
| Bs CFU2 | 70 | | 175 | | |
| At2 | 71 | | 175 | | |
| At2 CFU2 | 72 | | 175 | | |
| As2 | 73 | | 175 | | |
| DW | 76 | Developmental | 95 | Dry weight of plants. | Plants were grown for 7 weeks at 23°C. |
| Silique 22 | 159 | Developmental | 95 | Silique length. | 22°C, 16 hrs daylight. |
| Germ 10 | 161 | Developmental | 177 | Days to germination. | Stratified for 3 days at 4°C in the dark, followed by growth at 10°C with 16 hrs daylight. |
| Germ 16 | 162 | Developmental | 176 | | Stratified for 3 days at 4°C in the dark, followed by growth at 16°C with 16 hrs daylight. |
| Width 10 | 164 | Developmental | 176 | Plant diameter. | 10°C, 16 hrs daylight. |
| Width 16 | 165 | Developmental | 175 | | 16°C, 16 hrs daylight. |
| Width 22 | 166 | Developmental | 175 | | 22°C, 16 hrs daylight. |
| Chlorosis 16 | 168 | Developmental | 176 | Visual chlorosis presence. | 16°C, 16 hrs daylight. |
| Anthocyanin 10 | 170 | Developmental | 177 | Visual anthocyanin presence. | 10°C, 16 hrs daylight. |
| Anthocyanin 16 | 171 | Developmental | 176 | | 16°C, 16 hrs daylight. |
| Anthocyanin 22 | 172 | Developmental | 177 | | 22°C, 16 hrs daylight. |
| Leaf serr 10 | 173 | Developmental | 174 | Level of leaf serration. | 10°C, 16 hrs daylight. |
| Leaf roll 16 | 177 | Developmental | 176 | Level of roll presence. | 16°C, 16 hrs daylight. |
| Rosette Erect 22 | 179 | Developmental | 176 | Presence of rosette erectness. | 22°C, 16 hrs daylight. |
| Seedling Growth | 272 | Developmental | 101 | Seedling growth rate. | Seeds were grown for one week in the greenhouse under long day (16 hours light). |

| Phenotype | ID | Type | Sample size | Description | Growth Conditions |
|---|---|---|---|---|---|
| Vern Growth | 273 | Developmental | 111 | Vegetative growth rate during vernalization. | Seeds were grown for one week in the greenhouse under long day (16 hours light), vernalized for 4 weeks (4°C, 16h light, 50% relative humidity). |
| After Vern Growth | 274 | Developmental | 111 | Vegetative growth rate after vernalization. | Seeds were grown for one week in the greenhouse under long day (16 hours light), vernalized for 4 weeks (4°C, 16 hrs light, 50% relative humidity)and then returned to greenhouse. |
| Secondary Dormancy | 277 | Developmental | 94 | Decrease in germination rate after prolonged exposure to cold temperature. | Fully after-ripened seeds were treated with a 1 and 6-week long exposure to 4°C. |
| Germ in dark | 278 | Developmental | 94 | Germination in the dark. | 4°C, in the dark. |
| DSDS50 | 279 | Developmental | 110 | Duration of seed dry storage required for 50% of the seeds to germinate. | Dry storage, followed by 25°C, 12 hrs day, 20°C, 12 hrs night for 1 week. |
| Storage 56 days | 283 | Developmental | 111 | Primary dormancy. | 56 days dry storage. |

**Supplementary Table 2** Summary of selected QTL for each analyzed trait and the heritability estimates compared to those by the whole genome.

h2     Narrow-sense heritability estimated as the mean of the squared correlation coefficients between the true and the predicted phenotype in a 5-fold cross validation.
    For h2_QTL, a random effects model using only the QTL as explanatory variables were fitted. For h2_GENOME, a whole-genome ridge regression using all the SNPs were fitted.
se     Standard error estimated via a 5-fold cross validation.

| TRAIT | TYPE | SELECTED SNPS (QTL) | | h2_QTL | se(h2_QTL) | h2_GENOME | se(h2_GENOME) |
|---|---|---|---|---|---|---|---|
| | | CHROMOSOME | POSITION | | | | |
| SD | Flowering | 1 | 4593289 | 0.6063 | 0.0690 | 0.6517 | 0.0355 |
| | | 1 | 18903090 | | | | |
| | | 3 | 18923922 | | | | |
| | | 3 | 18929030 | | | | |
| | | 4 | 16084919 | | | | |
| FT10 | Flowering | 2 | 13151174 | 0.5796 | 0.0432 | 0.5825 | 0.0370 |
| | | 4 | 16017869 | | | | |
| | | 5 | 6534392 | | | | |
| | | 5 | 18607728 | | | | |
| FT16 | Flowering | 1 | 6369609 | 0.6448 | 0.0771 | 0.6328 | 0.0758 |
| | | 2 | 9611587 | | | | |
| | | 3 | 23090917 | | | | |
| | | 4 | 12519944 | | | | |
| FT22 | Flowering | 1 | 6369765 | 0.5255 | 0.0701 | 0.6045 | 0.0690 |
| | | 4 | 12519944 | | | | |
| | | 5 | 2551768 | | | | |
| | | 5 | 2554284 | | | | |
| | | 5 | 6844135 | | | | |
| Emco5 | Defense | 1 | 1430178 | 0.4539 | 0.0138 | 0.0768 | 0.0355 |
| | | 2 | 4185247 | | | | |
| Emwa1 | Defense | 1 | 17250538 | 0.6452 | 0.0632 | 0.1221 | 0.0316 |
| | | 2 | 13008747 | | | | |
| | | 2 | 17934073 | | | | |
| | | 4 | 8196803 | | | | |
| | | 5 | 25721325 | | | | |
| Hiks1 | Defense | 1 | 22583408 | 0.4195 | 0.0971 | 0.1581 | 0.0652 |
| | | 5 | 9299223 | | | | |
| | | 5 | 10841701 | | | | |
| | | 5 | 17477817 | | | | |
| Li7 | Ionomics | 1 | 11096840 | 0.4292 | 0.1103 | 0.1824 | 0.0621 |
| | | 3 | 10620051 | | | | |
| | | 4 | 15226225 | | | | |
| S34 | Ionomics | 2 | 621979 | 0.4796 | 0.0691 | 0.0358 | 0.0143 |
| | | 2 | 5755893 | | | | |
| | | 3 | 22362360 | | | | |
| | | 4 | 7788807 | | | | |
| | | 5 | 10133357 | | | | |
| K39 | Ionomics | 1 | 10146885 | 0.5541 | 0.0468 | 0.1006 | 0.0493 |
| | | 1 | 17666204 | | | | |
| | | 1 | 28516934 | | | | |
| | | 4 | 6786084 | | | | |
| | | 5 | 11291662 | | | | |
| Mn55 | Ionomics | 1 | 8502187 | 0.5565 | 0.0329 | 0.1146 | 0.0414 |
| | | 1 | 13853615 | | | | |
| | | 1 | 29996840 | | | | |
| | | 3 | 20599509 | | | | |
| Fe56 | Ionomics | 3 | 7092529 | 0.5229 | 0.0522 | 0.0425 | 0.0142 |
| | | 3 | 7918007 | | | | |
| | | 3 | 16718549 | | | | |
| | | 4 | 795370 | | | | |
| | | 5 | 5611234 | | | | |
| Co59 | Ionomics | 3 | 13135745 | 0.5556 | 0.0998 | 0.0600 | 0.0251 |
| | | 3 | 20321272 | | | | |
| | | 4 | 5514273 | | | | |
| | | 4 | 7592626 | | | | |
| | | 4 | 8569114 | | | | |
| Zn66 | Ionomics | 1 | 21179549 | 0.6129 | 0.0855 | 0.0305 | 0.0067 |
| | | 2 | 6367225 | | | | |
| | | 2 | 7419526 | | | | |
| | | 3 | 9969573 | | | | |
| | | 5 | 26078291 | | | | |
| avrRpm1 | Defense | 2 | 17504634 | 0.3796 | 0.0636 | 0.1607 | 0.0729 |
| | | 3 | 2270902 | | | | |
| FLC | Flowering | 1 | 19790829 | 0.6054 | 0.1047 | 0.3710 | 0.0579 |
| | | 4 | 1507838 | | | | |
| | | 5 | 5883775 | | | | |
| | | 5 | 10172996 | | | | |
| | | 5 | 24786228 | | | | |
| FRI | Flowering | 1 | 5989995 | 0.4252 | 0.0966 | 0.1771 | 0.0630 |
| | | 4 | 268809 | | | | |
| 8W GH LN | Flowering | 1 | 2005921 | 0.5711 | 0.0653 | 0.4772 | 0.0918 |
| | | 2 | 1977590 | | | | |
| | | 3 | 12358261 | | | | |
| 0W GH LN | Flowering | 1 | 2005921 | 0.6312 | 0.0361 | 0.5086 | 0.0639 |
| | | 3 | 14131141 | | | | |
| | | 4 | 16309006 | | | | |
| | | 5 | 18625726 | | | | |
| FT Diameter Field | Flowering | 2 | 8405178 | 0.3911 | 0.0424 | 0.1886 | 0.0262 |
| | | 2 | 14616766 | | | | |
| | | 4 | 15770883 | | | | |
| | | 5 | 433959 | | | | |
| | | 5 | 26809133 | | | | |
| At1 | Defense | 1 | 16010365 | 0.3588 | 0.0551 | 0.0457 | 0.0218 |
| | | 3 | 2613557 | | | | |
| | | 4 | 10057494 | | | | |
| | | 5 | 19958648 | | | | |
| At1 CFU2 | Defense | 1 | 6629169 | 0.3444 | 0.0756 | 0.0375 | 0.0194 |
| | | 1 | 7898750 | | | | |
| | | 1 | 8237125 | | | | |
| | | 4 | 12080070 | | | | |
| As CFU2 | Defense | 1 | 2275779 | 0.3674 | 0.0308 | 0.0063 | 0.0045 |
| | | 1 | 22984248 | | | | |
| | | 4 | 5814807 | | | | |
| | | 4 | 17180545 | | | | |
| | | 5 | 10058335 | | | | |
| Bs | Defense | 1 | 8298611 | 0.3664 | 0.0642 | 0.0251 | 0.0102 |
| | | 1 | 28757586 | | | | |
| | | 3 | 23000304 | | | | |
| | | 4 | 18459798 | | | | |
| Bs CFU2 | Defense | 1 | 17212115 | 0.3688 | 0.0847 | 0.0547 | 0.0230 |
| | | 1 | 19333698 | | | | |
| | | 2 | 17078909 | | | | |
| | | 4 | 10857336 | | | | |
| At2 | Defense | 1 | 16776084 | 0.3845 | 0.1021 | 0.0613 | 0.0158 |
| | | 1 | 18397234 | | | | |
| | | 4 | 6262290 | | | | |
| | | 5 | 555655 | | | | |
| At2 CFU2 | Defense | 1 | 25379336 | 0.2893 | 0.0569 | 0.0124 | 0.0059 |
| | | 2 | 8414639 | | | | |
| | | 3 | 12697805 | | | | |
| | | 5 | 19565066 | | | | |
| As2 | Defense | 1 | 18040347 | 0.3115 | 0.0324 | 0.0173 | 0.0046 |
| | | 2 | 16474284 | | | | |
| | | 4 | 5004139 | | | | |
| DW | Developmental | 3 | 1801701 | 0.6587 | 0.0513 | 0.1186 | 0.0204 |
| | | 3 | 6145352 | | | | |
| | | 3 | 14997936 | | | | |
| | | 3 | 20154975 | | | | |
| | | 4 | 17026991 | | | | |
| Silique 22 | Developmental | 1 | 2968159 | 0.5579 | 0.0354 | 0.2079 | 0.0453 |
| | | 2 | 17801496 | | | | |
| | | 4 | 403634 | | | | |
| | | 5 | 17364116 | | | | |
| | | 5 | 18640009 | | | | |
| Germ 10 | Developmental | 1 | 1429372 | 0.4449 | 0.0381 | 0.0581 | 0.0039 |
| | | 1 | 9638902 | | | | |
| | | 3 | 6487689 | | | | |
| | | 3 | 15305943 | | | | |
| | | 4 | 179908 | | | | |
| Germ 16 | Developmental | 1 | 1551963 | 0.4554 | 0.0778 | 0.1082 | 0.0439 |
| | | 1 | 10488901 | | | | |
| | | 2 | 10188094 | | | | |
| | | 4 | 9061476 | | | | |
| | | 5 | 17648491 | | | | |
| Width 10 | Developmental | 1 | 21041405 | 0.5254 | 0.0641 | 0.2292 | 0.0627 |
| | | 3 | 20976454 | | | | |
| | | 5 | 6373912 | | | | |
| | | 5 | 14131512 | | | | |
| | | 5 | 16952385 | | | | |
| Width 16 | Developmental | 1 | 12615860 | 0.4068 | 0.0359 | 0.1812 | 0.0651 |
| | | 3 | 15719656 | | | | |
| | | 3 | 20882629 | | | | |
| | | 5 | 18262951 | | | | |
| Width 22 | Developmental | 1 | 21752821 | 0.4021 | 0.0610 | 0.0991 | 0.0513 |
| | | 1 | 24461138 | | | | |
| | | 3 | 15753112 | | | | |
| | | 4 | 15083851 | | | | |
| | | 4 | 17570674 | | | | |
| Chlorosis 16 | Developmental | 1 | 23272710 | 0.4038 | 0.0305 | 0.0115 | 0.0048 |
| | | 1 | 30244136 | | | | |
| | | 2 | 15088120 | | | | |
| | | 4 | 9996734 | | | | |
| | | 5 | 6416385 | | | | |
| Anthocyanin 10 | Developmental | 1 | 1921764 | 0.4588 | 0.0871 | 0.1232 | 0.0755 |
| | | 1 | 4865222 | | | | |
| | | 2 | 1937020 | | | | |
| | | 2 | 17731129 | | | | |
| | | 3 | 8594331 | | | | |
| Anthocyanin 16 | Developmental | 3 | 7931982 | 0.3999 | 0.0442 | 0.0407 | 0.0188 |
| | | 3 | 17584494 | | | | |
| | | 3 | 18237858 | | | | |
| | | 5 | 14776227 | | | | |
| | | 5 | 18408198 | | | | |
| Anthocyanin 22 | Developmental | 1 | 16933062 | 0.3841 | 0.0630 | 0.1058 | 0.0583 |
| | | 2 | 18684705 | | | | |
| | | 3 | 18230944 | | | | |
| | | 5 | 26305400 | | | | |
| Leaf serr 10 | Developmental | 1 | 21866684 | 0.3992 | 0.0317 | 0.1721 | 0.0534 |
| | | 1 | 25545686 | | | | |
| | | 3 | 711663 | | | | |
| | | 5 | 809032 | | | | |
| Leaf roll 16 | Developmental | 1 | 1543644 | 0.3579 | 0.0836 | 0.1223 | 0.0593 |
| | | 1 | 12541124 | | | | |
| | | 5 | 9691412 | | | | |
| Rosette Erect 22 | Developmental | 1 | 10702954 | 0.4682 | 0.0760 | 0.1833 | 0.0574 |
| | | 4 | 158958 | | | | |
| | | 4 | 5423725 | | | | |
| | | 5 | 15438762 | | | | |
| | | 5 | 17474995 | | | | |
| Seedling Growth | Developmental | 1 | 1409102 | 0.5260 | 0.0797 | 0.0298 | 0.0190 |
| | | 1 | 3351283 | | | | |
| | | 2 | 5985892 | | | | |
| | | 3 | 1193580 | | | | |
| Vern Growth | Developmental | 1 | 22861979 | 0.6022 | 0.0722 | 0.1161 | 0.0399 |
| | | 1 | 25496457 | | | | |
| | | 3 | 10027171 | | | | |
| | | 4 | 16443666 | | | | |
| | | 5 | 21775600 | | | | |
| After Vern Growth | Developmental | 3 | 11799463 | 0.3852 | 0.0926 | 0.0788 | 0.0207 |
| | | 4 | 7543367 | | | | |
| | | 4 | 17416904 | | | | |
| Secondary Dormancy | Developmental | 1 | 22945590 | 0.6968 | 0.0445 | 0.2840 | 0.0689 |
| | | 3 | 15137506 | | | | |
| | | 3 | 16164636 | | | | |
| | | 3 | 22951949 | | | | |
| | | 5 | 6097616 | | | | |
| Germ in dark | Developmental | 1 | 10725637 | 0.6287 | 0.0516 | 0.1960 | 0.0418 |
|  |  | 4 | 7403647 |  |  |  |  |
|  |  | 5 | 10511334 |  |  |  |  |
|  |  | 5 | 22310661 |  |  |  |  |
| DSDS50 | Developmental | 1 | 1045551 | 0.6611 | 0.0555 | 0.2830 | 0.0569 |
|  |  | 1 | 11593466 |  |  |  |  |
|  |  | 2 | 10750002 |  |  |  |  |
|  |  | 3 | 21285974 |  |  |  |  |
|  |  | 4 | 14688343 |  |  |  |  |
| Storage 56 days | Developmental | 1 | 863771 | 0.6353 | 0.0230 | 0.3598 | 0.0424 |
|  |  | 1 | 19520347 |  |  |  |  |
|  |  | 2 | 883192 |  |  |  |  |
|  |  | 2 | 5713096 |  |  |  |  |
|  |  | 5 | 15859708 |  |  |  |  |

**Comparison of *p*-values and predictive ability**

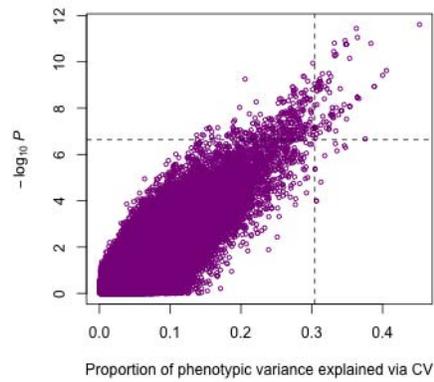

**Genome-wide association mapping via Wilcoxon test**

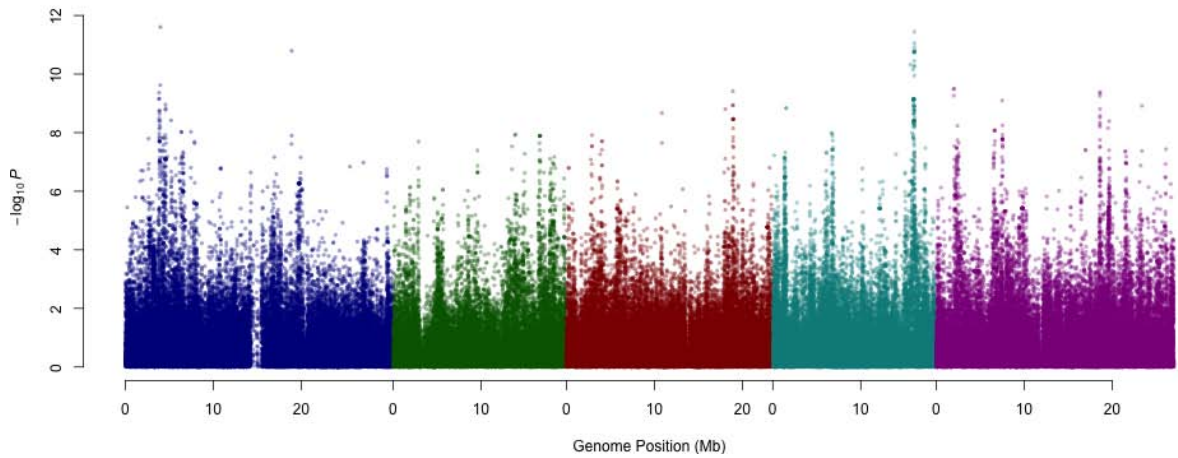

**Predictive ability assessed by cross validation**

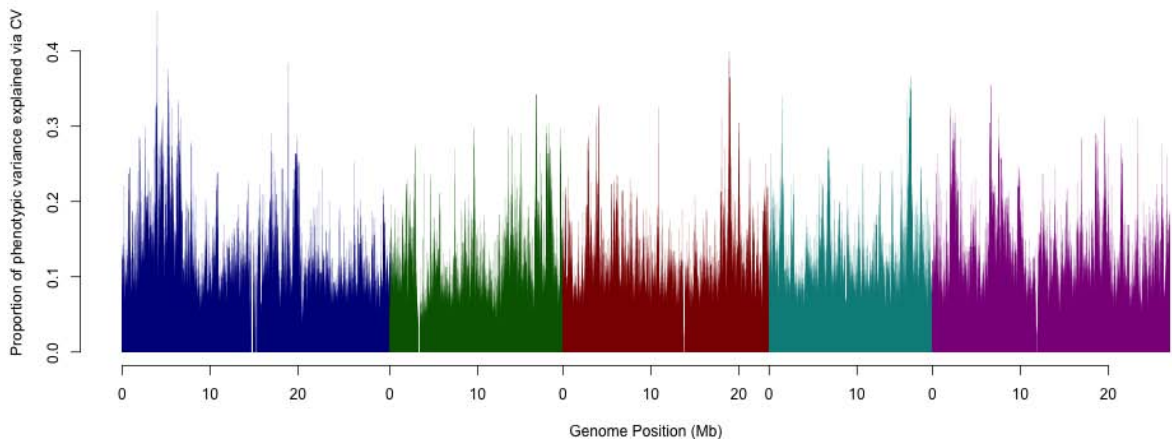

**Supplementary Figure 1** - Results of GWAS *p*-values and cross-validated predictive ability for SD

**Comparison of *p*-values and predictive ability**

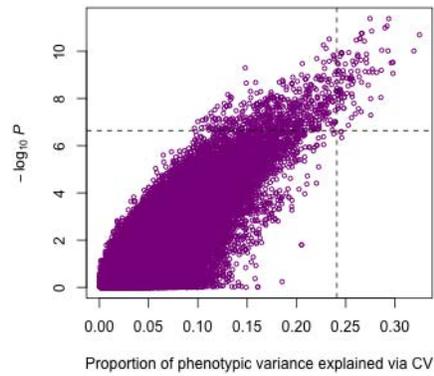

**Genome-wide association mapping via Wilcoxon test**

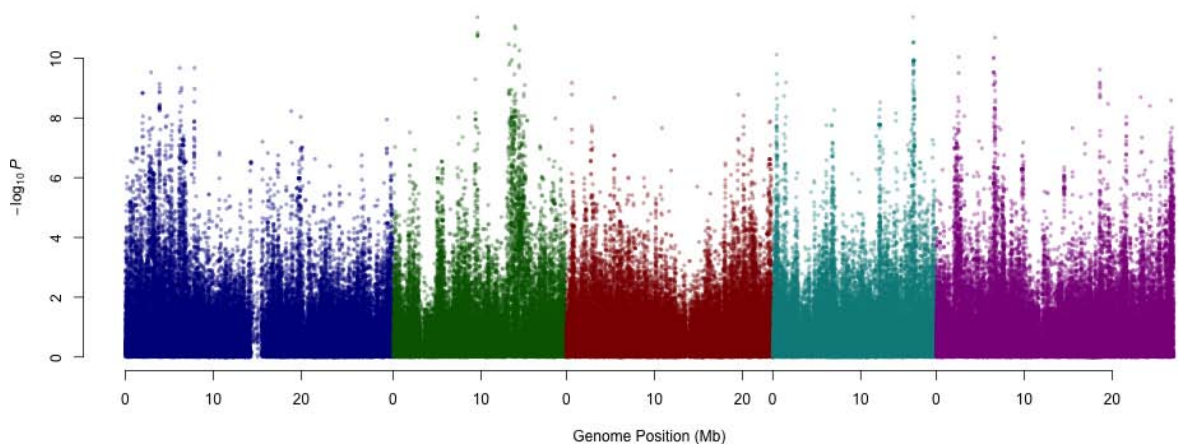

**Predictive ability assessed by cross validation**

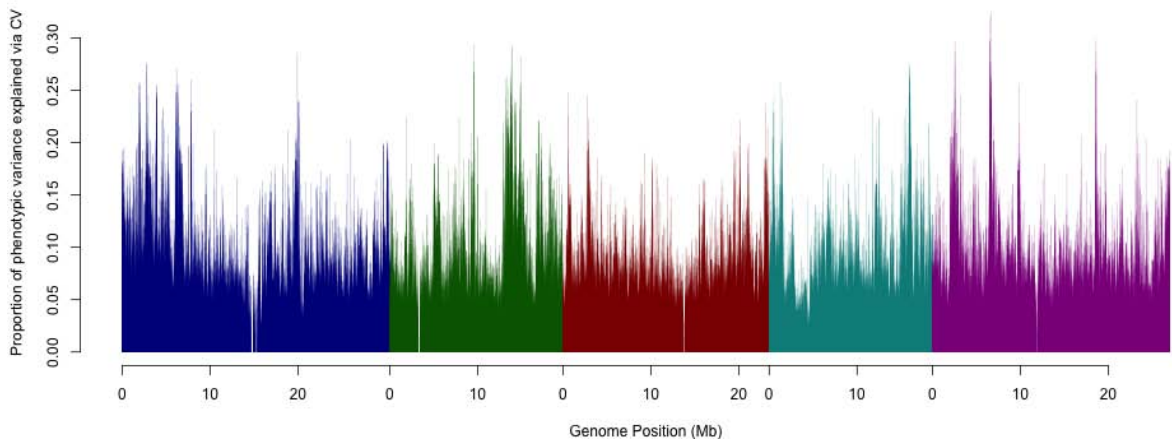

**Supplementary Figure 2** - Results of GWAS *p*-values and cross-validated predictive ability for FT10

**Comparison of *p*-values and predictive ability**

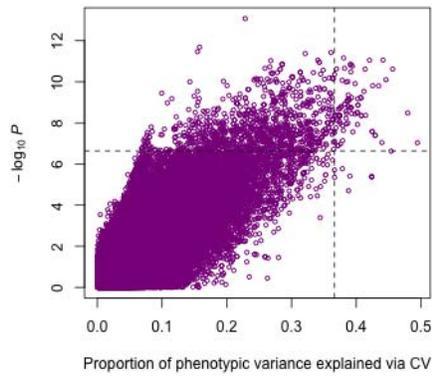

**Genome-wide association mapping via Wilcoxon test**

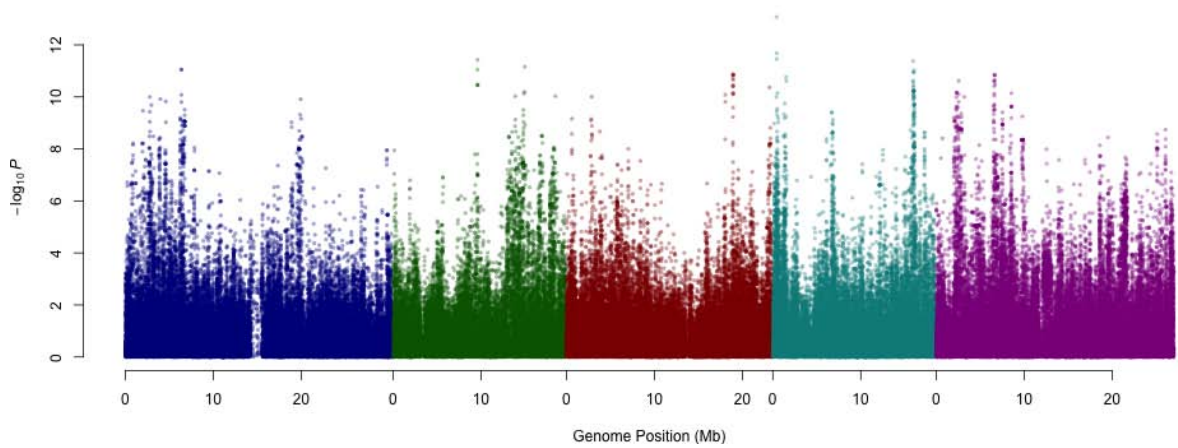

**Predictive ability assessed by cross validation**

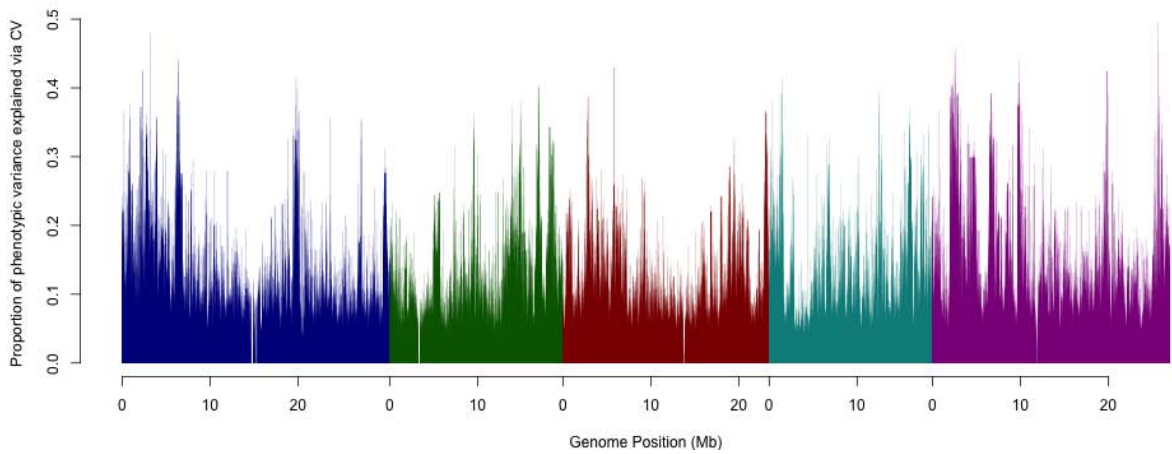

**Supplementary Figure 3** - Results of GWAS *p*-values and cross-validated predictive ability for FT16

**Comparison of *p*-values and predictive ability**

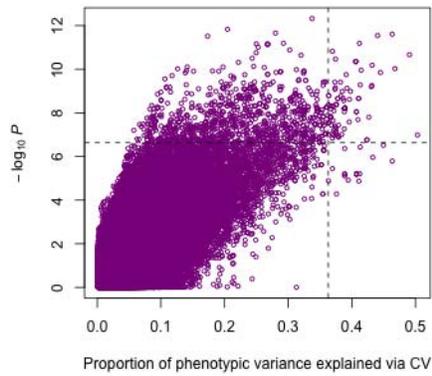

**Genome-wide association mapping via Wilcoxon test**

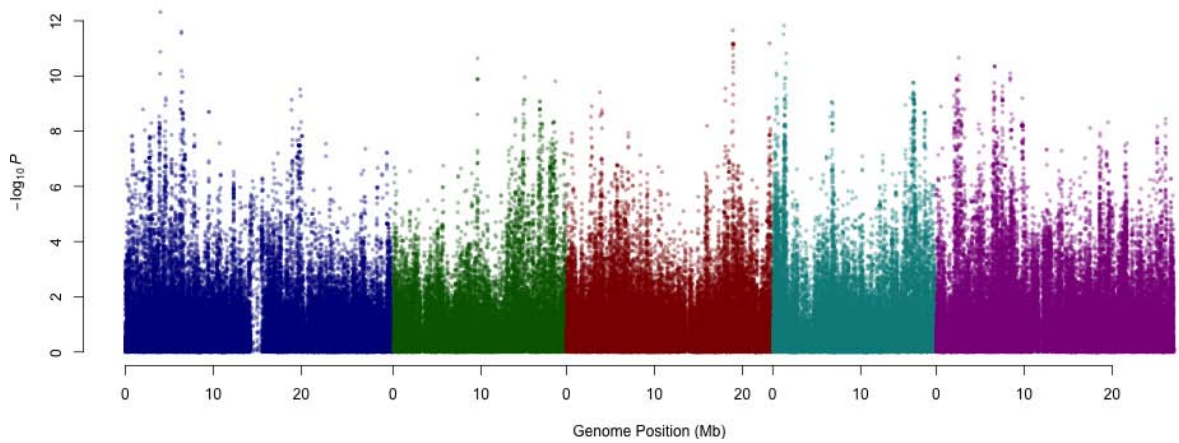

**Predictive ability assessed by cross validation**

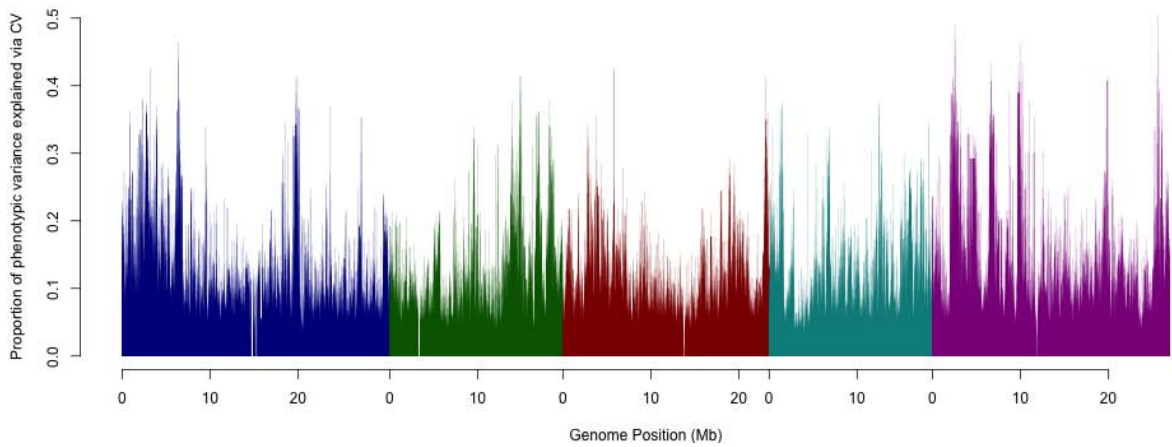

**Supplementary Figure 4** - Results of GWAS *p*-values and cross-validated predictive ability for FT22

**Comparison of *p*-values and predictive ability**

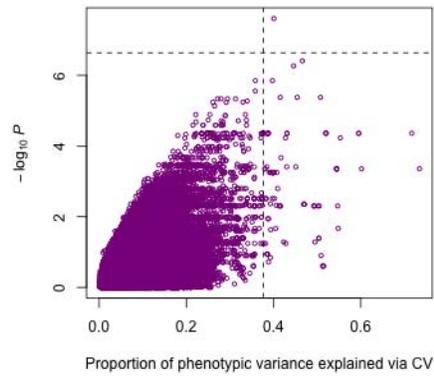

**Genome-wide association mapping via Wilcoxon test**

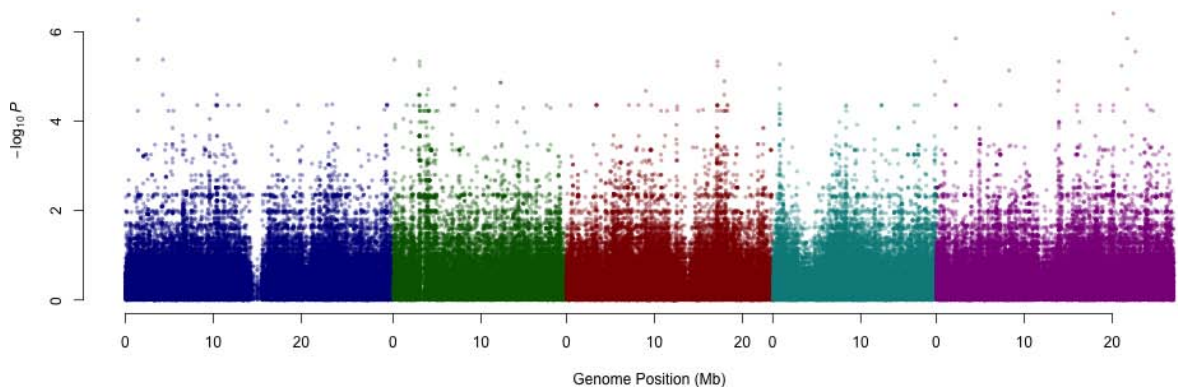

**Predictive ability assessed by cross validation**

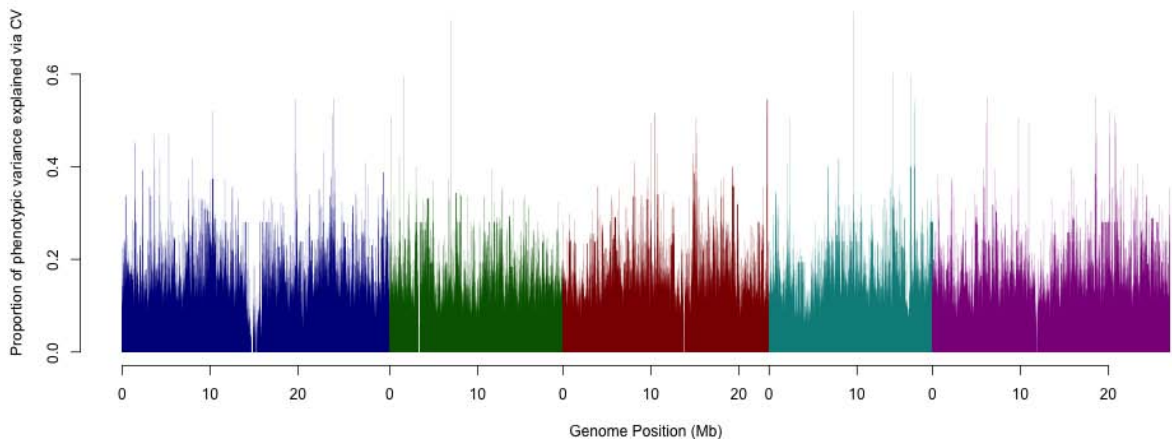

**Supplementary Figure 5** - Results of GWAS *p*-values and cross-validated predictive ability for Emco5

**Comparison of *p*-values and predictive ability**

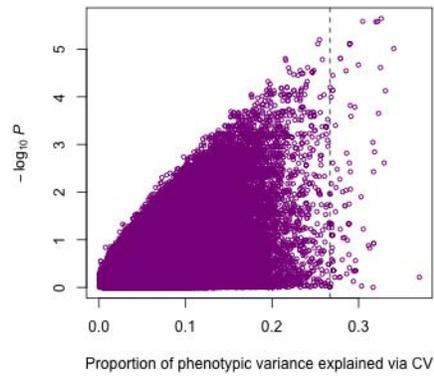

**Genome-wide association mapping via Wilcoxon test**

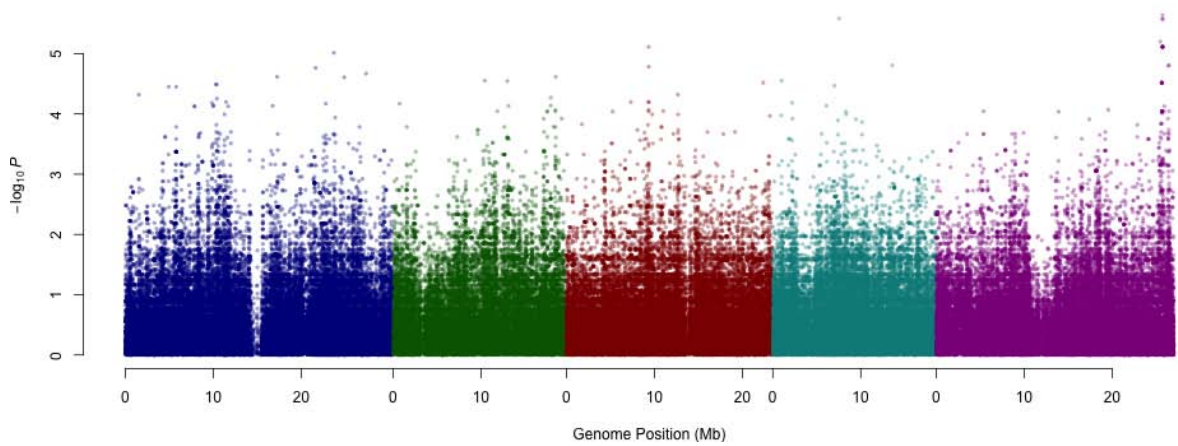

**Predictive ability assessed by cross validation**

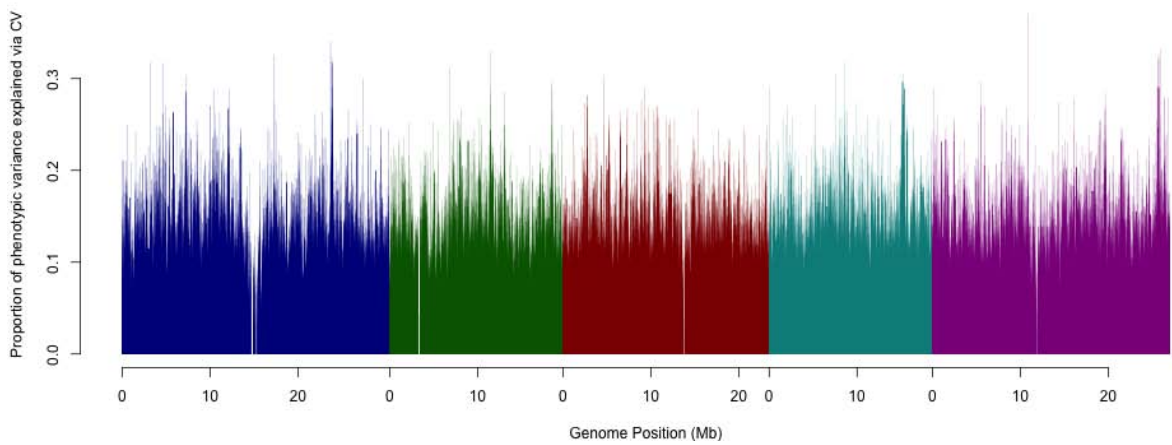

**Supplementary Figure 6** - Results of GWAS *p*-values and cross-validated predictive ability for Emwa1

**Comparison of *p*-values and predictive ability**

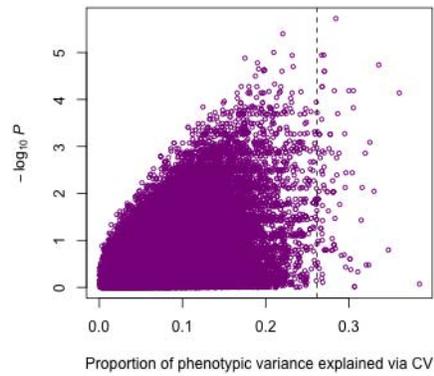

**Genome-wide association mapping via Wilcoxon test**

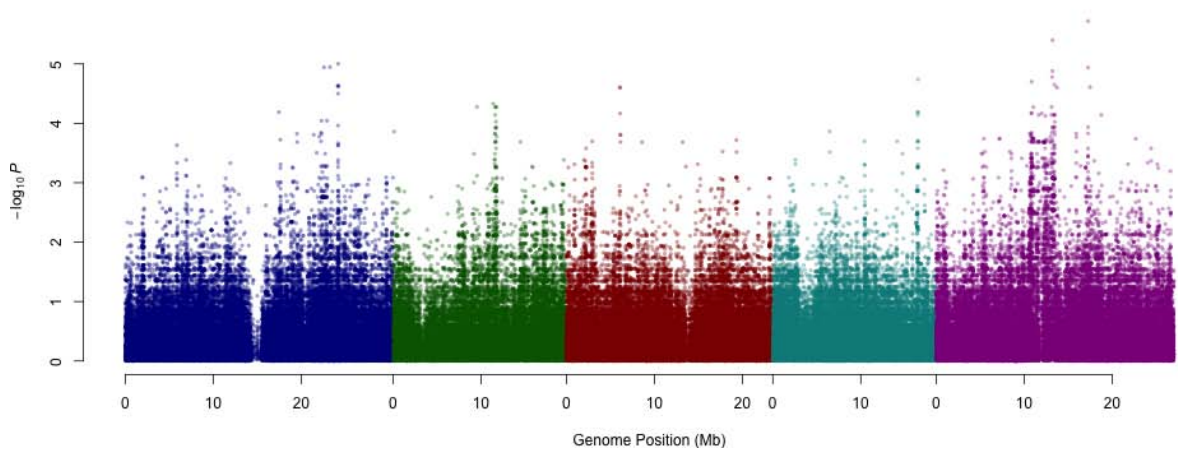

**Predictive ability assessed by cross validation**

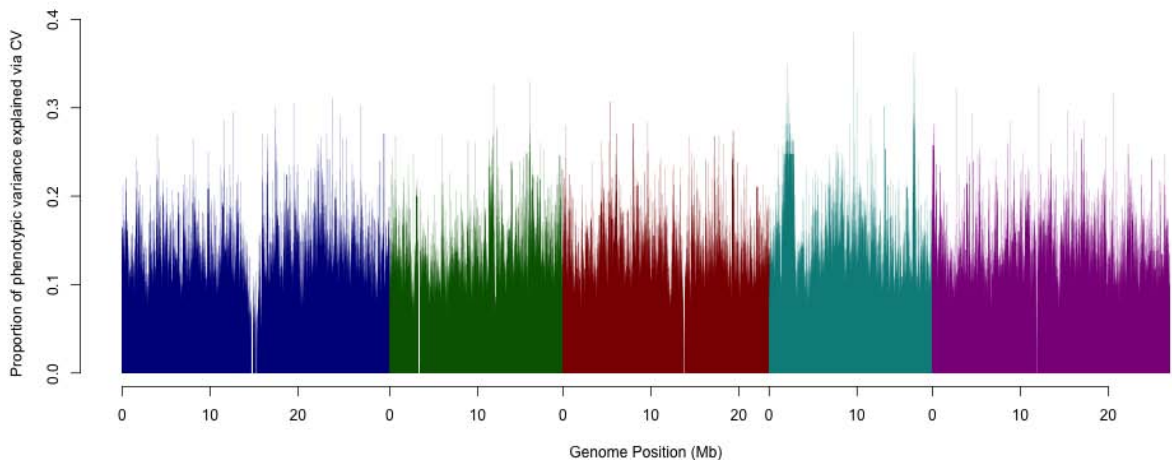

**Supplementary Figure 7** - Results of GWAS *p*-values and cross-validated predictive ability for Hiks1

**Comparison of *p*-values and predictive ability**

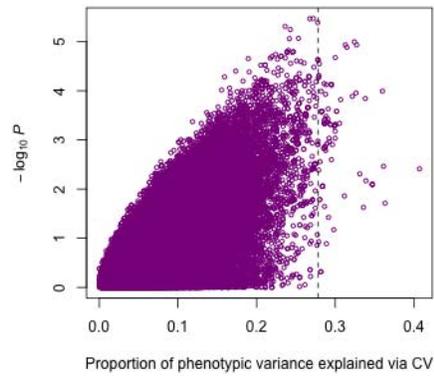

**Genome-wide association mapping via Wilcoxon test**

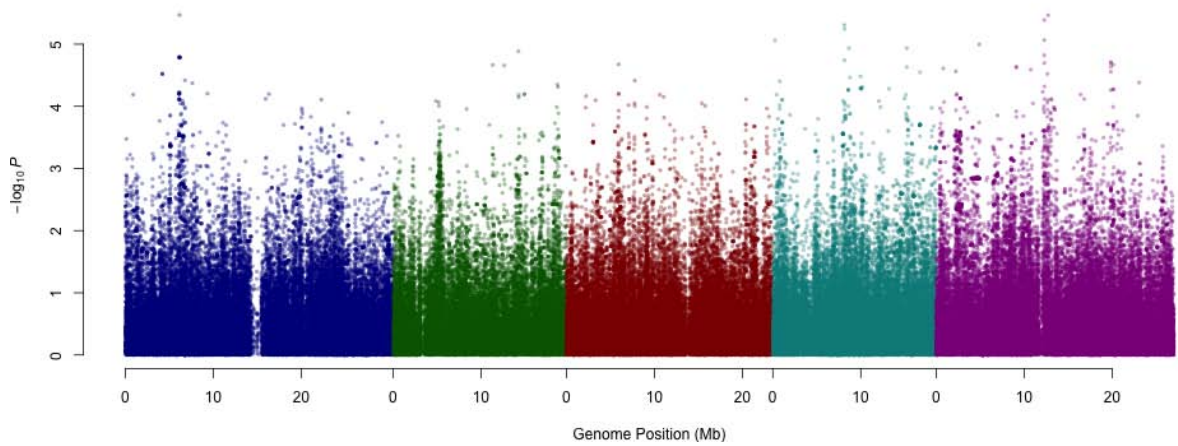

**Predictive ability assessed by cross validation**

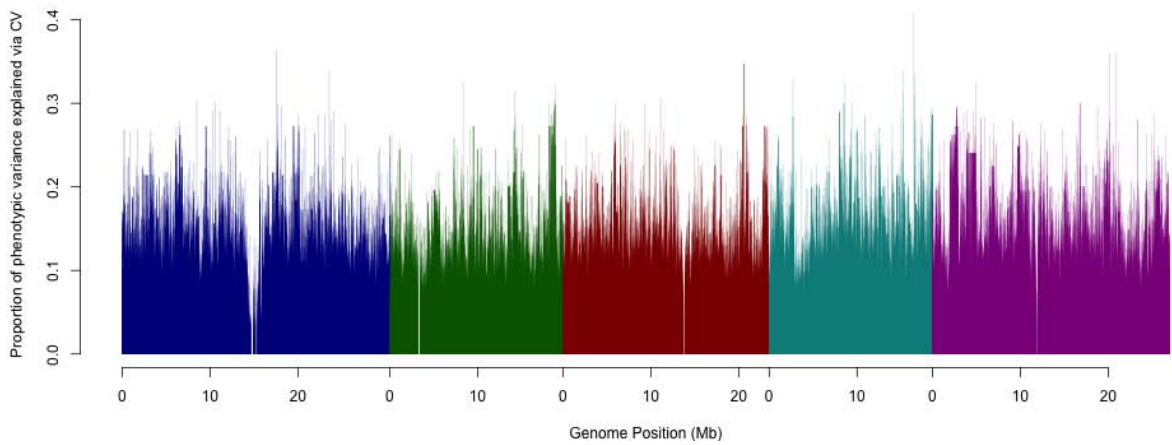

**Supplementary Figure 8** - Results of GWAS *p*-values and cross-validated predictive ability for Li7

**Comparison of *p*-values and predictive ability**

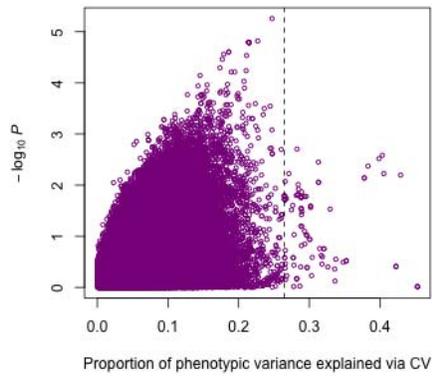

**Genome-wide association mapping via Wilcoxon test**

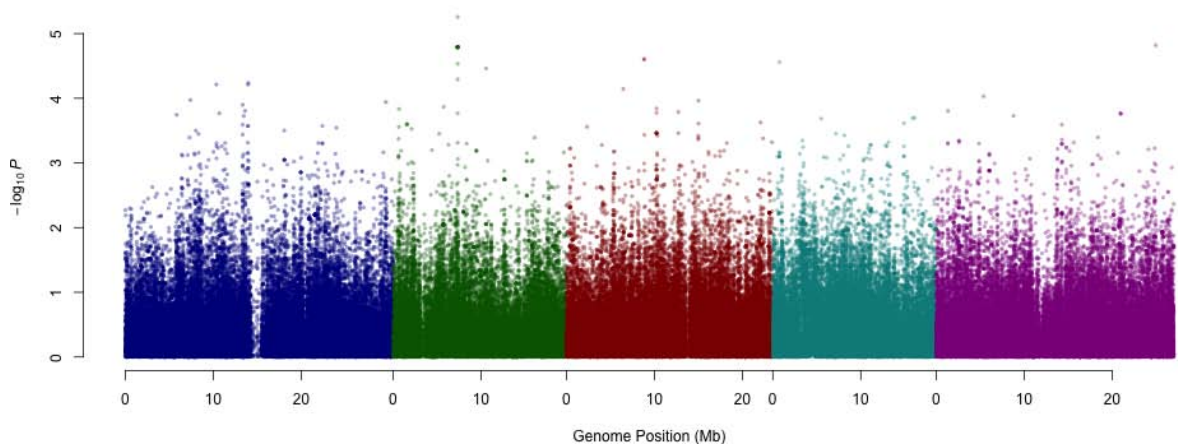

**Predictive ability assessed by cross validation**

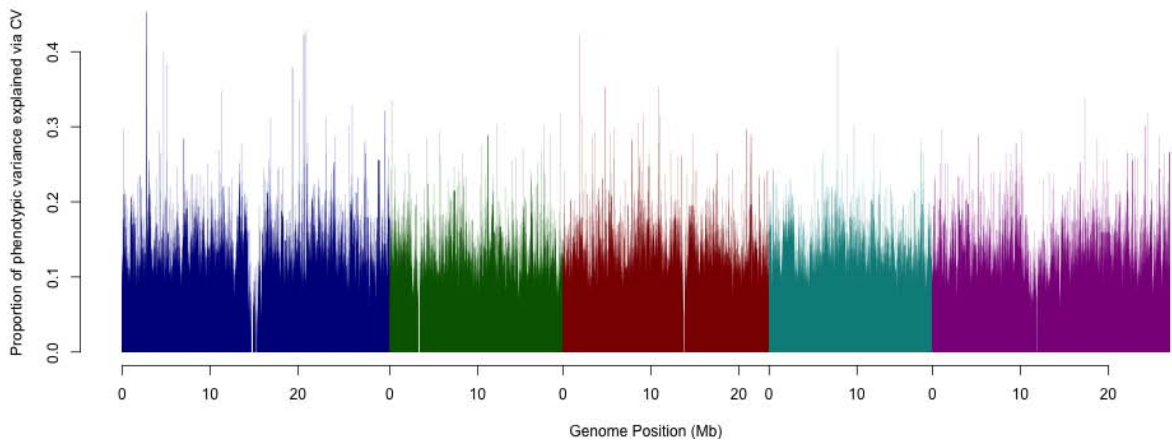

**Supplementary Figure 9** - Results of GWAS *p*-values and cross-validated predictive ability for S34

**Comparison of *p*-values and predictive ability**

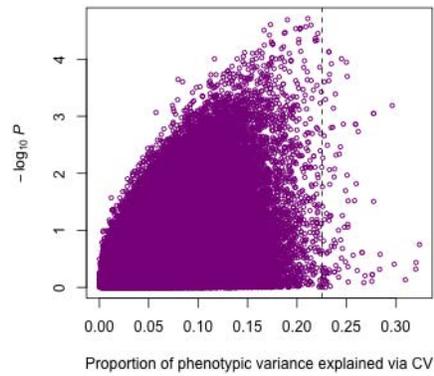

**Genome-wide association mapping via Wilcoxon test**

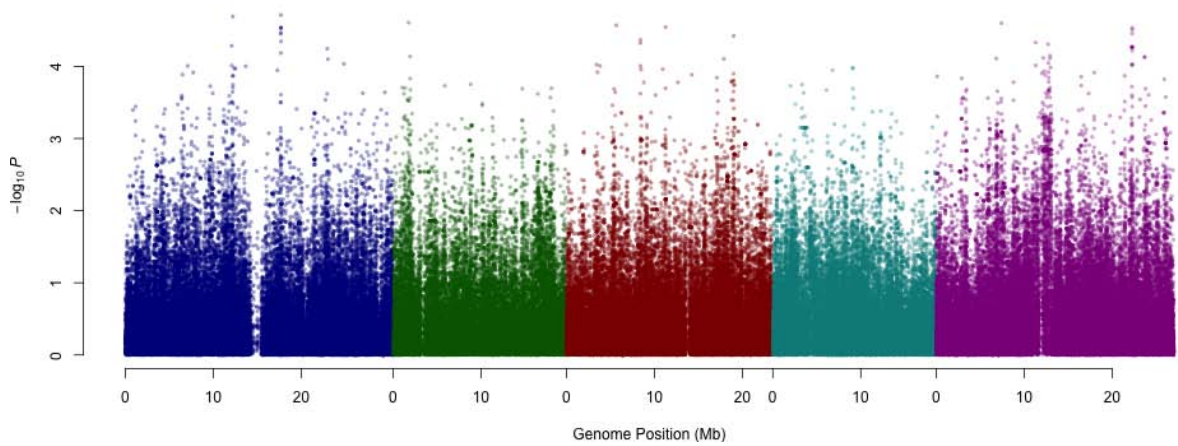

**Predictive ability assessed by cross validation**

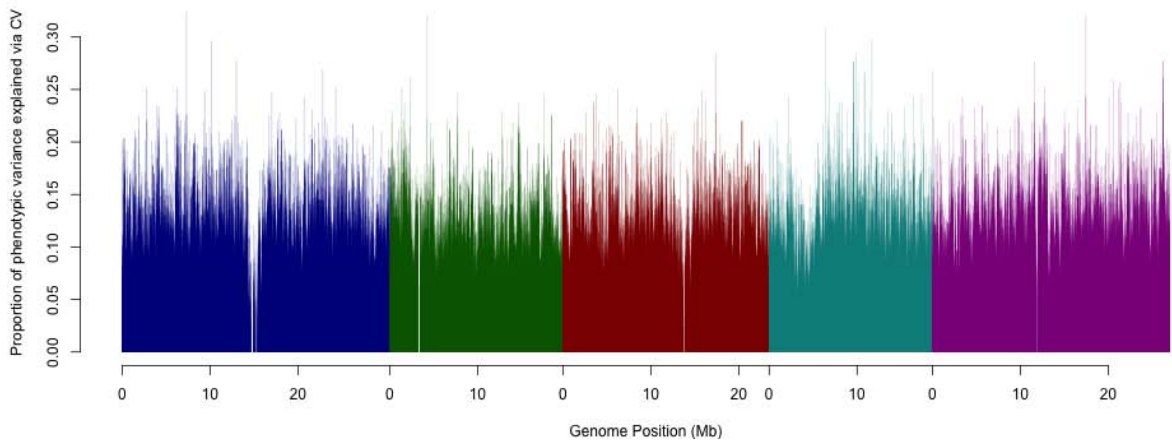

**Supplementary Figure 10** - Results of GWAS *p*-values and cross-validated predictive ability for K39

**Comparison of *p*-values and predictive ability**

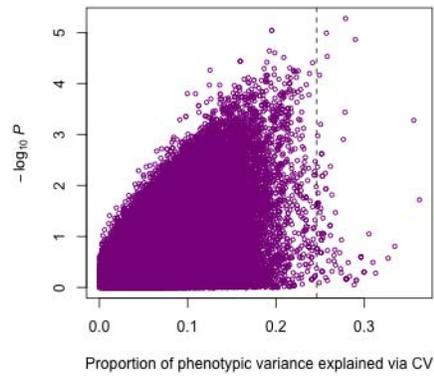

**Genome-wide association mapping via Wilcoxon test**

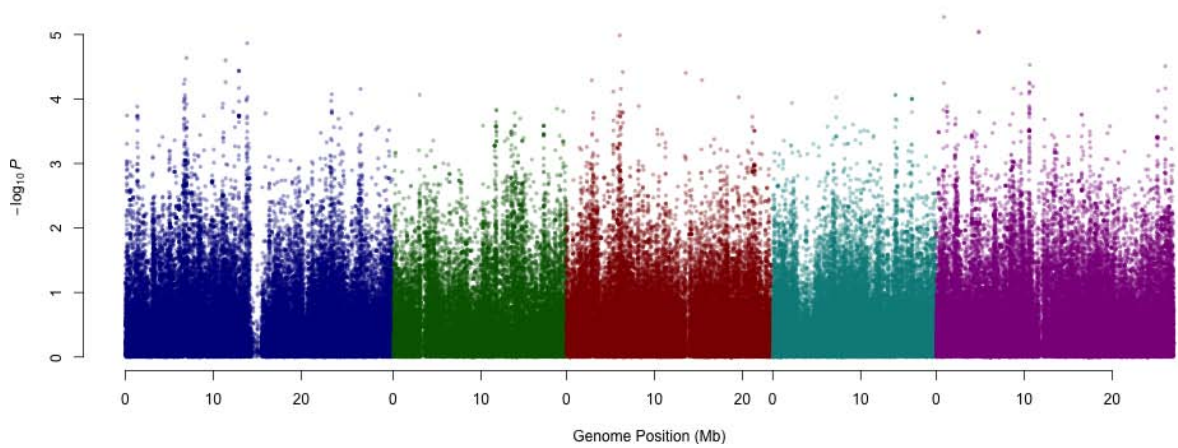

**Predictive ability assessed by cross validation**

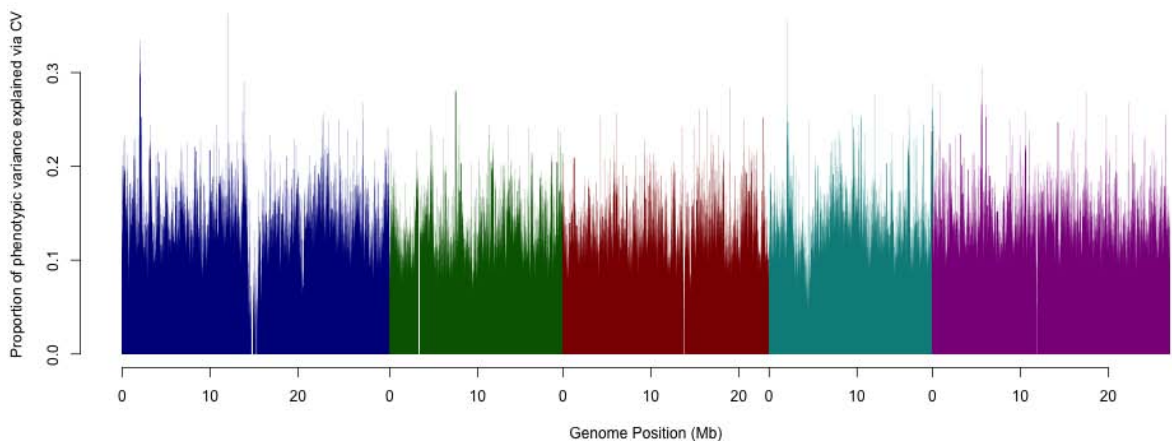

**Supplementary Figure 11** - Results of GWAS *p*-values and cross-validated predictive ability for Mn55

**Comparison of *p*-values and predictive ability**

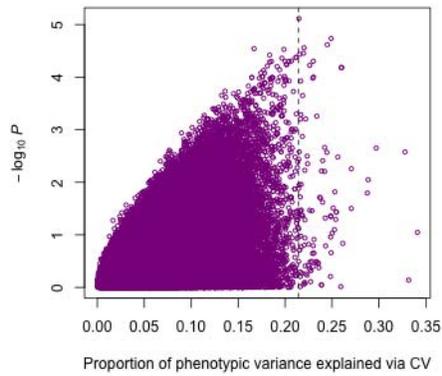

**Genome-wide association mapping via Wilcoxon test**

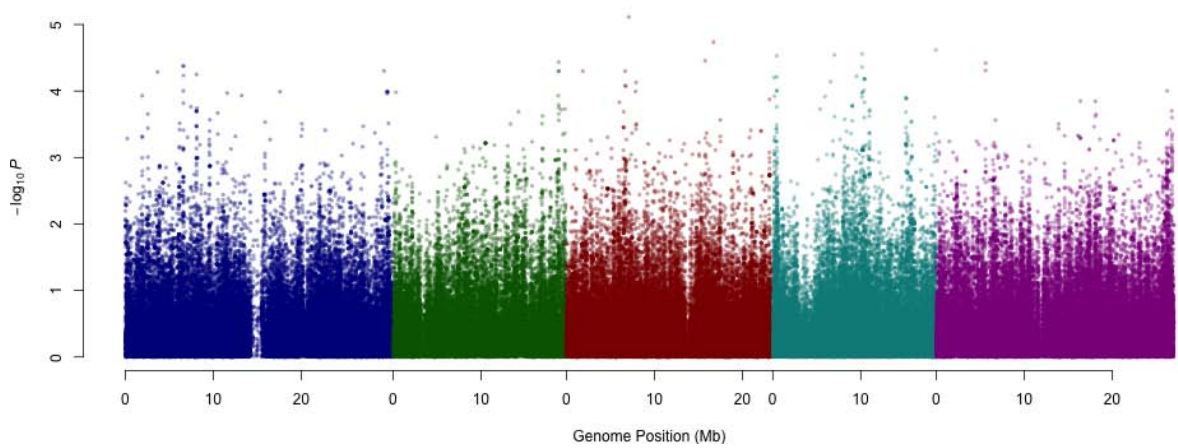

**Predictive ability assessed by cross validation**

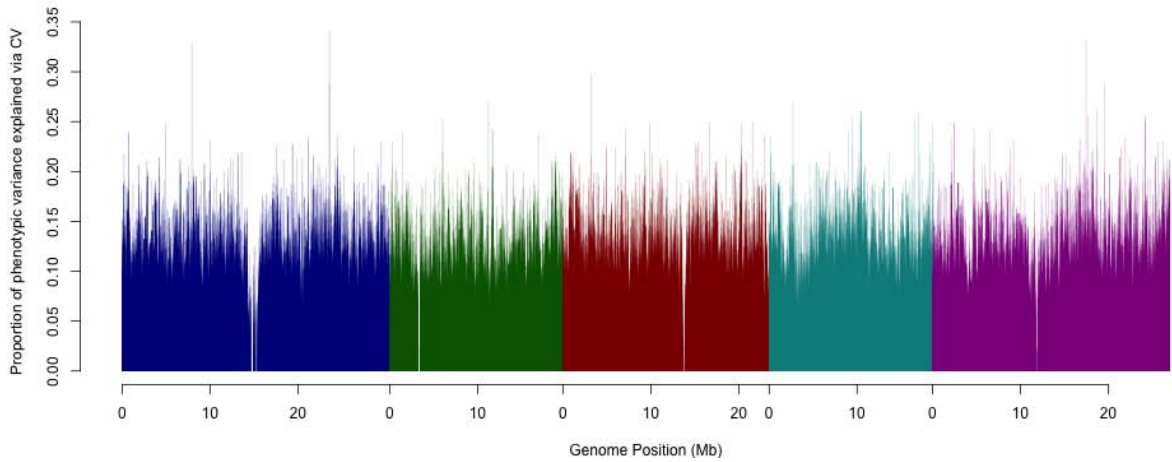

**Supplementary Figure 12** - Results of GWAS *p*-values and cross-validated predictive ability for Fe56

**Comparison of *p*-values and predictive ability**

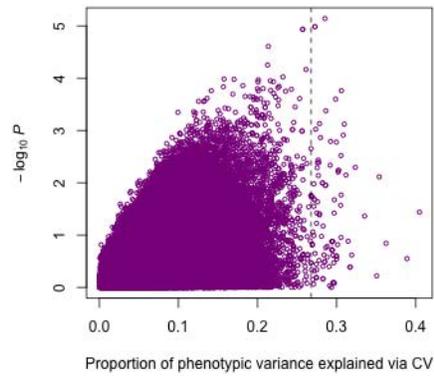

**Genome-wide association mapping via Wilcoxon test**

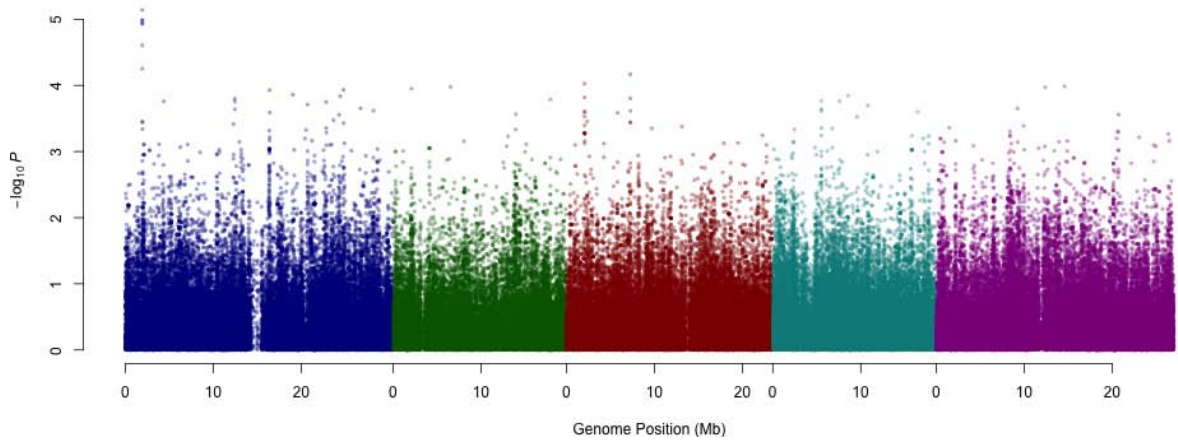

**Predictive ability assessed by cross validation**

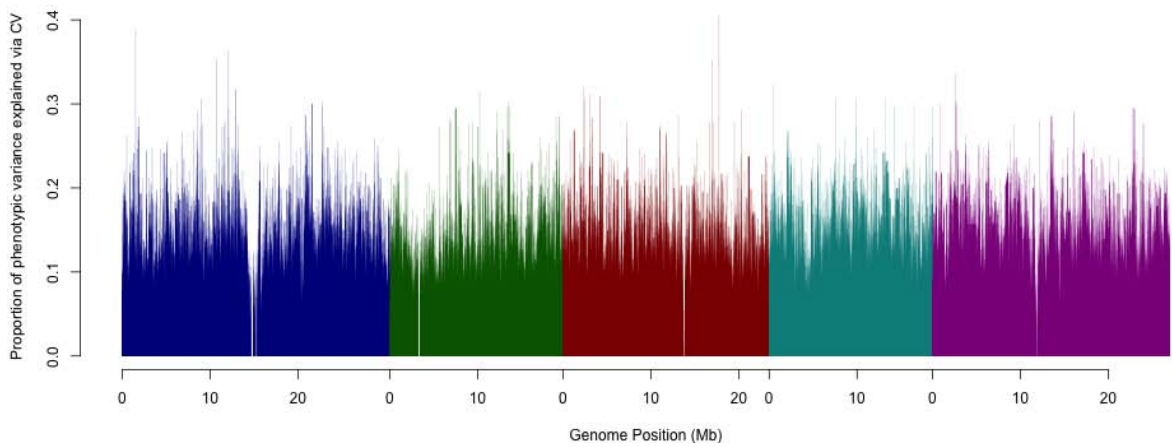

**Supplementary Figure 13** - Results of GWAS *p*-values and cross-validated predictive ability for Co59

**Comparison of *p*-values and predictive ability**

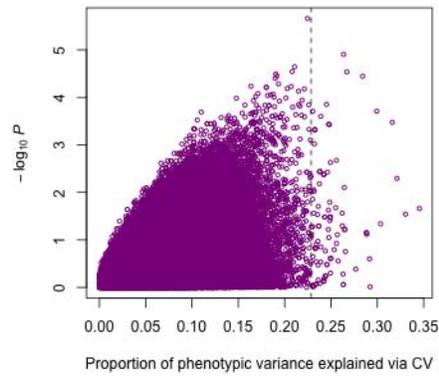

**Genome-wide association mapping via Wilcoxon test**

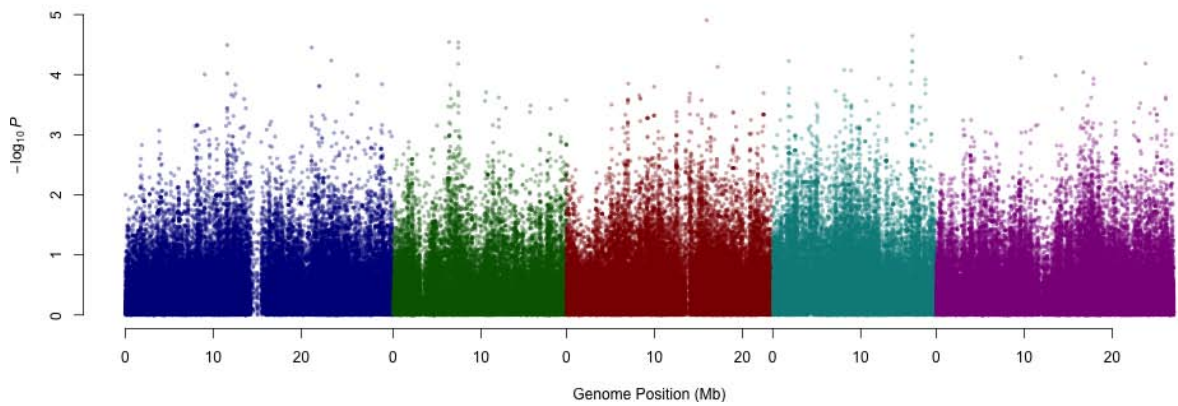

**Predictive ability assessed by cross validation**

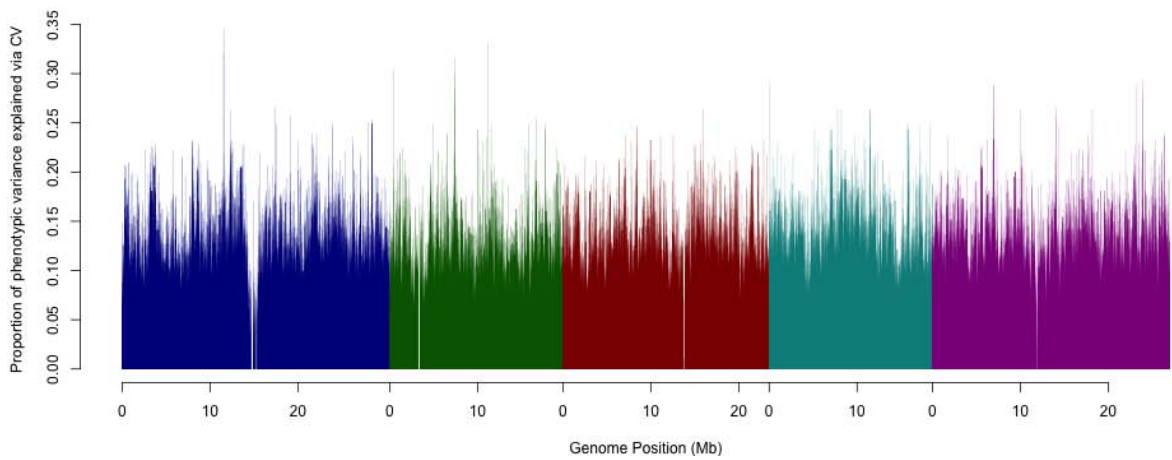

**Supplementary Figure 14** - Results of GWAS *p*-values and cross-validated predictive ability for Zn66

**Comparison of *p*-values and predictive ability**

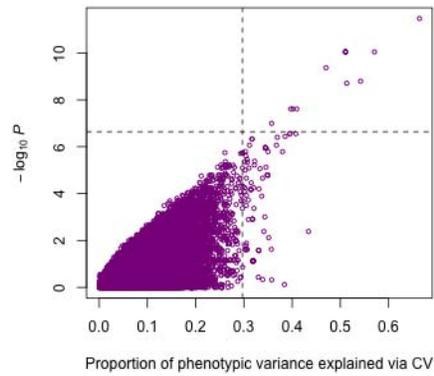

**Genome-wide association mapping via Wilcoxon test**

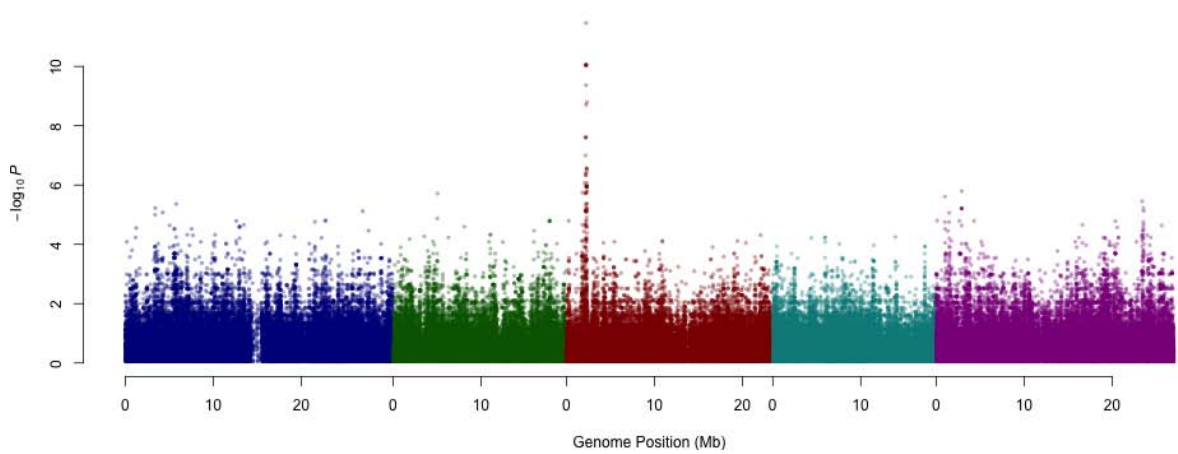

**Predictive ability assessed by cross validation**

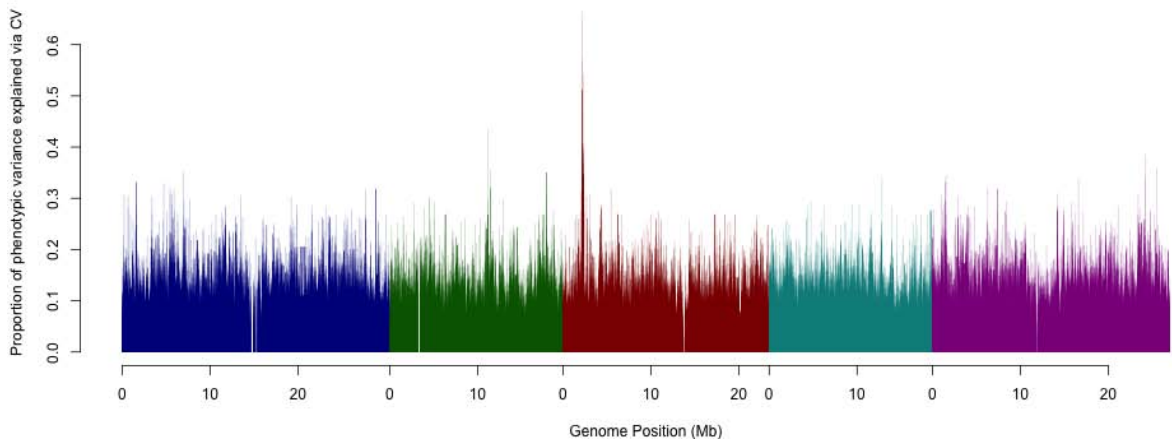

**Supplementary Figure 15** - Results of GWAS *p*-values and cross-validated predictive ability for avrRpm1

**Comparison of *p*-values and predictive ability**

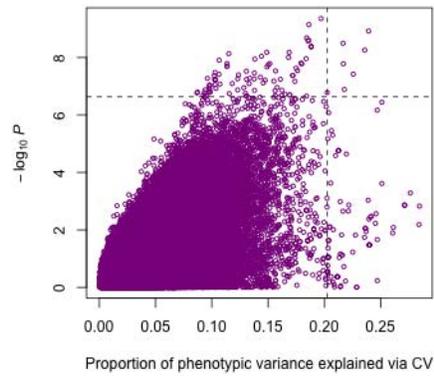

**Genome-wide association mapping via Wilcoxon test**

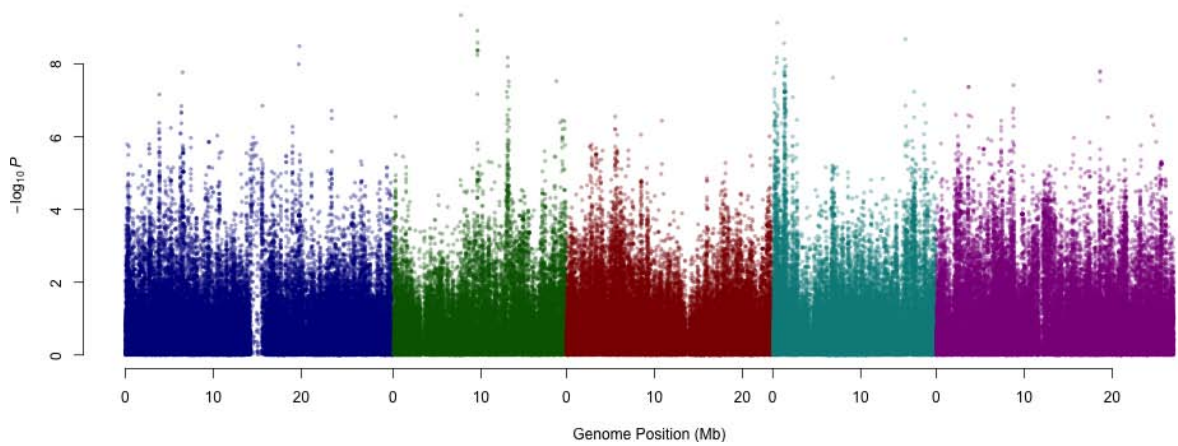

**Predictive ability assessed by cross validation**

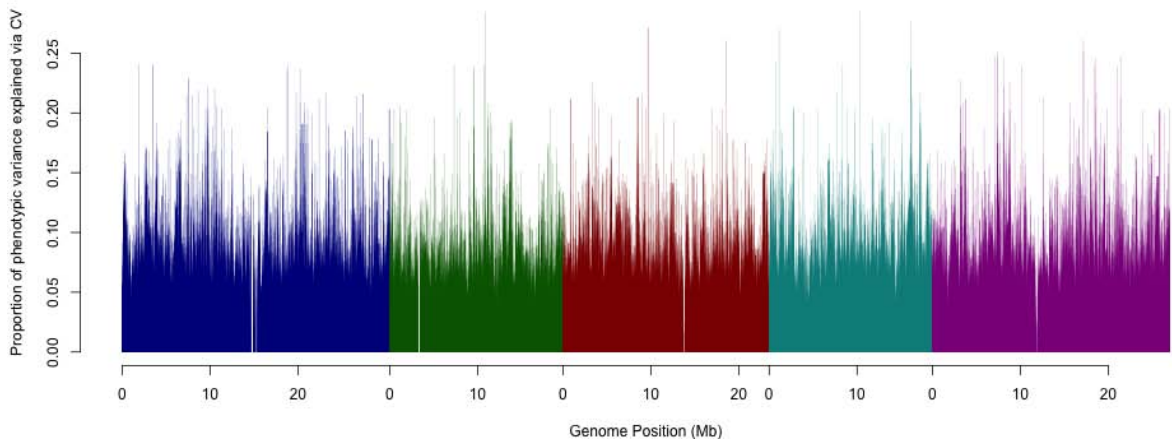

**Supplementary Figure 16** - Results of GWAS *p*-values and cross-validated predictive ability for FLC

**Comparison of *p*-values and predictive ability**

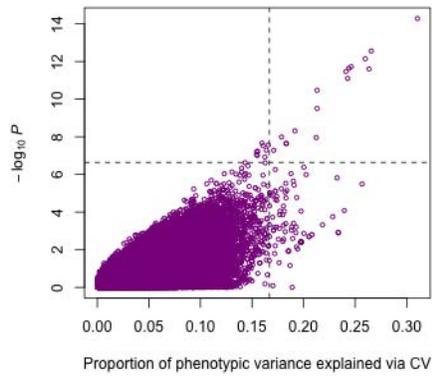

**Genome-wide association mapping via Wilcoxon test**

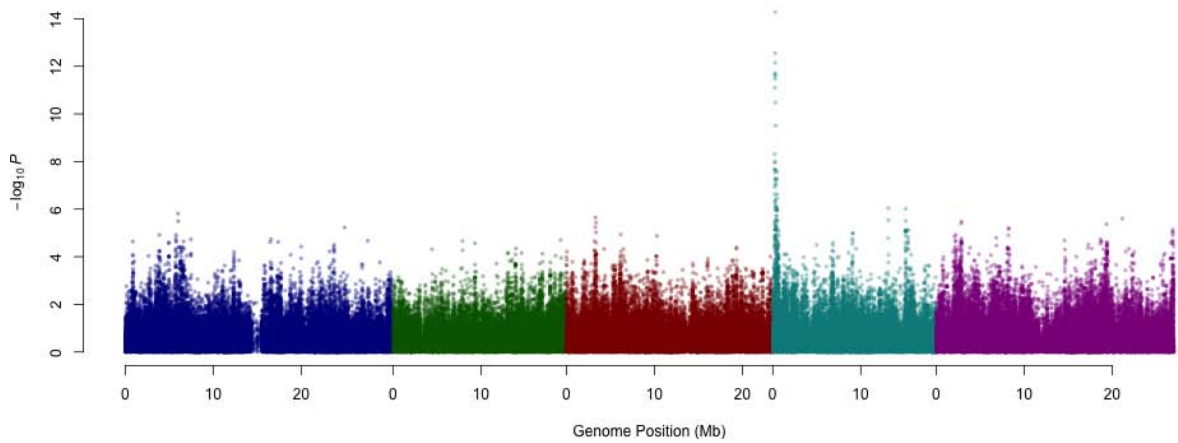

**Predictive ability assessed by cross validation**

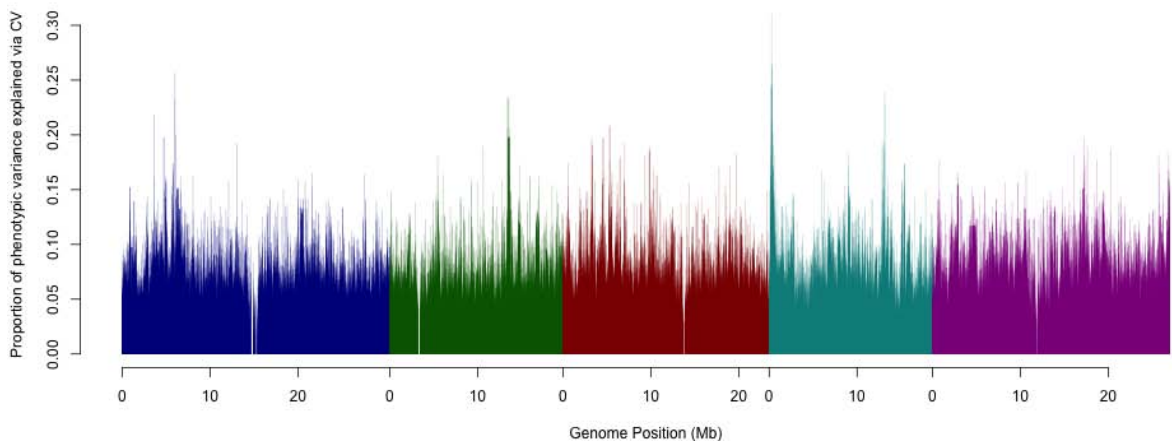

**Supplementary Figure 17** - Results of GWAS *p*-values and cross-validated predictive ability for FRI

**Comparison of *p*-values and predictive ability**

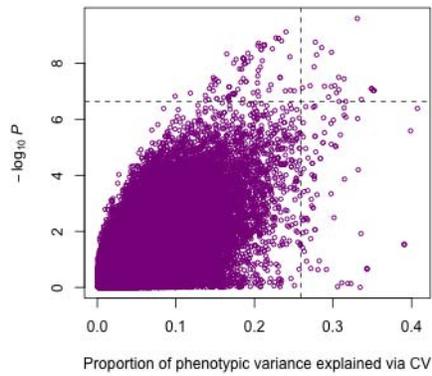

**Genome-wide association mapping via Wilcoxon test**

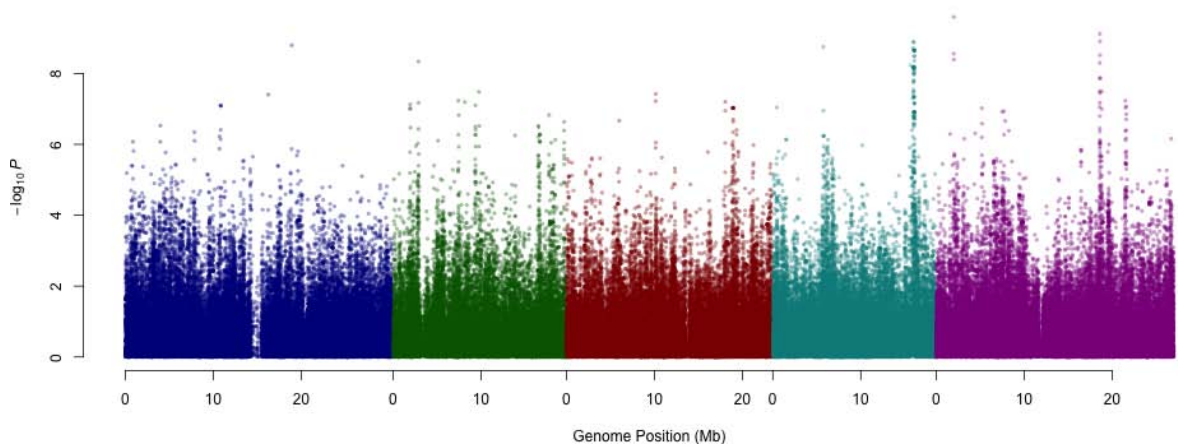

**Predictive ability assessed by cross validation**

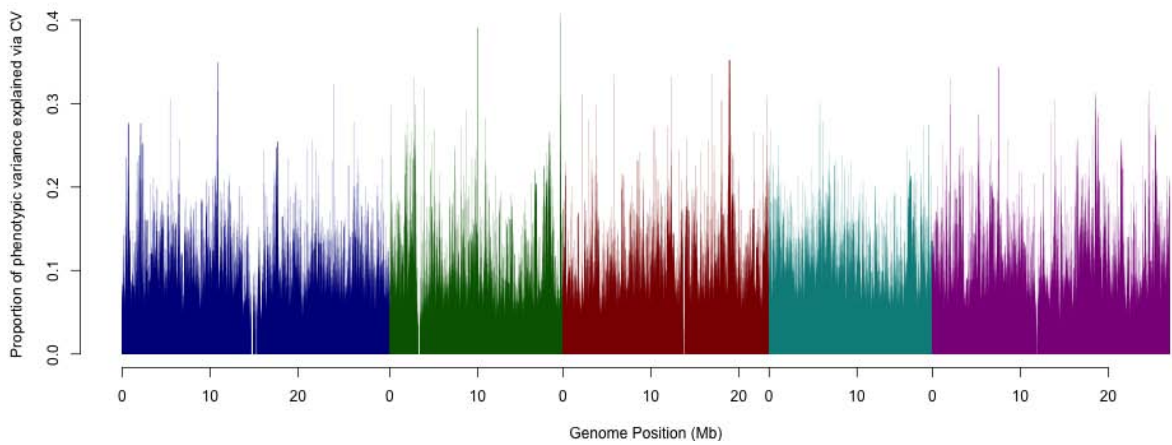

**Supplementary Figure 18** - Results of GWAS *p*-values and cross-validated predictive ability for 8W GH LN

**Comparison of *p*-values and predictive ability**

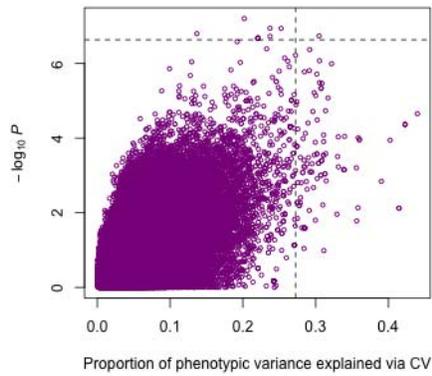

**Genome-wide association mapping via Wilcoxon test**

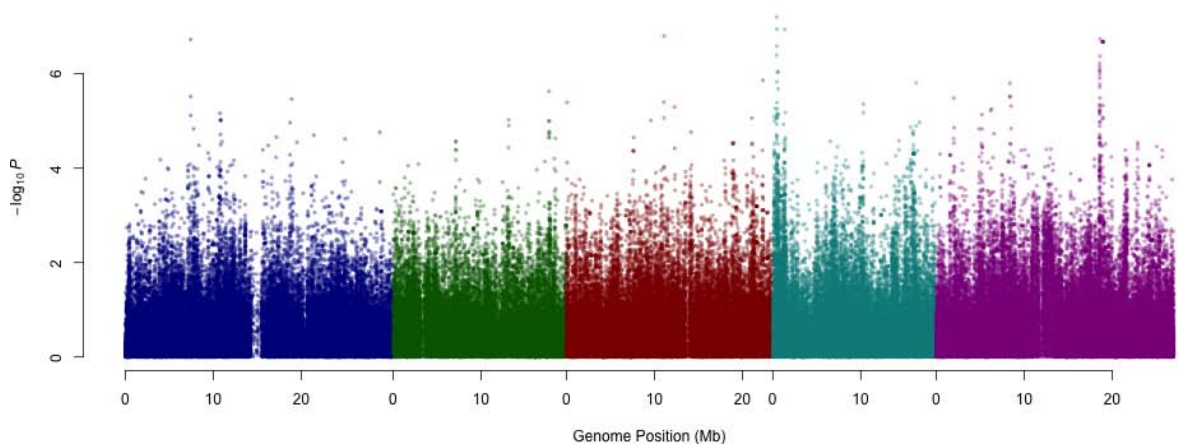

**Predictive ability assessed by cross validation**

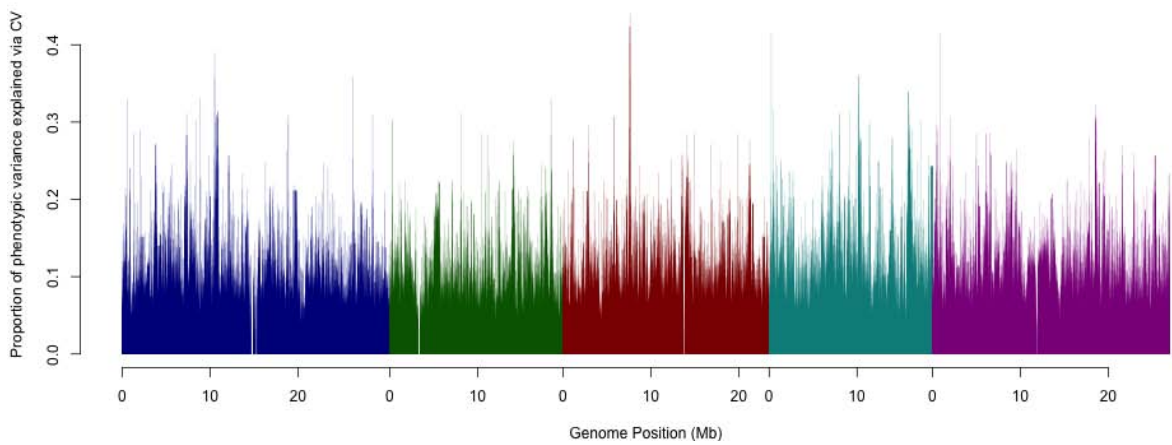

**Supplementary Figure 19** - Results of GWAS *p*-values and cross-validated predictive ability for 0W GH LN

**Comparison of *p*-values and predictive ability**

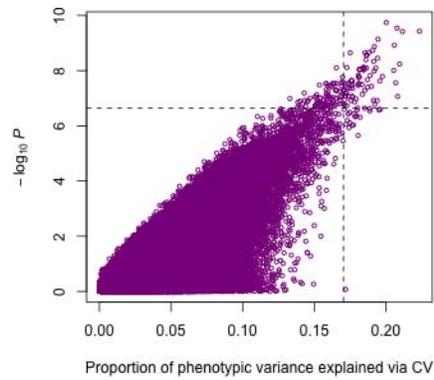

**Genome-wide association mapping via Wilcoxon test**

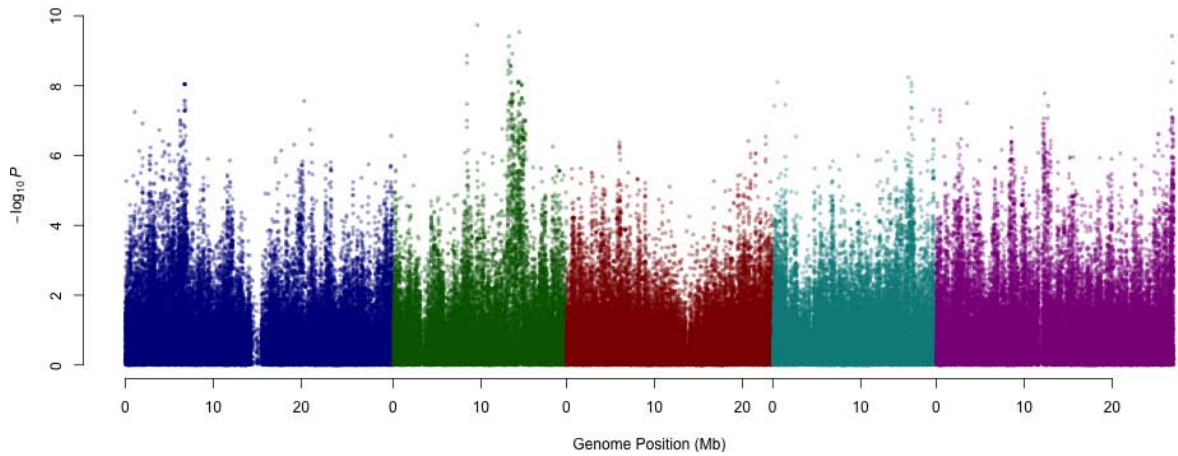

**Predictive ability assessed by cross validation**

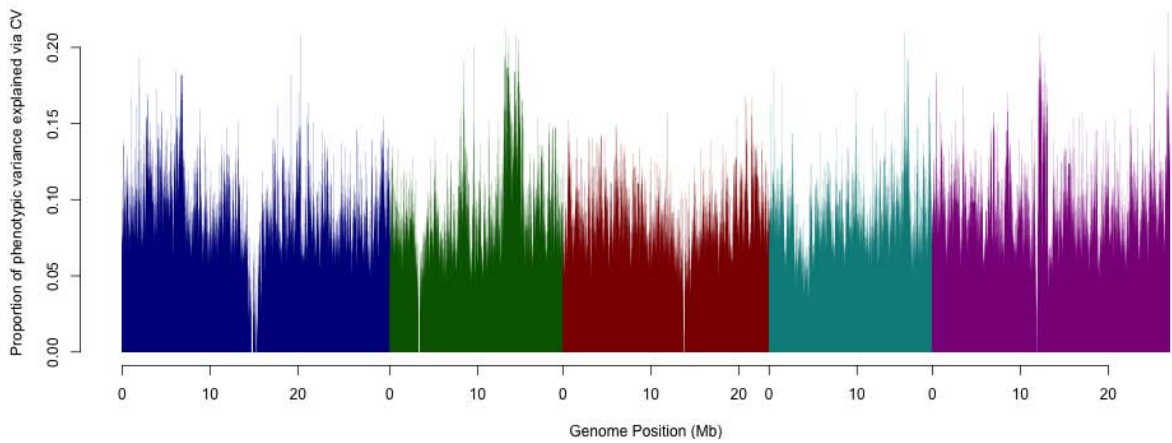

**Supplementary Figure 20** - Results of GWAS *p*-values and cross-validated predictive ability for FT Diameter Field

**Comparison of *p*-values and predictive ability**

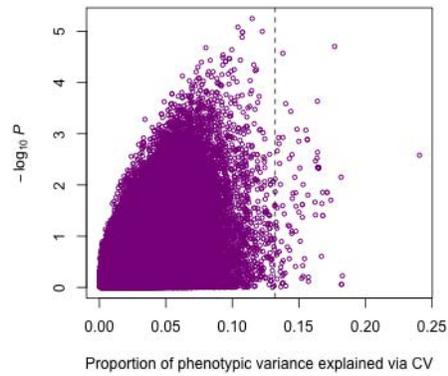

**Genome-wide association mapping via Wilcoxon test**

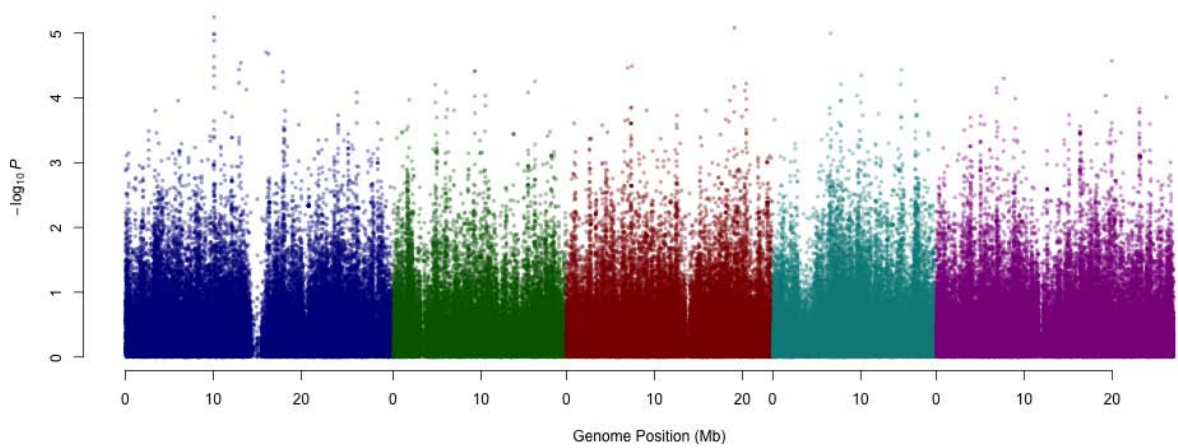

**Predictive ability assessed by cross validation**

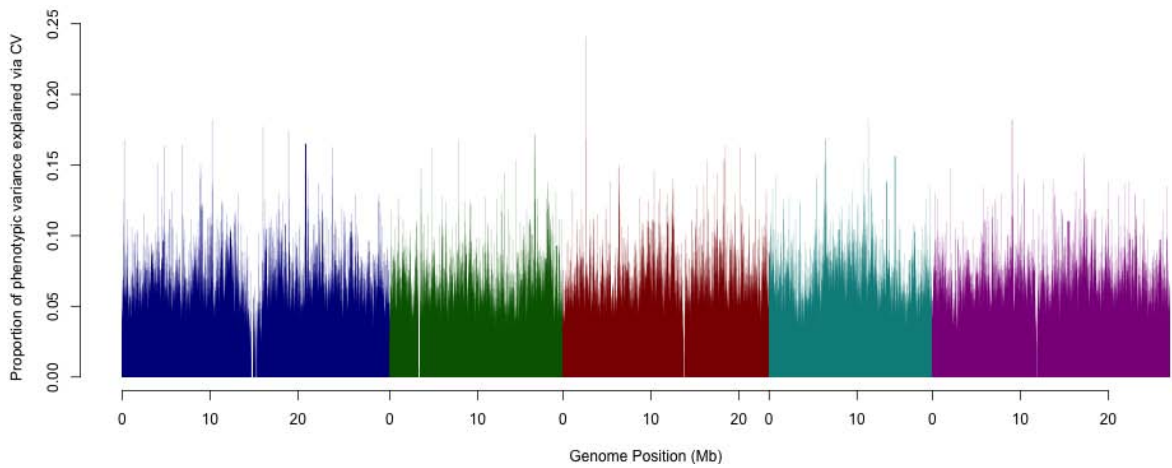

**Supplementary Figure 21** - Results of GWAS *p*-values and cross-validated predictive ability for At1

**Comparison of *p*-values and predictive ability**

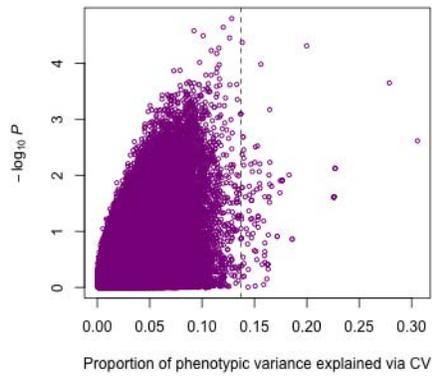

**Genome-wide association mapping via Wilcoxon test**

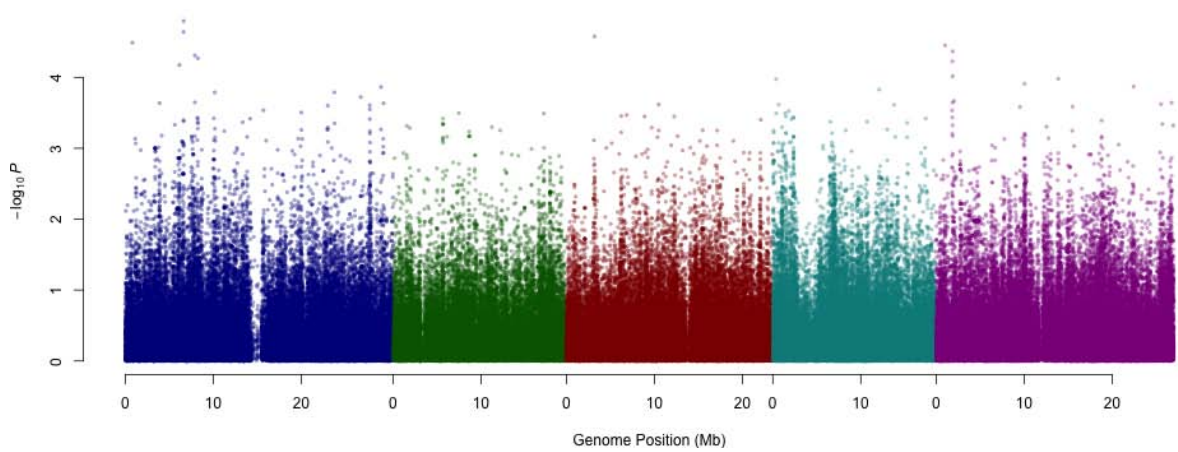

**Predictive ability assessed by cross validation**

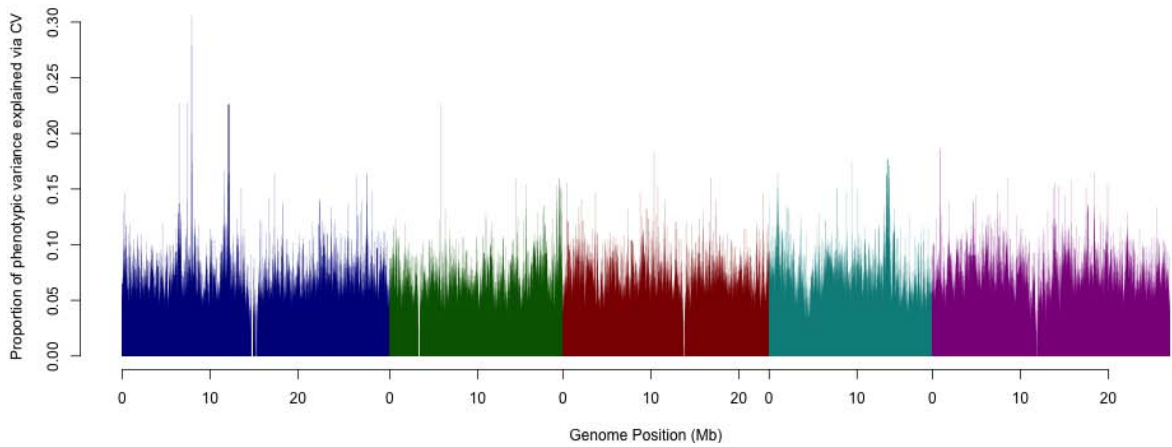

**Supplementary Figure 22** - Results of GWAS *p*-values and cross-validated predictive ability for At1 CFU2

**Comparison of *p*-values and predictive ability**

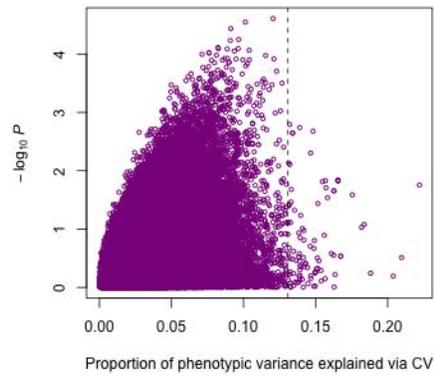

**Genome-wide association mapping via Wilcoxon test**

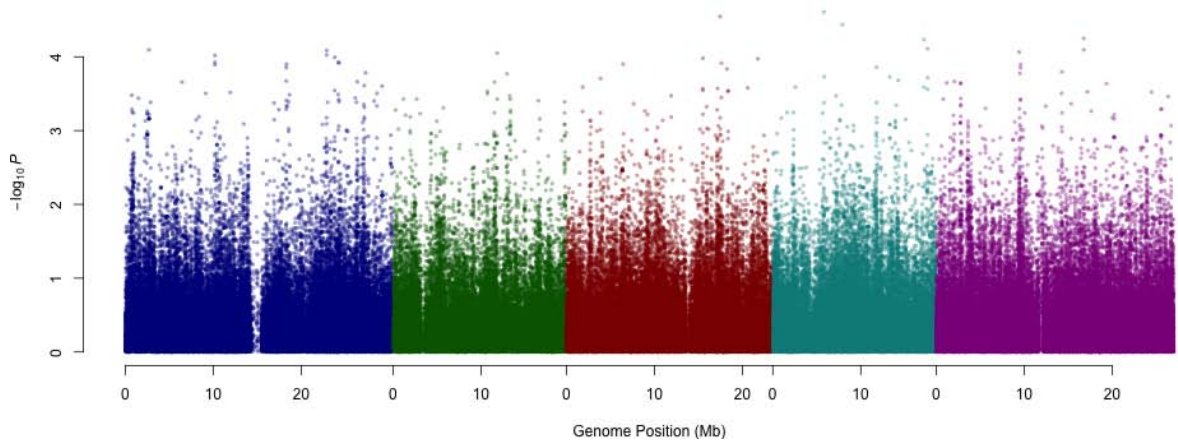

**Predictive ability assessed by cross validation**

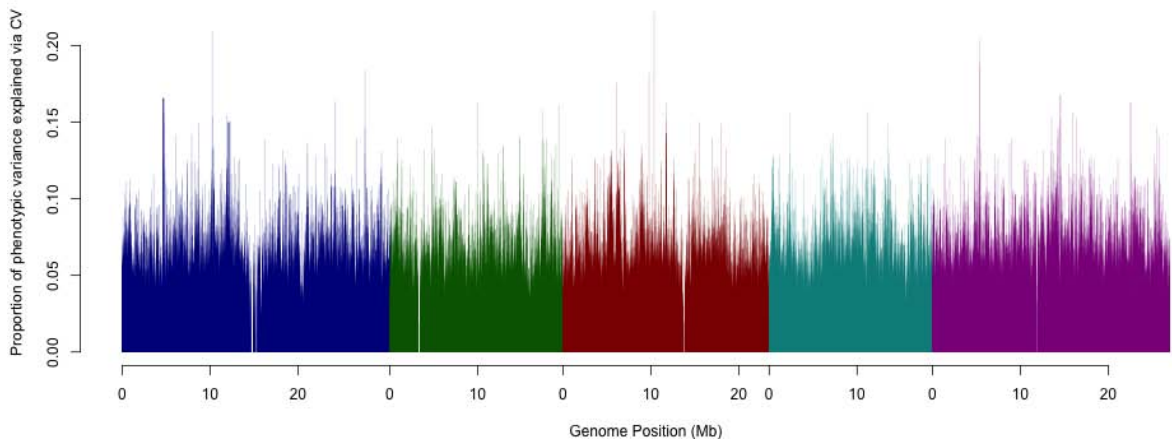

**Supplementary Figure 23** - Results of GWAS *p*-values and cross-validated predictive ability for As CFU2

**Comparison of *p*-values and predictive ability**

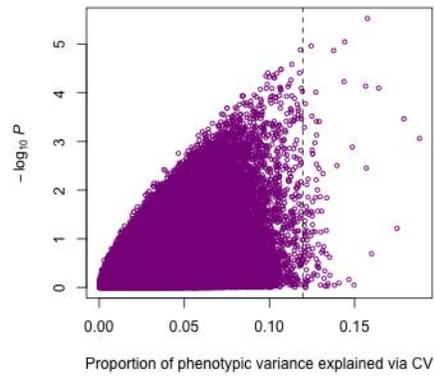

**Genome-wide association mapping via Wilcoxon test**

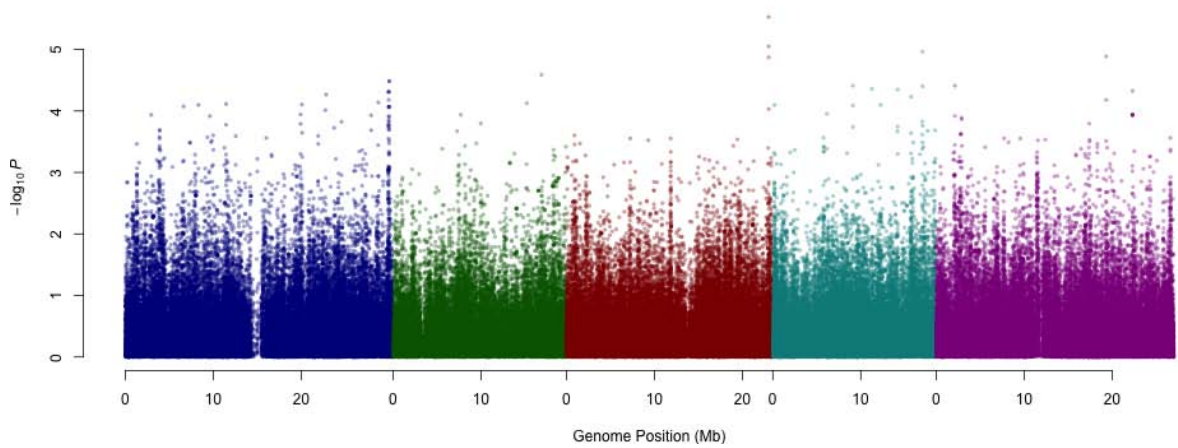

**Predictive ability assessed by cross validation**

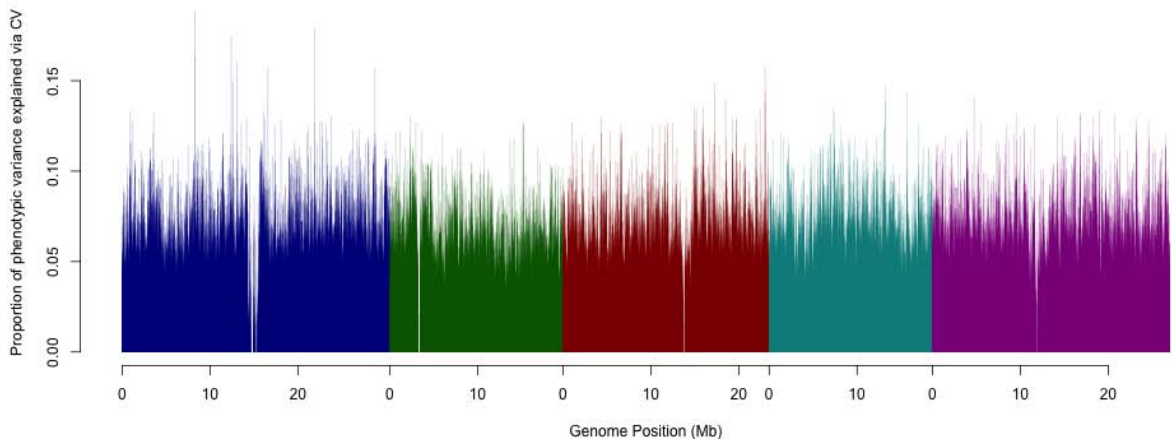

**Supplementary Figure 24** - Results of GWAS *p*-values and cross-validated predictive ability for Bs

**Comparison of *p*-values and predictive ability**

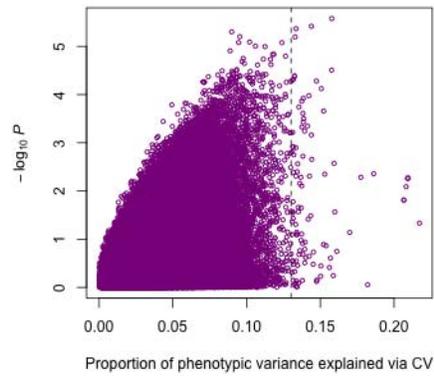

**Genome-wide association mapping via Wilcoxon test**

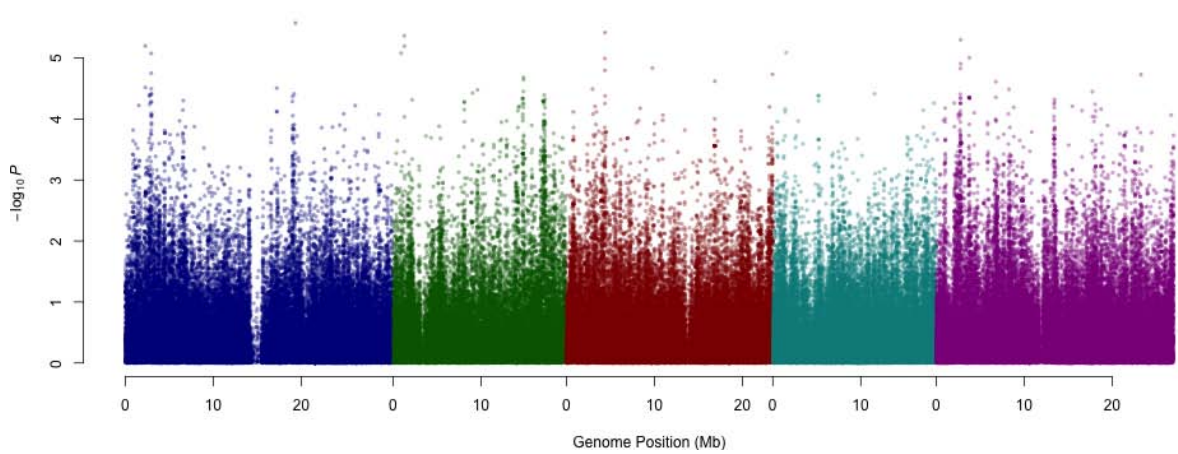

**Predictive ability assessed by cross validation**

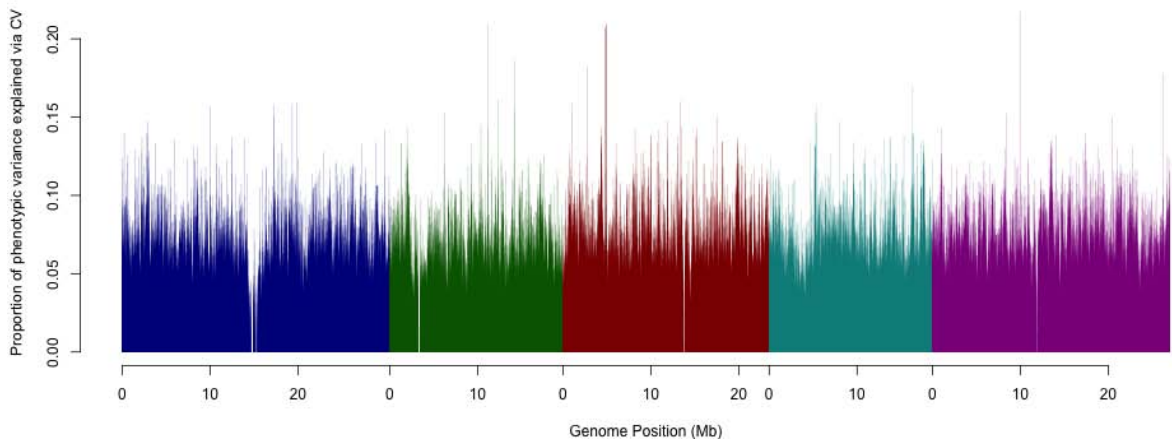

**Supplementary Figure 25** - Results of GWAS *p*-values and cross-validated predictive ability for Bs CFU2

**Comparison of *p*-values and predictive ability**

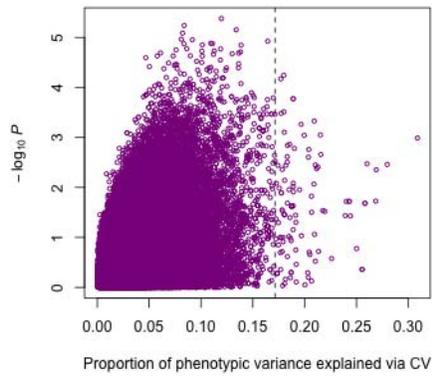

**Genome-wide association mapping via Wilcoxon test**

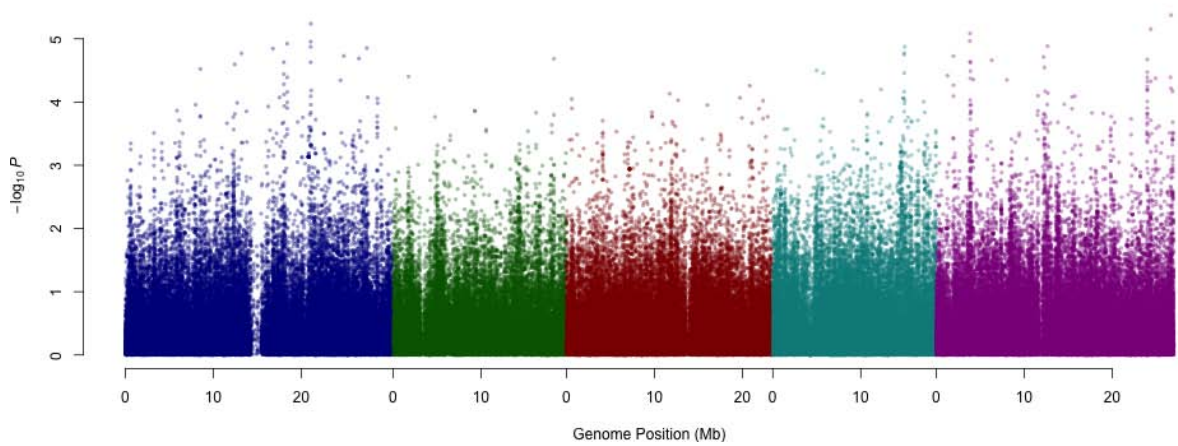

**Predictive ability assessed by cross validation**

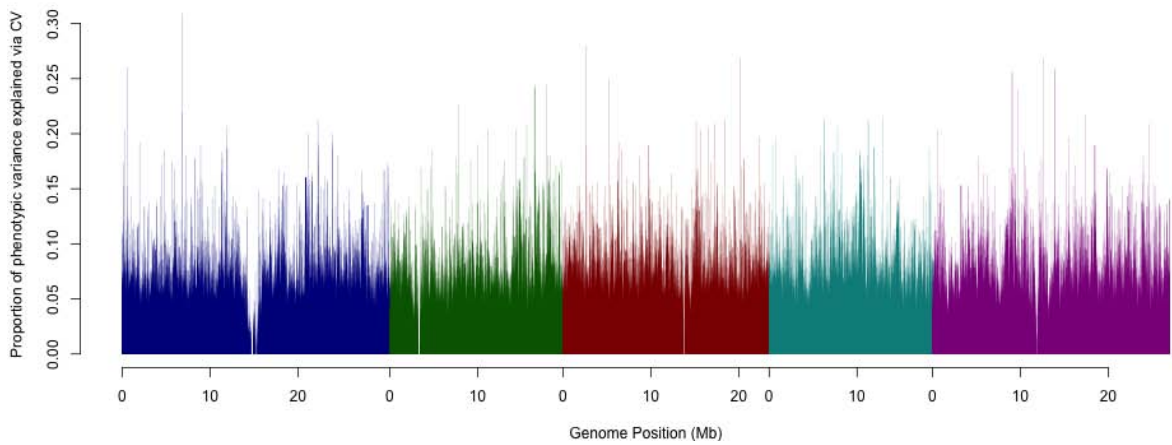

**Supplementary Figure 26** - Results of GWAS *p*-values and cross-validated predictive ability for At2

**Comparison of *p*-values and predictive ability**

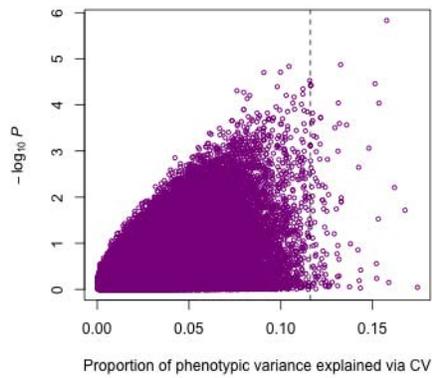

**Genome-wide association mapping via Wilcoxon test**

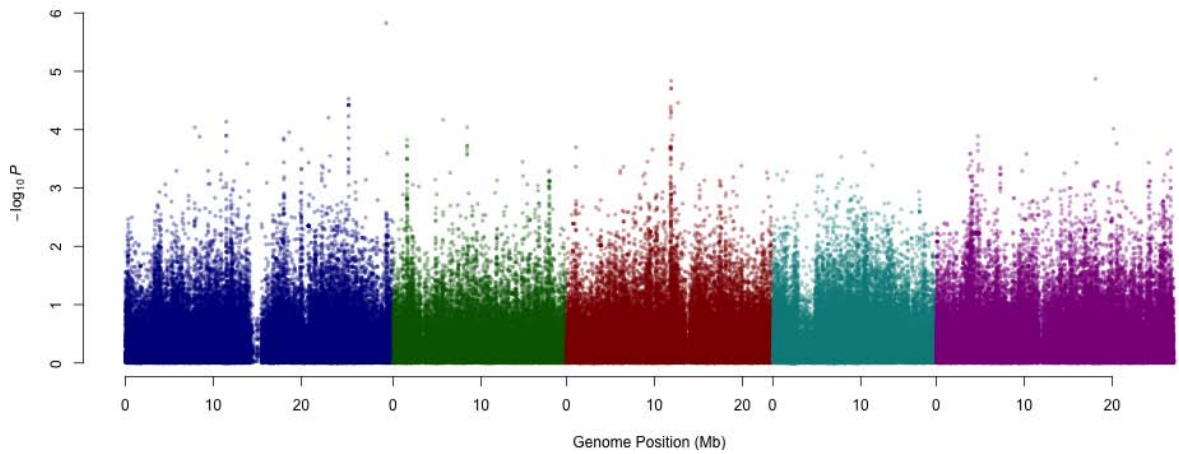

**Predictive ability assessed by cross validation**

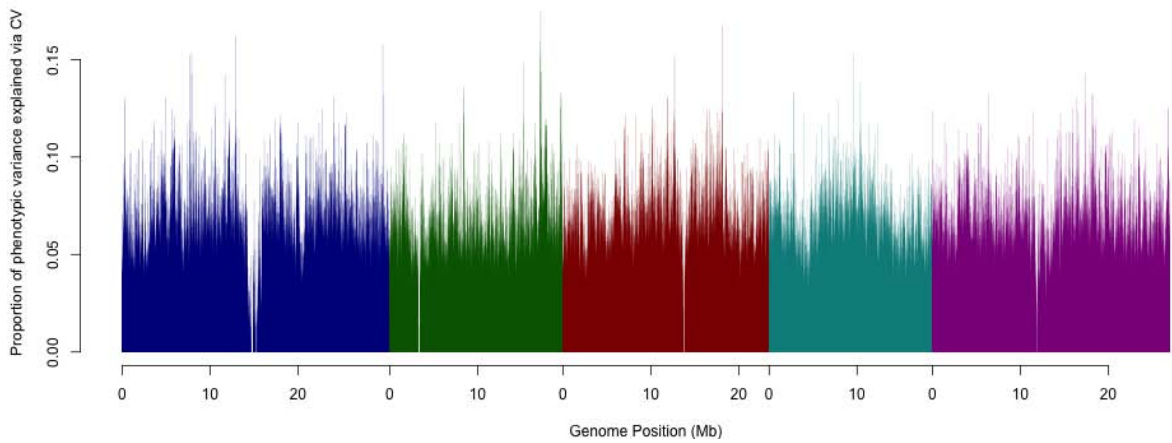

**Supplementary Figure 27** - Results of GWAS *p*-values and cross-validated predictive ability for At2 CFU2

**Comparison of *p*-values and predictive ability**

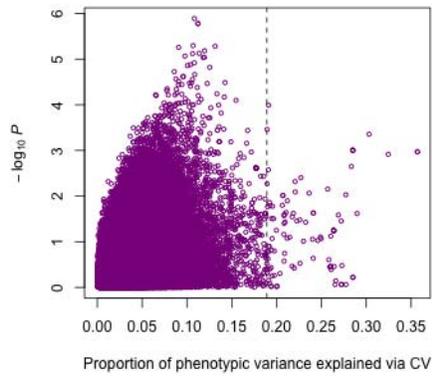

**Genome-wide association mapping via Wilcoxon test**

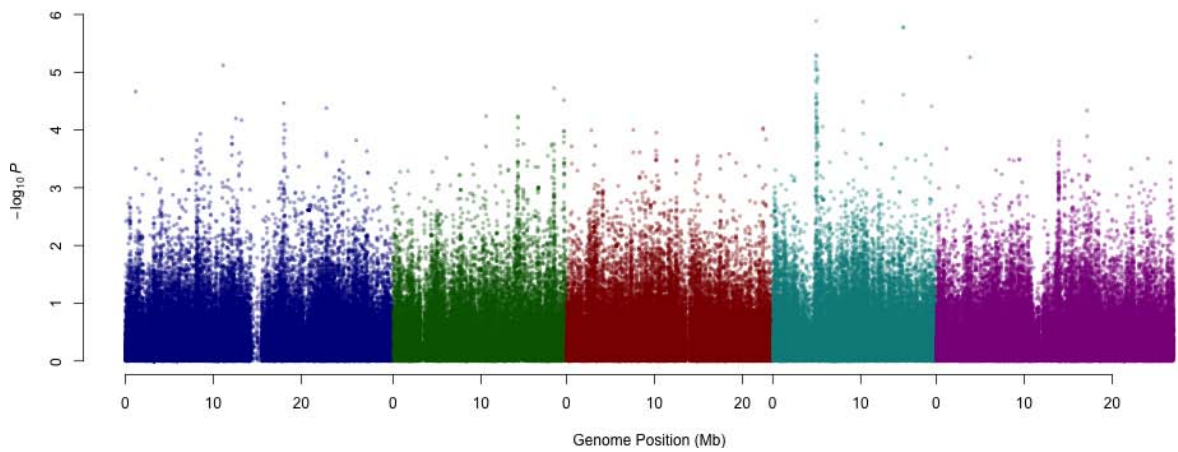

**Predictive ability assessed by cross validation**

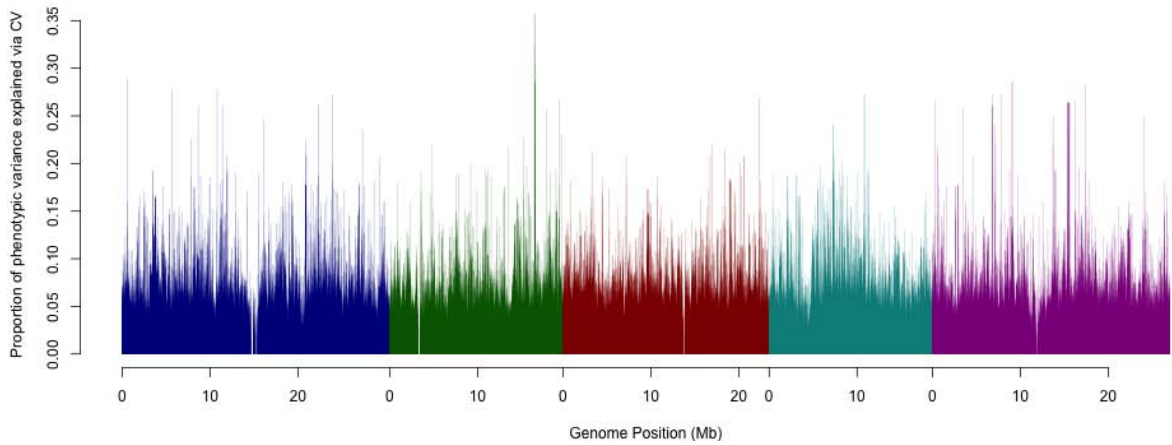

**Supplementary Figure 28** - Results of GWAS *p*-values and cross-validated predictive ability for As2

**Comparison of *p*-values and predictive ability**

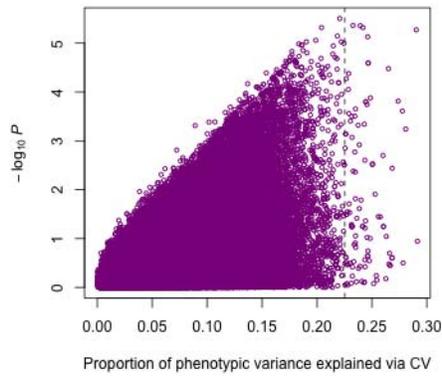

**Genome-wide association mapping via Wilcoxon test**

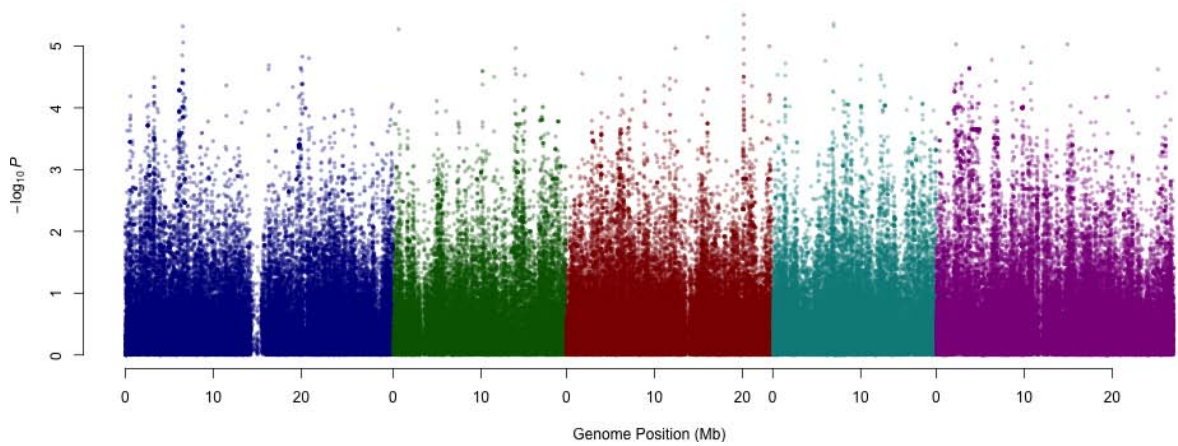

**Predictive ability assessed by cross validation**

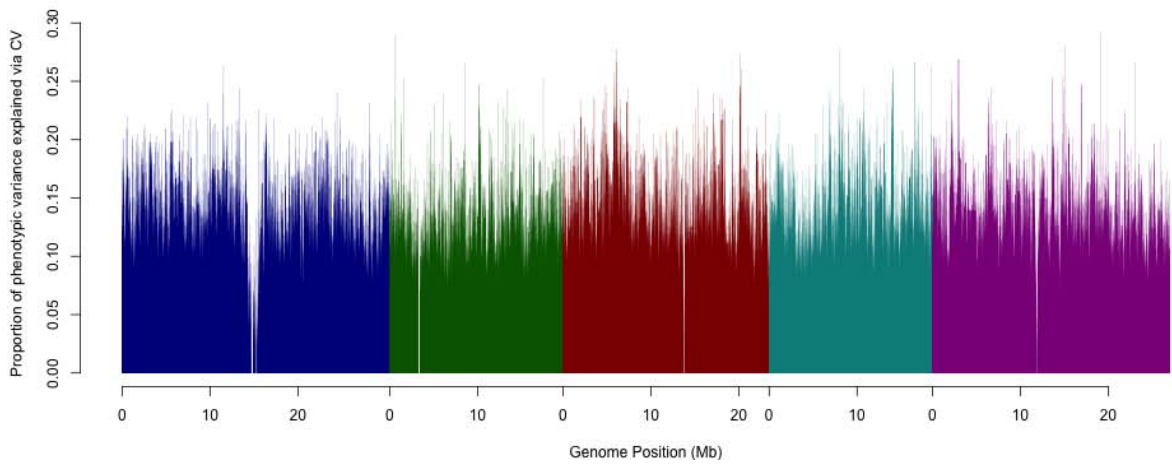

**Supplementary Figure 29** - Results of GWAS *p*-values and cross-validated predictive ability for DW

**Comparison of *p*-values and predictive ability**

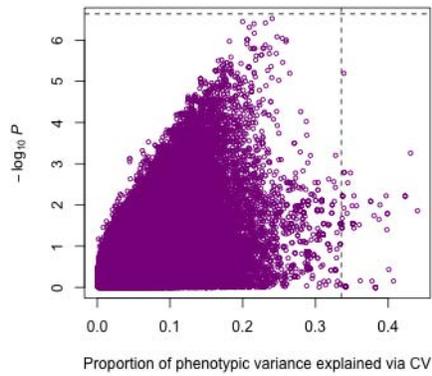

**Genome-wide association mapping via Wilcoxon test**

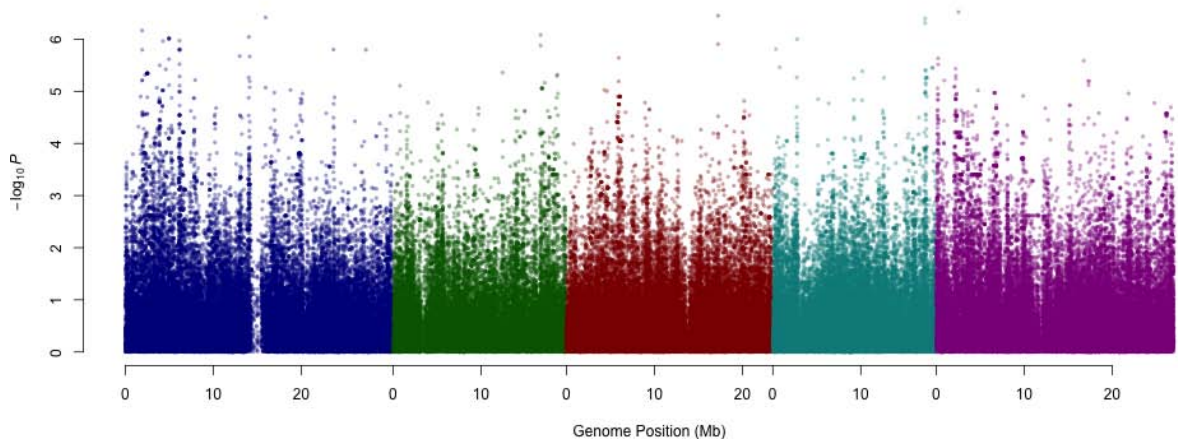

**Predictive ability assessed by cross validation**

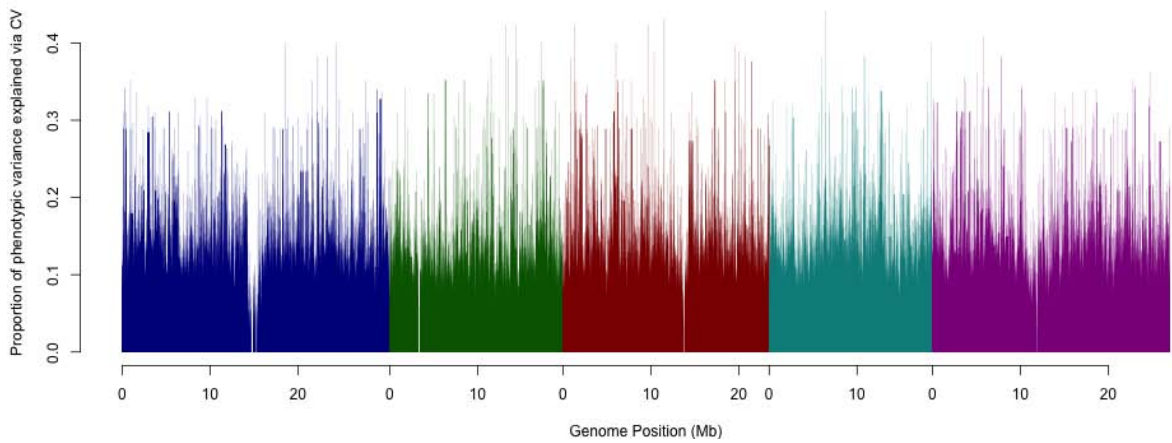

**Supplementary Figure 30** - Results of GWAS *p*-values and cross-validated predictive ability for Silique 22

## Comparison of *p*-values and predictive ability

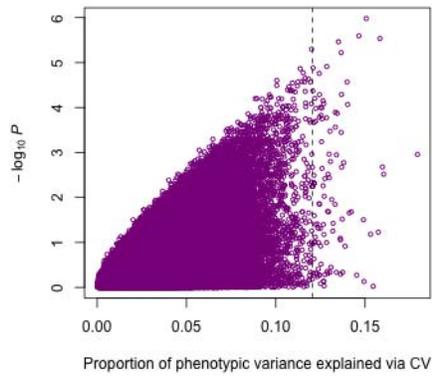

## Genome-wide association mapping via Wilcoxon test

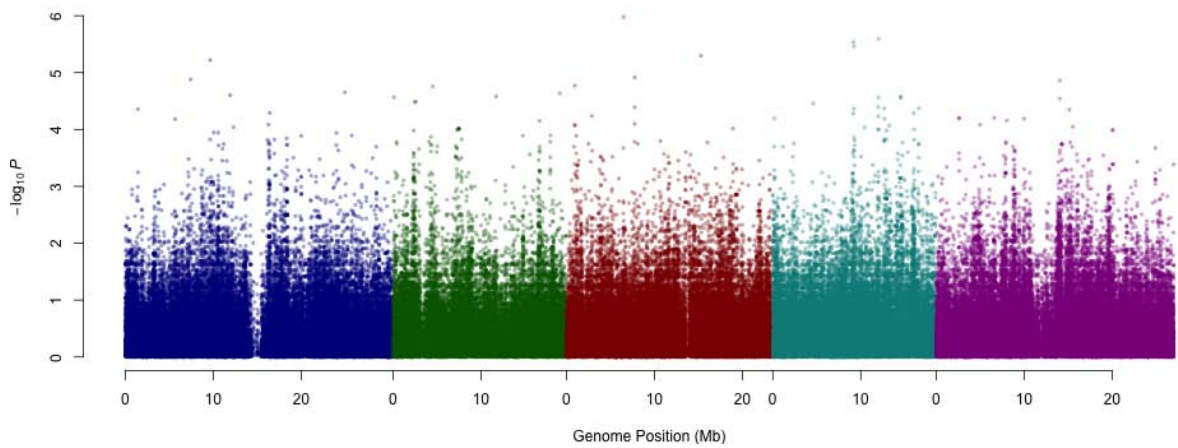

## Predictive ability assessed by cross validation

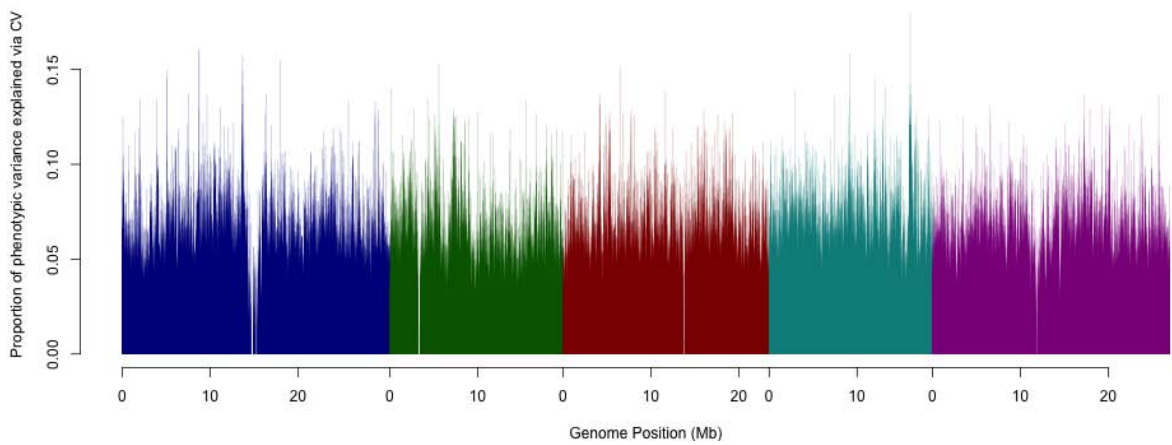

**Supplementary Figure 31** - Results of GWAS *p*-values and cross-validated predictive ability for Germ 10

**Comparison of *p*-values and predictive ability**

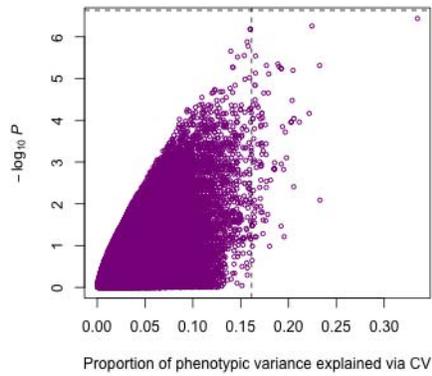

**Genome-wide association mapping via Wilcoxon test**

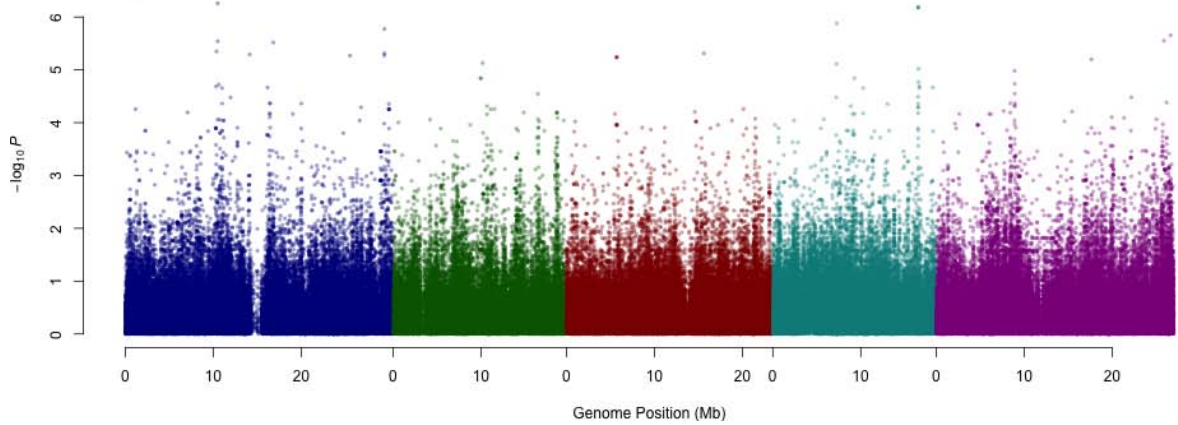

**Predictive ability assessed by cross validation**

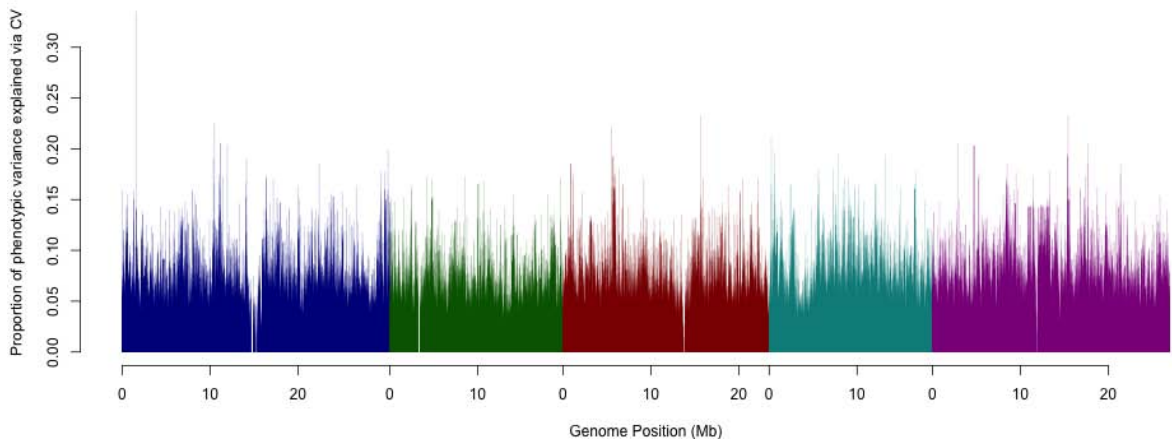

**Supplementary Figure 32** - Results of GWAS *p*-values and cross-validated predictive ability for Germ 16

**Comparison of *p*-values and predictive ability**

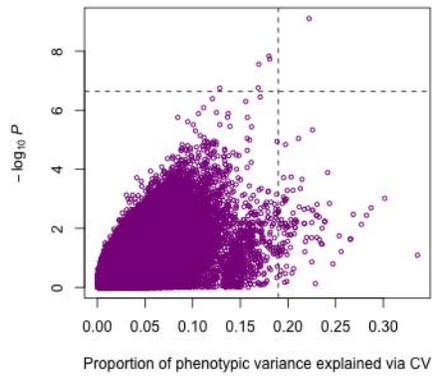

**Genome-wide association mapping via Wilcoxon test**

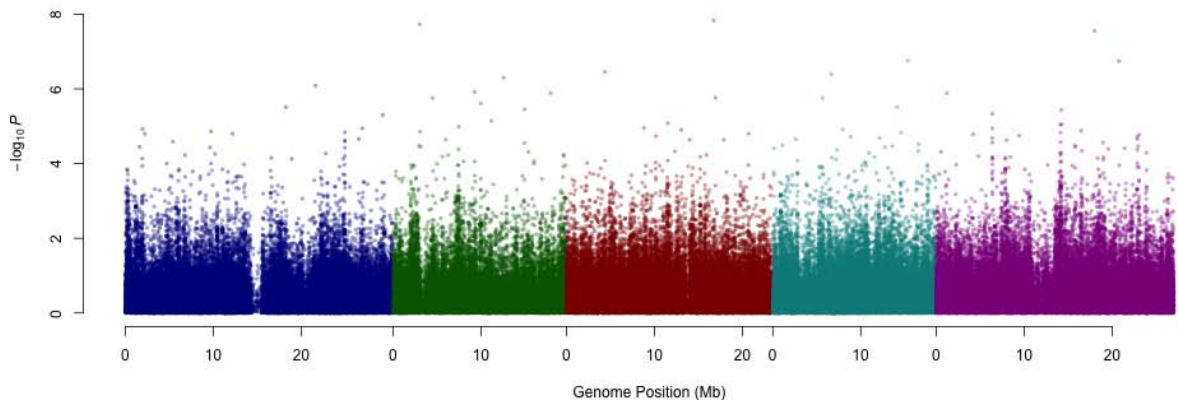

**Predictive ability assessed by cross validation**

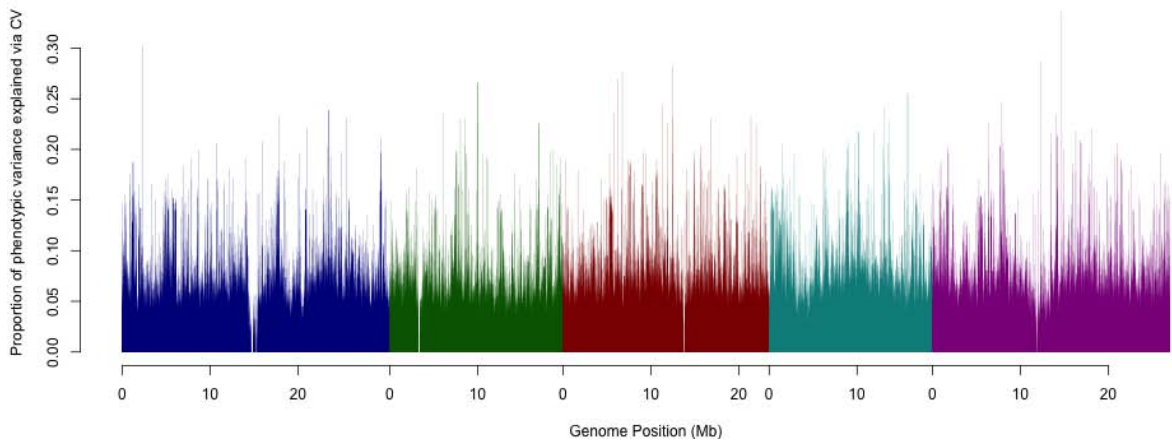

**Supplementary Figure 33** - Results of GWAS *p*-values and cross-validated predictive ability for Width 10

**Comparison of *p*-values and predictive ability**

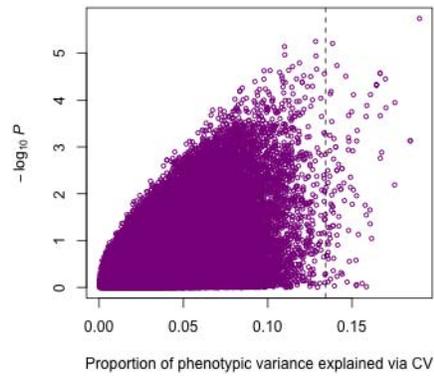

**Genome-wide association mapping via Wilcoxon test**

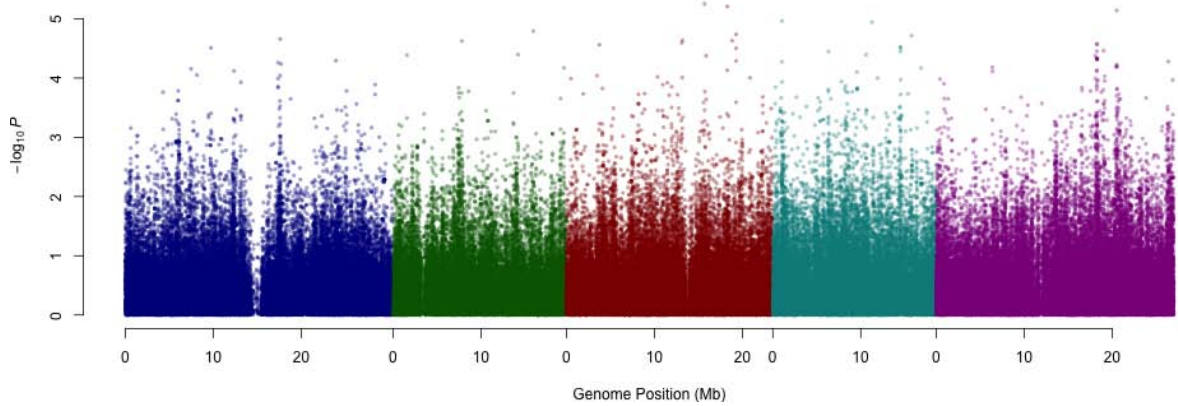

**Predictive ability assessed by cross validation**

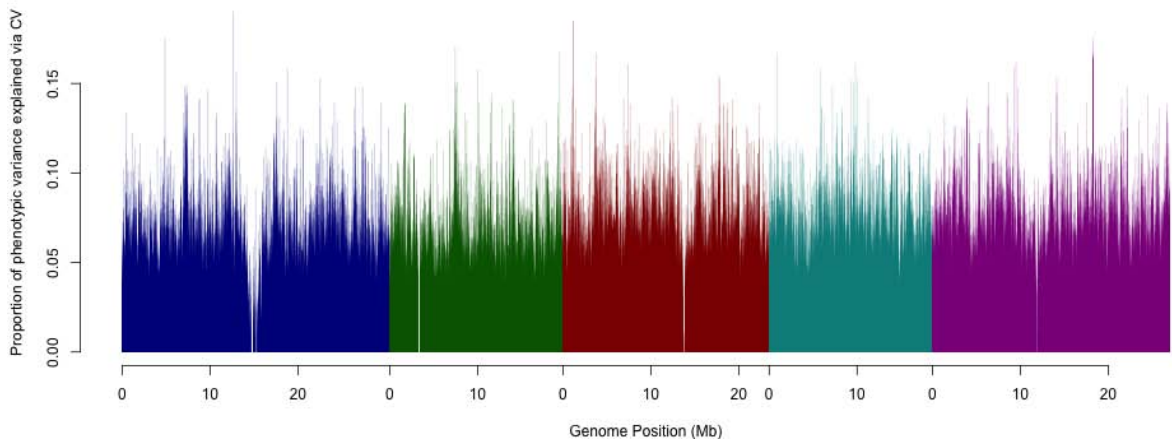

**Supplementary Figure 34** - Results of GWAS *p*-values and cross-validated predictive ability for Width 16

**Comparison of *p*-values and predictive ability**

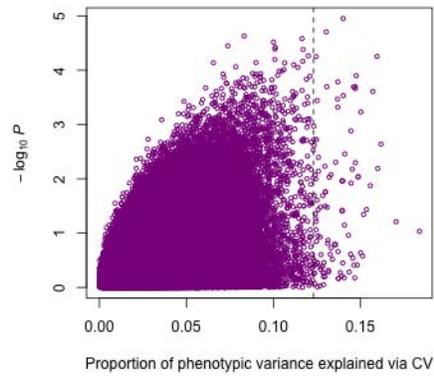

**Genome-wide association mapping via Wilcoxon test**

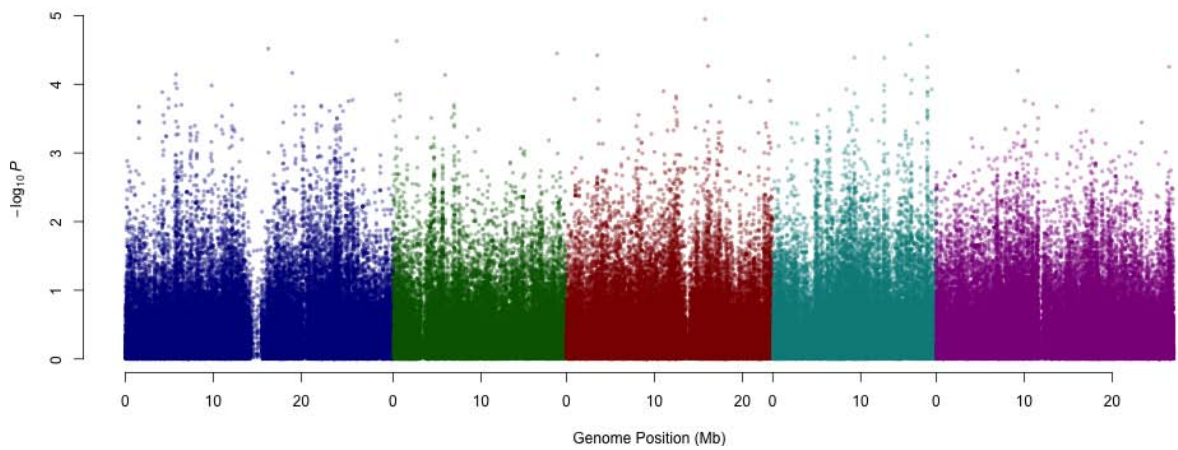

**Predictive ability assessed by cross validation**

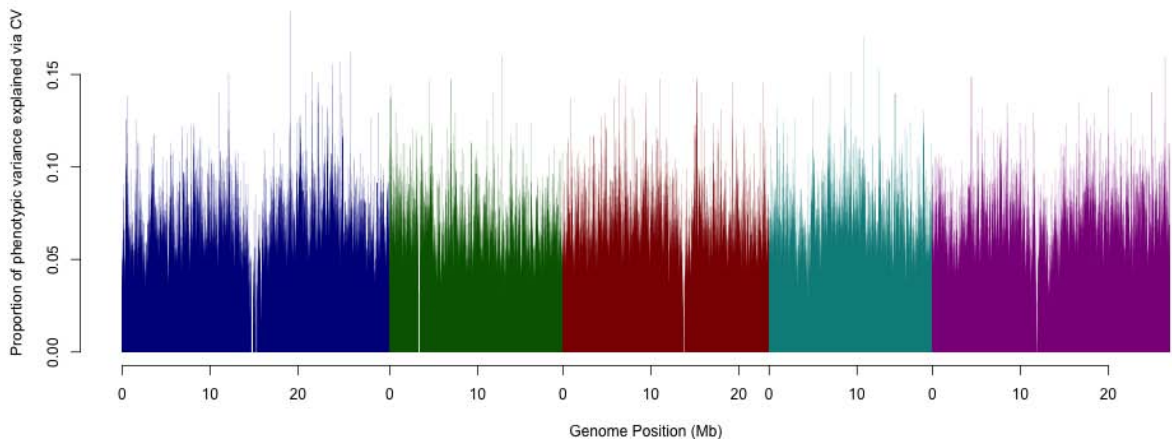

**Supplementary Figure 35** - Results of GWAS *p*-values and cross-validated predictive ability for Width 22

**Comparison of *p*-values and predictive ability**

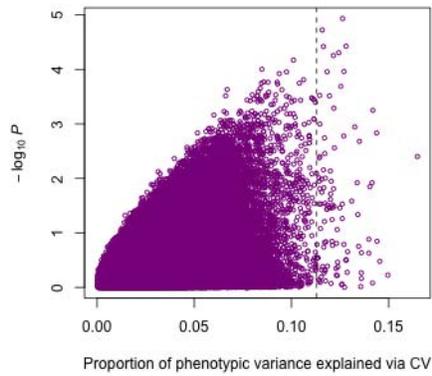

**Genome-wide association mapping via Wilcoxon test**

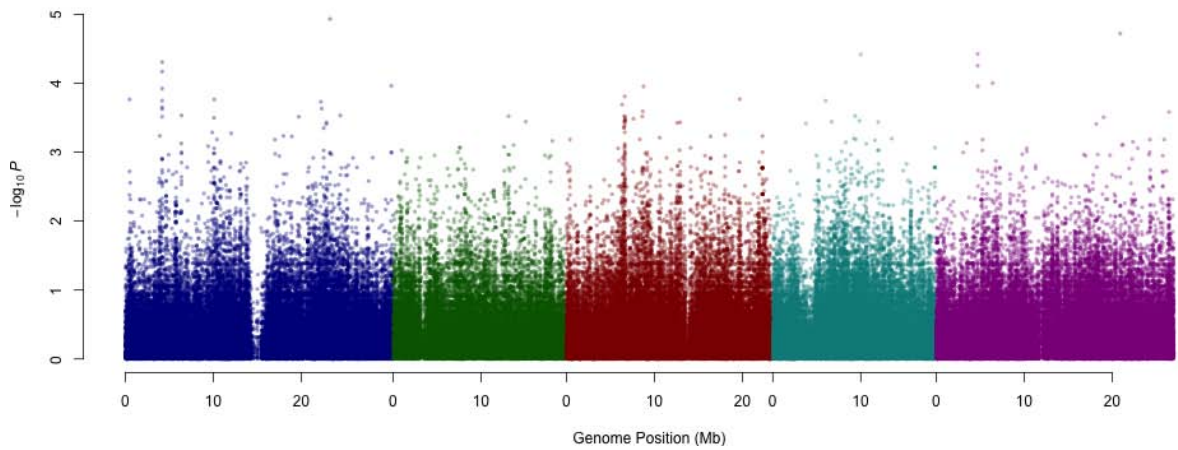

**Predictive ability assessed by cross validation**

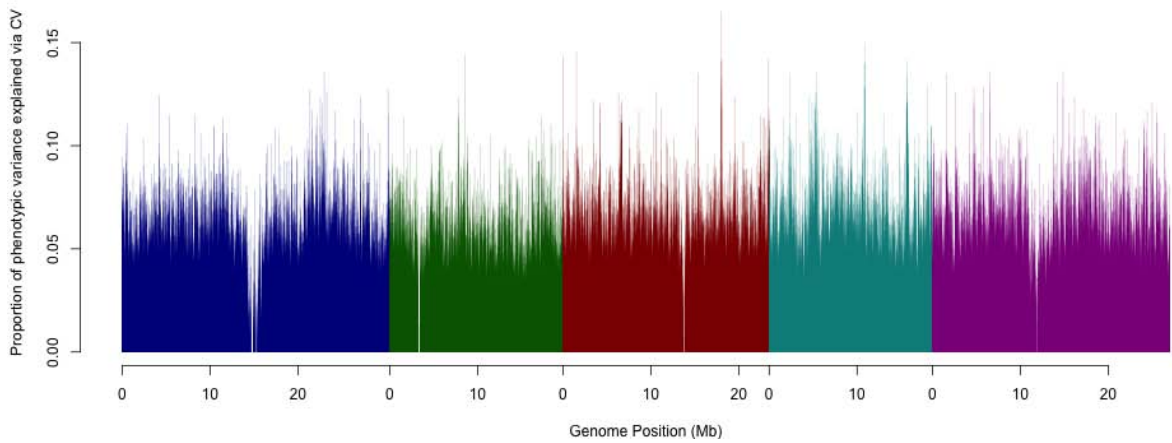

**Supplementary Figure 36** - Results of GWAS *p*-values and cross-validated predictive ability for Chlorosis 16

**Comparison of *p*-values and predictive ability**

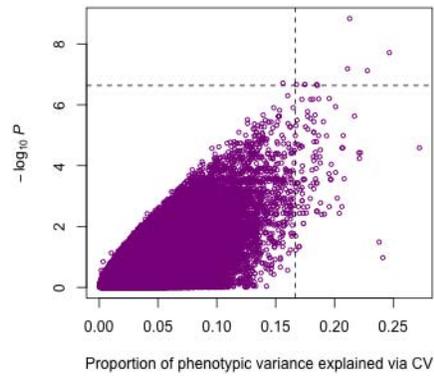

**Genome-wide association mapping via Wilcoxon test**

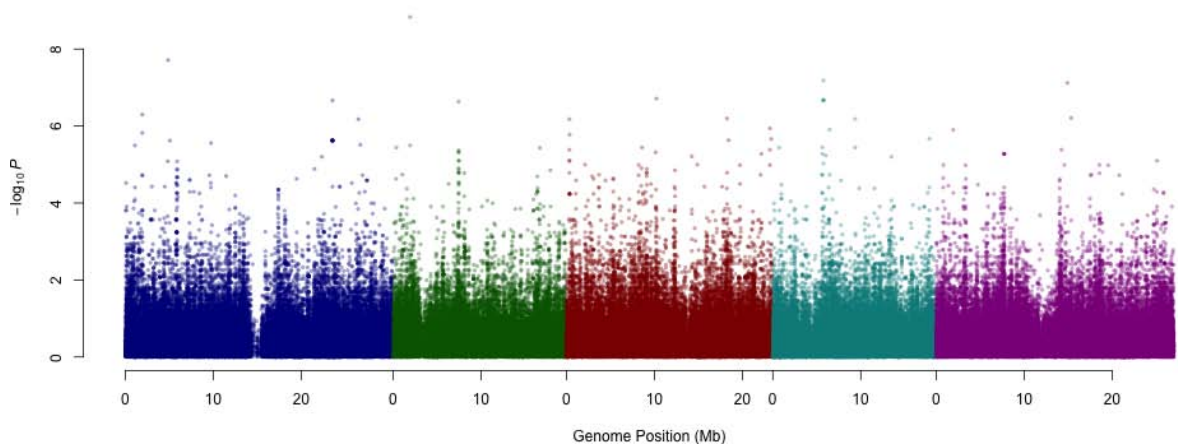

**Predictive ability assessed by cross validation**

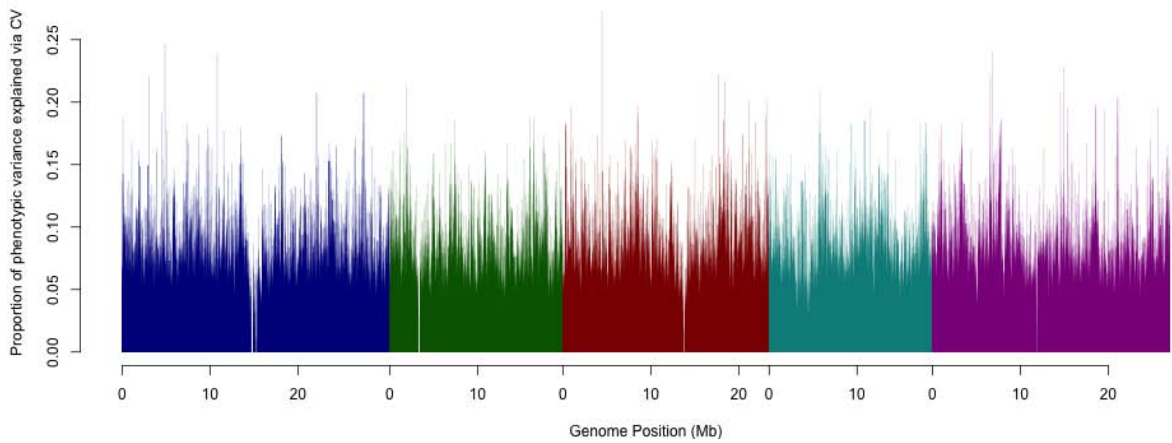

**Supplementary Figure 37** - Results of GWAS *p*-values and cross-validated predictive ability for Anthocyanin 10

**Comparison of *p*-values and predictive ability**

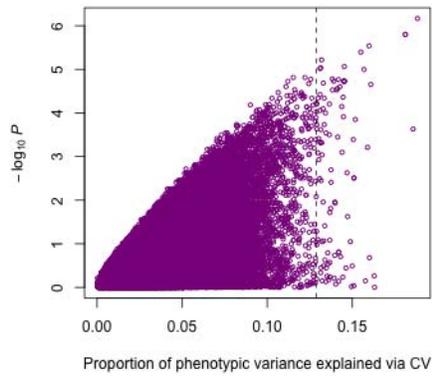

**Genome-wide association mapping via Wilcoxon test**

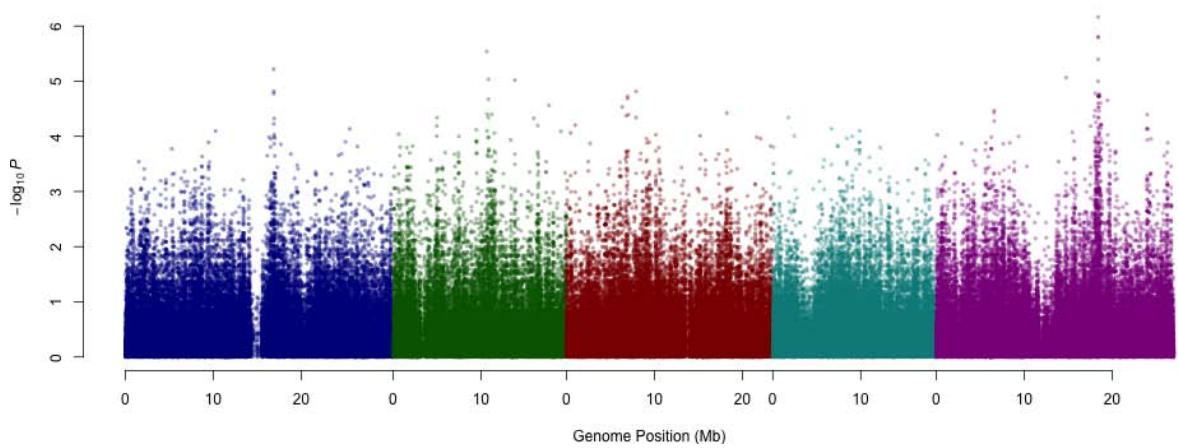

**Predictive ability assessed by cross validation**

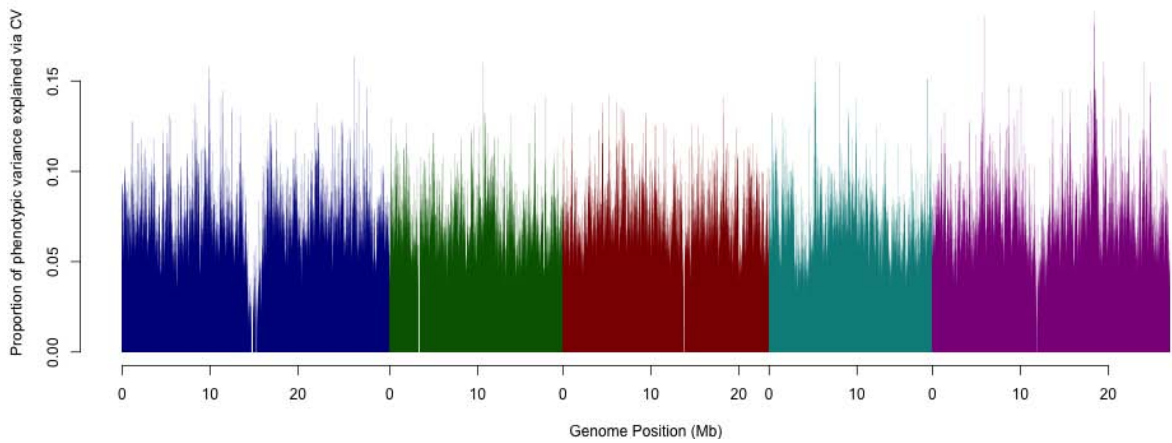

**Supplementary Figure 38** - Results of GWAS *p*-values and cross-validated predictive ability for Anthocyanin 16

**Comparison of *p*-values and predictive ability**

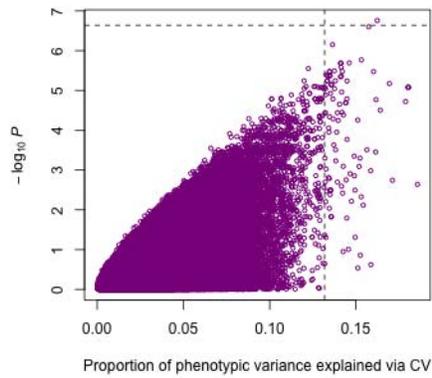

**Genome-wide association mapping via Wilcoxon test**

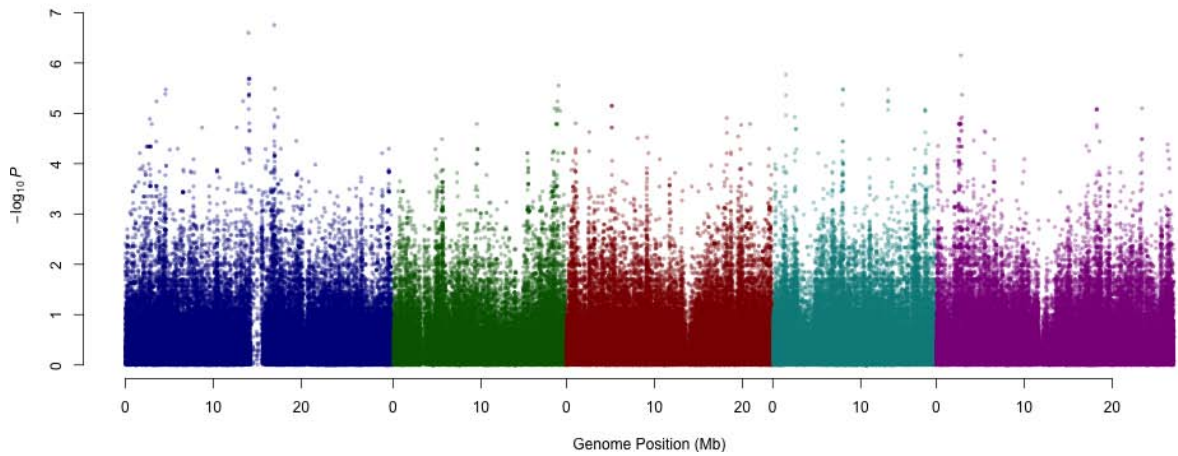

**Predictive ability assessed by cross validation**

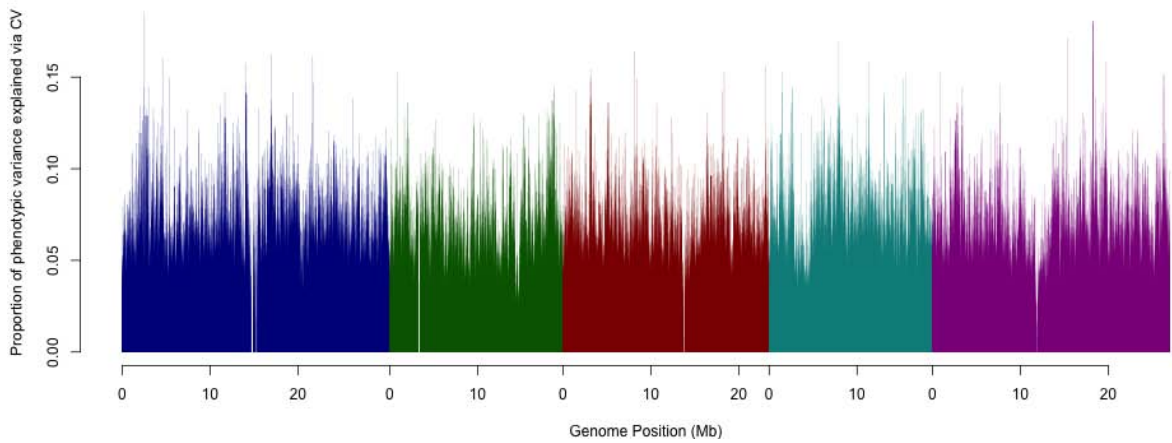

**Supplementary Figure 39** - Results of GWAS *p*-values and cross-validated predictive ability for Anthocyanin 22

**Comparison of *p*-values and predictive ability**

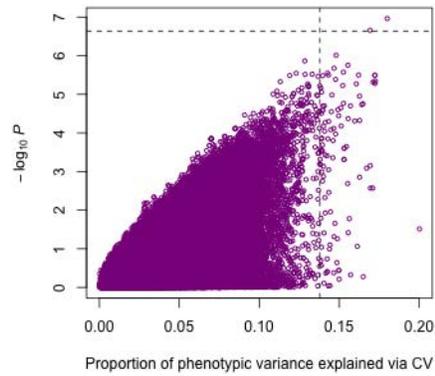

**Genome-wide association mapping via Wilcoxon test**

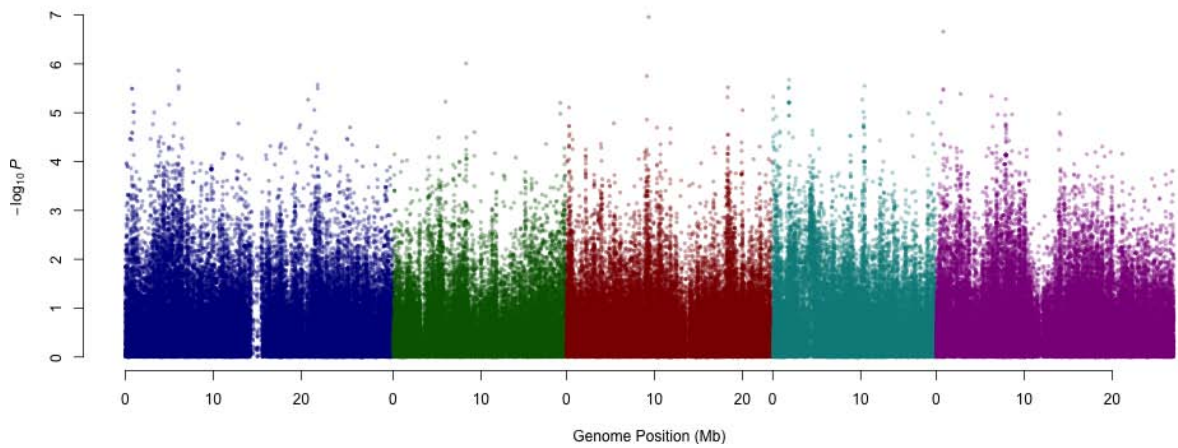

**Predictive ability assessed by cross validation**

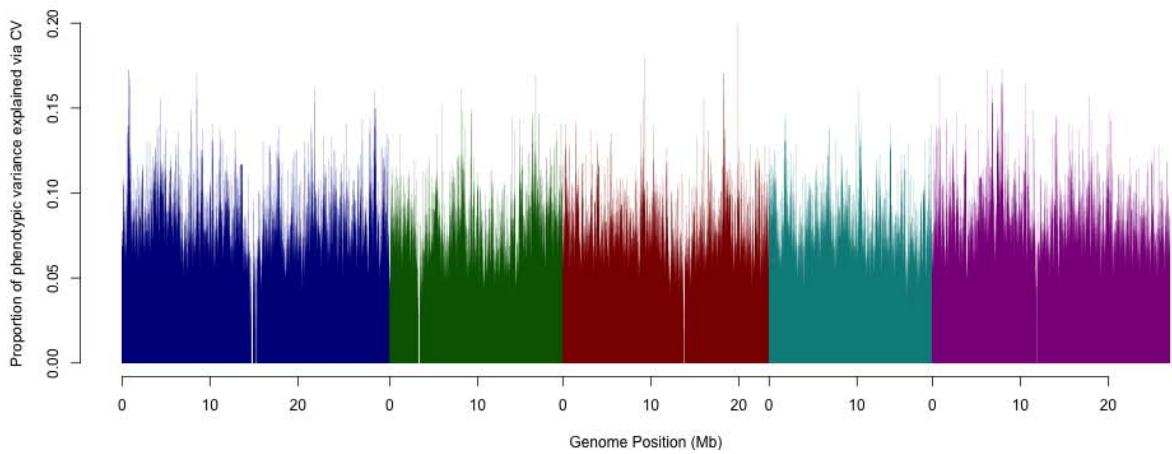

**Supplementary Figure 40** - Results of GWAS *p*-values and cross-validated predictive ability for Leaf serr 10

**Comparison of *p*-values and predictive ability**

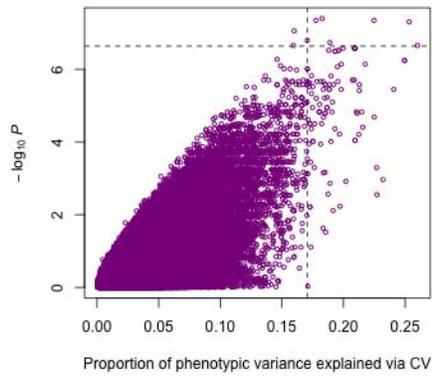

**Genome-wide association mapping via Wilcoxon test**

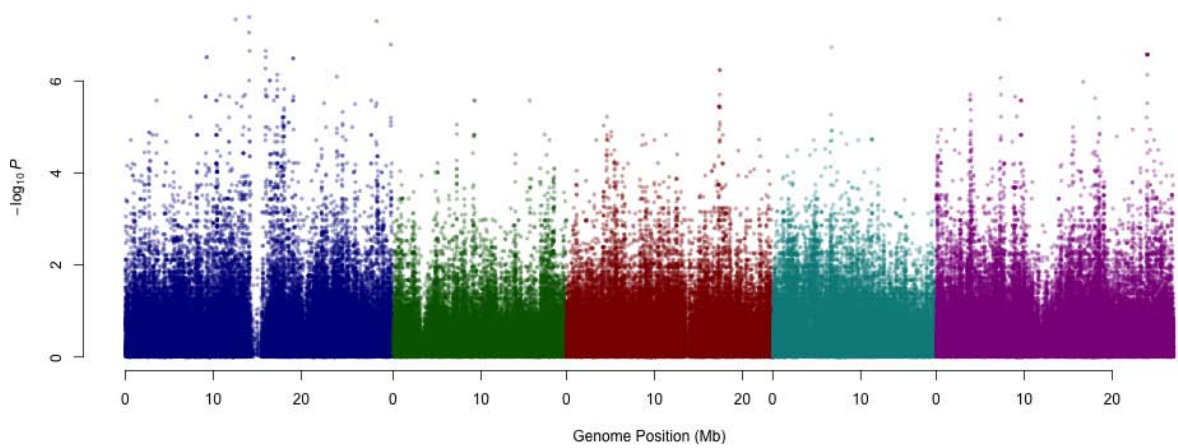

**Predictive ability assessed by cross validation**

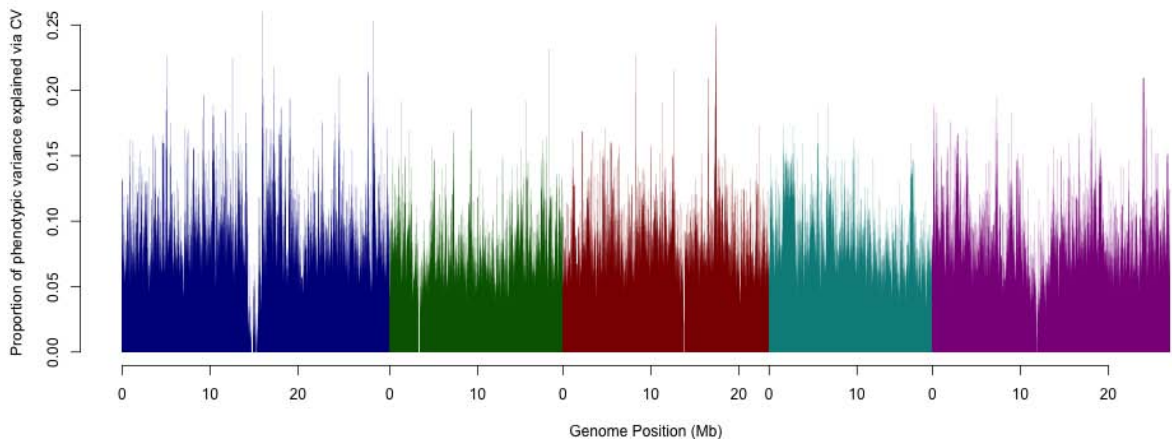

**Supplementary Figure 41** - Results of GWAS *p*-values and cross-validated predictive ability for Leaf roll 16

**Comparison of *p*-values and predictive ability**

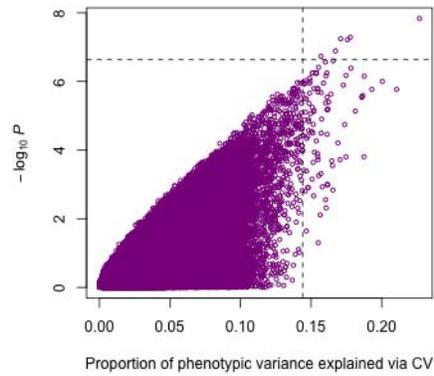

**Genome-wide association mapping via Wilcoxon test**

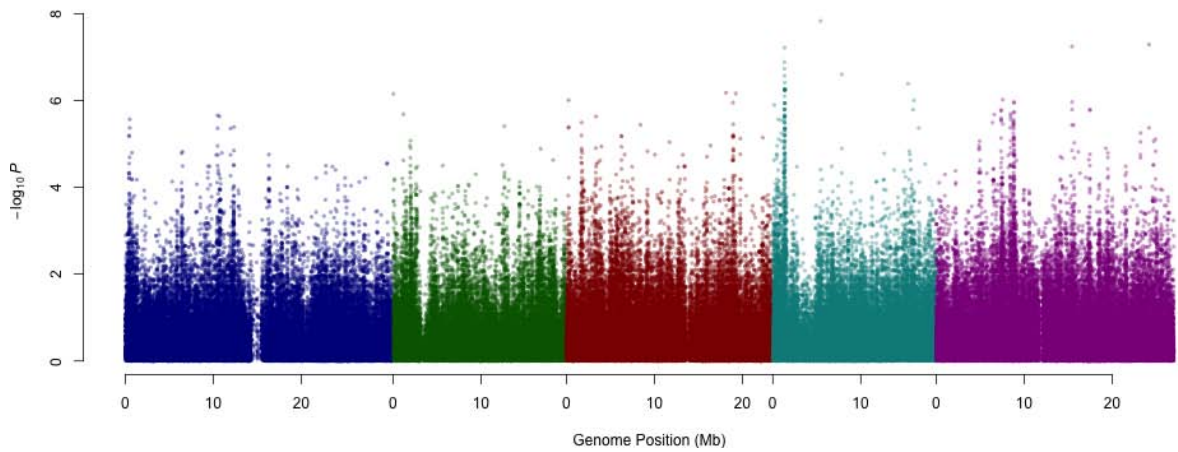

**Predictive ability assessed by cross validation**

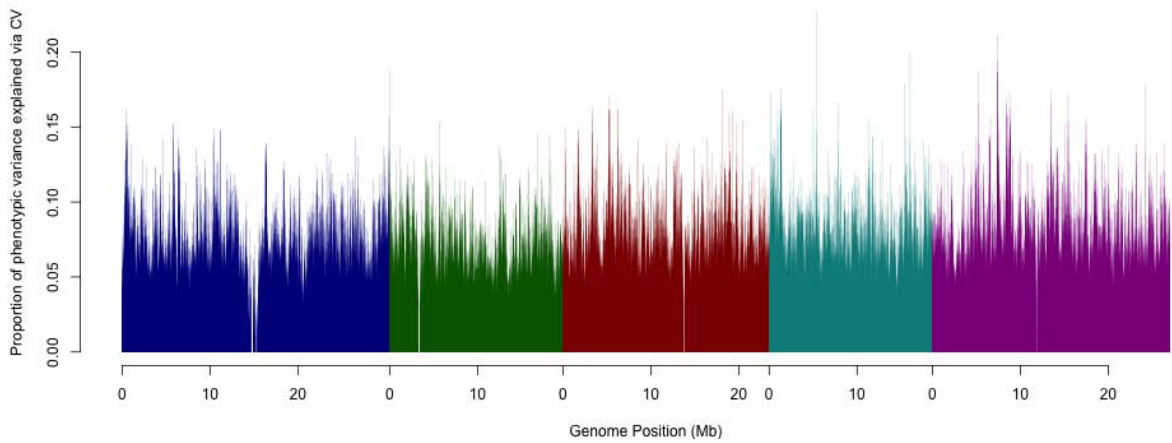

**Supplementary Figure 42** - Results of GWAS *p*-values and cross-validated predictive ability for Rosette Erect 22

**Comparison of *p*-values and predictive ability**

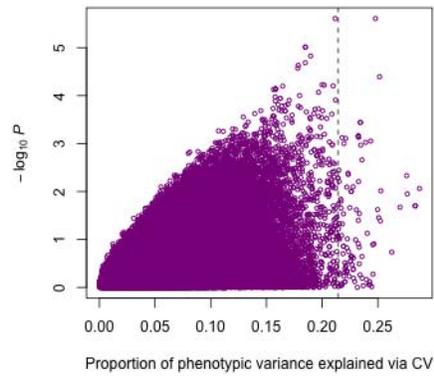

**Genome-wide association mapping via Wilcoxon test**

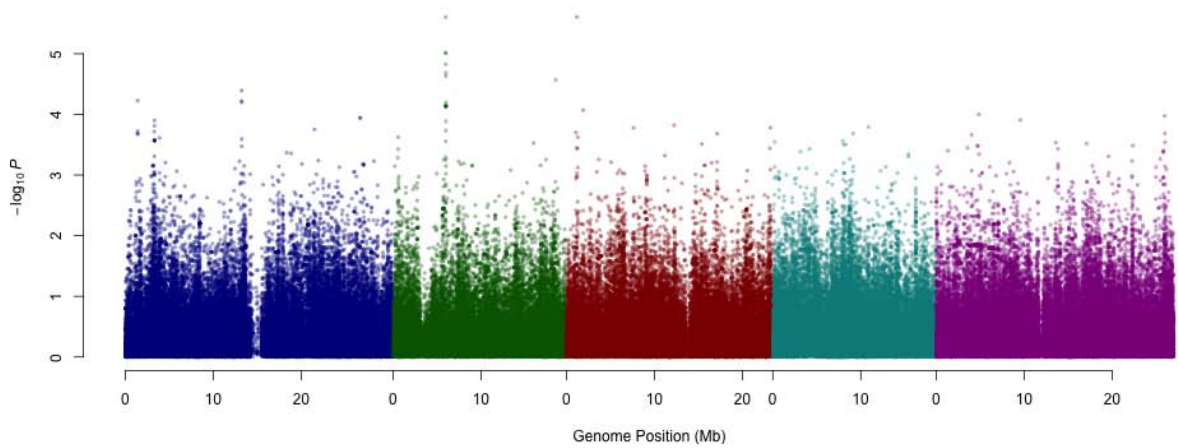

**Predictive ability assessed by cross validation**

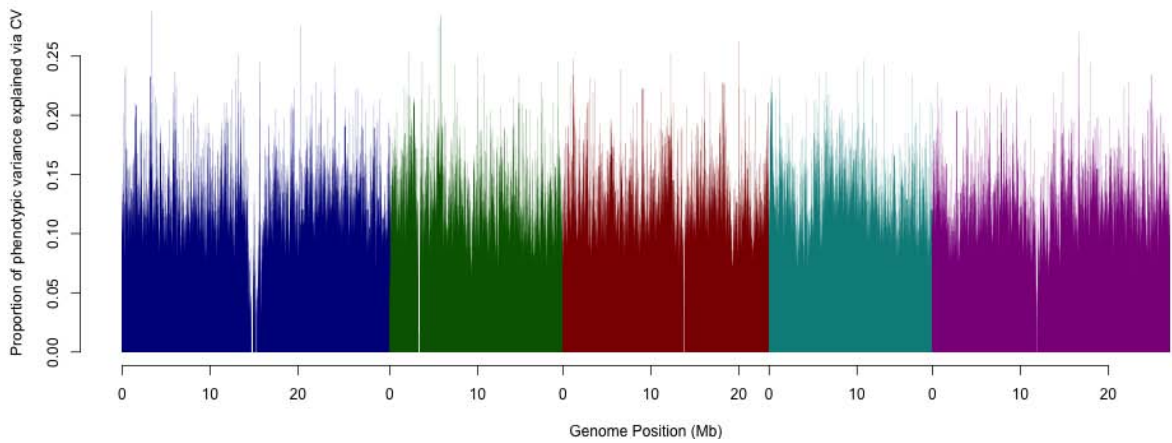

**Supplementary Figure 43** - Results of GWAS *p*-values and cross-validated predictive ability for Seedling Growth

**Comparison of *p*-values and predictive ability**

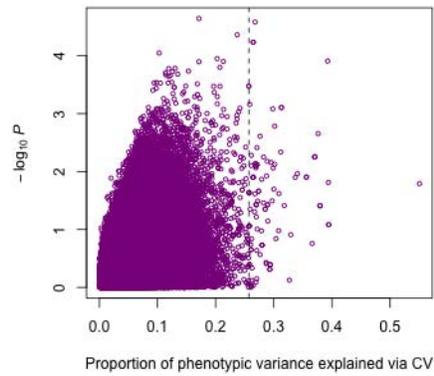

**Genome-wide association mapping via Wilcoxon test**

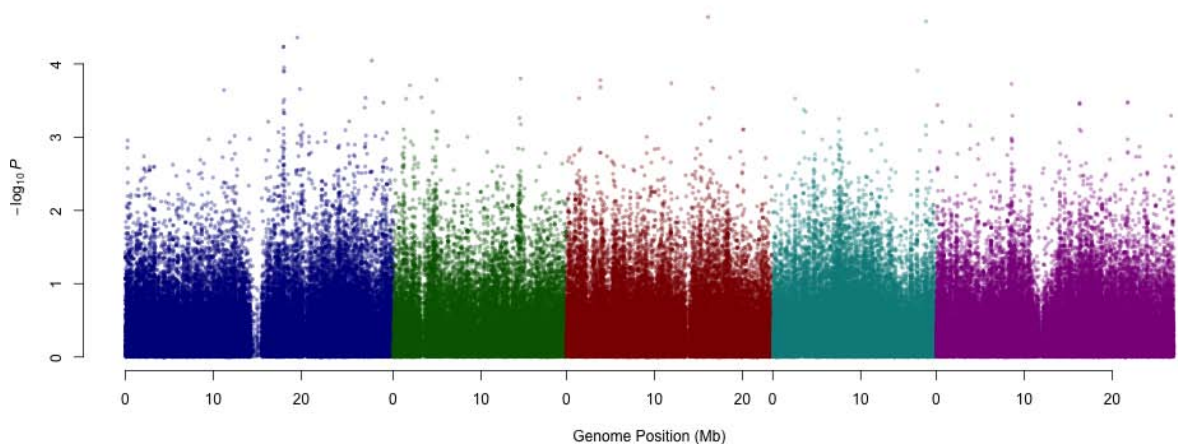

**Predictive ability assessed by cross validation**

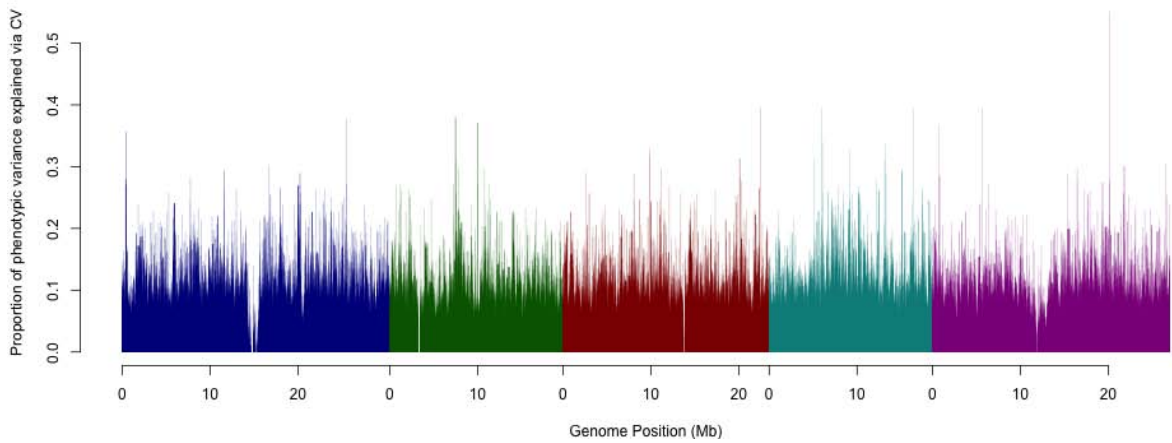

**Supplementary Figure 44** - Results of GWAS *p*-values and cross-validated predictive ability for Vern Growth

## Comparison of *p*-values and predictive ability

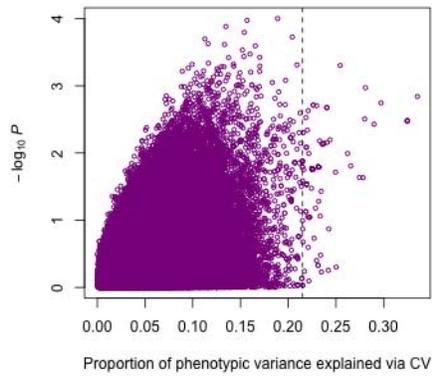

## Genome-wide association mapping via Wilcoxon test

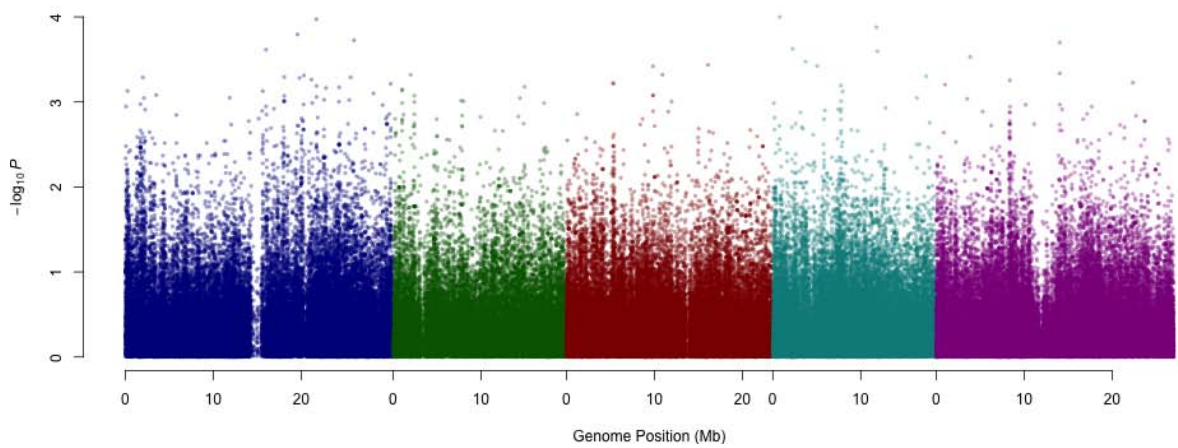

## Predictive ability assessed by cross validation

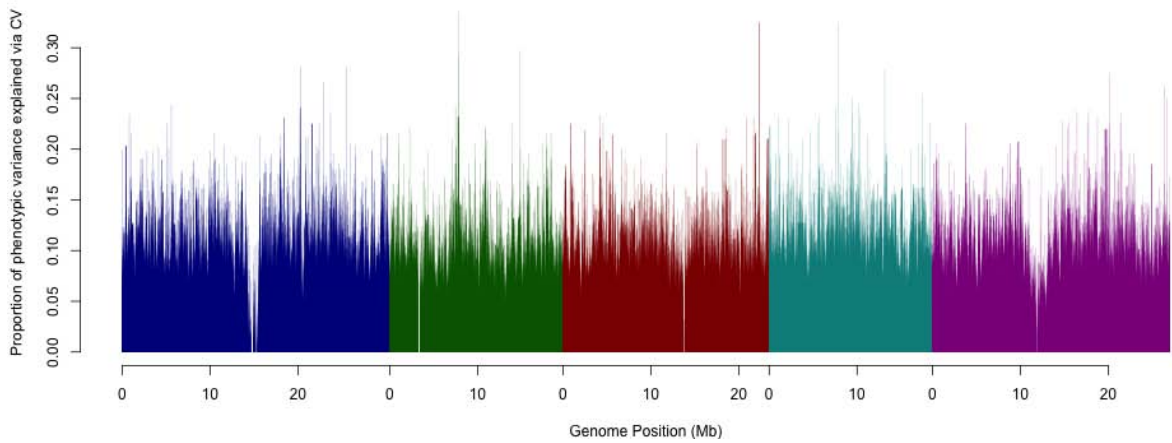

**Supplementary Figure 45** - Results of GWAS *p*-values and cross-validated predictive ability for After Vern Growth

**Comparison of *p*-values and predictive ability**

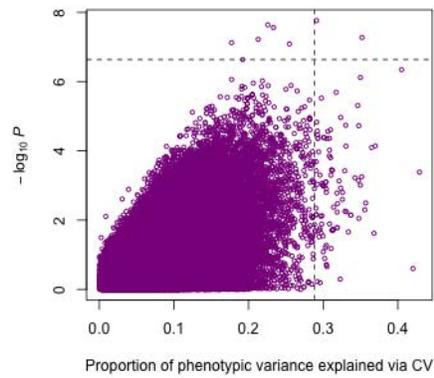

**Genome-wide association mapping via Wilcoxon test**

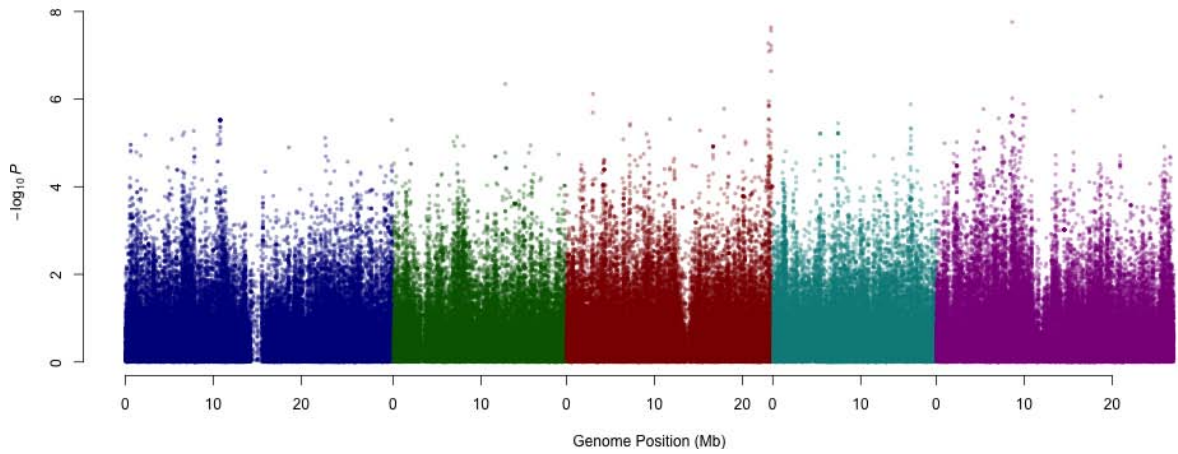

**Predictive ability assessed by cross validation**

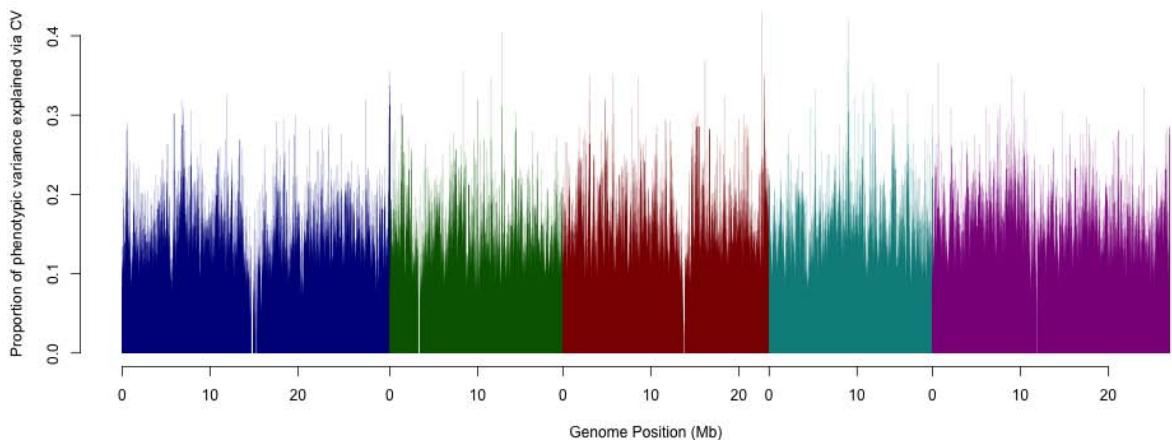

**Supplementary Figure 46** - Results of GWAS *p*-values and cross-validated predictive ability for Secondary Dormancy

**Comparison of *p*-values and predictive ability**

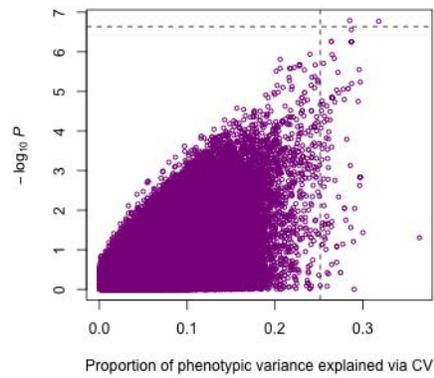

**Genome-wide association mapping via Wilcoxon test**

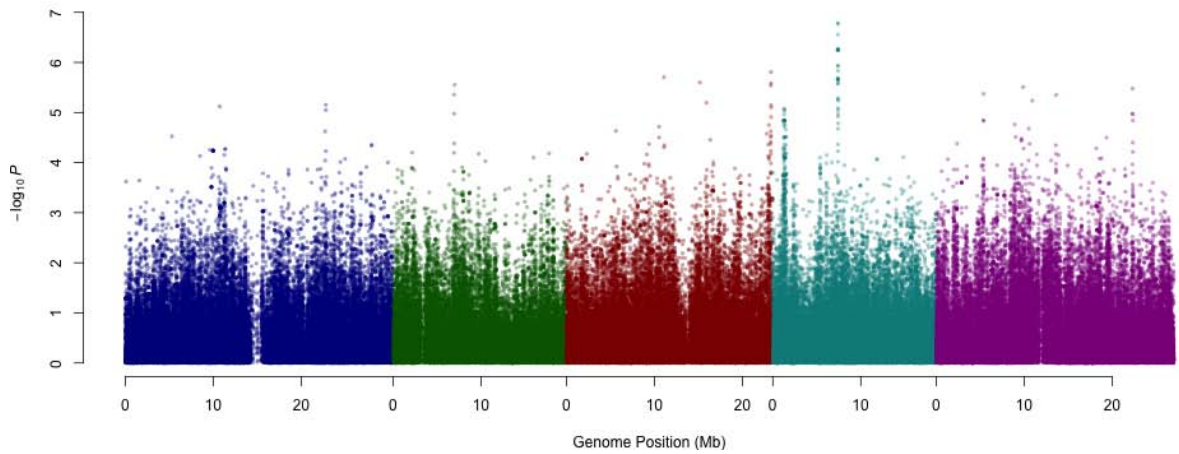

**Predictive ability assessed by cross validation**

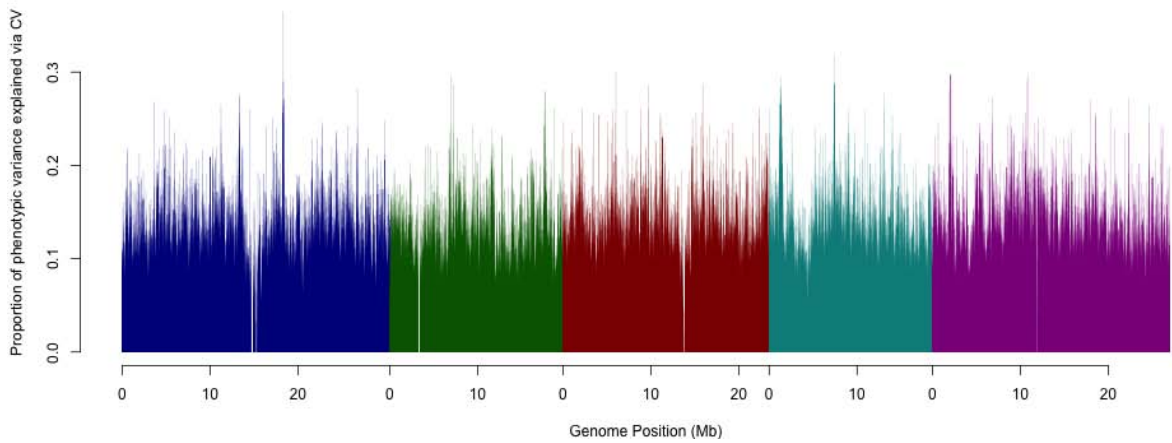

**Supplementary Figure 47** - Results of GWAS *p*-values and cross-validated predictive ability for Germ in dark

**Comparison of *p*-values and predictive ability**

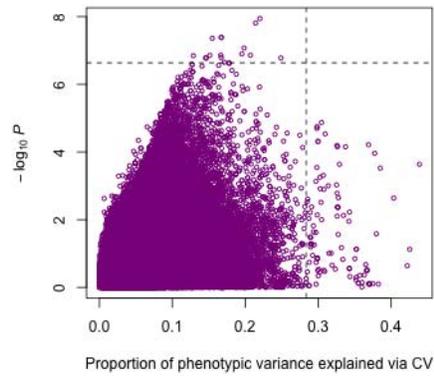

**Genome-wide association mapping via Wilcoxon test**

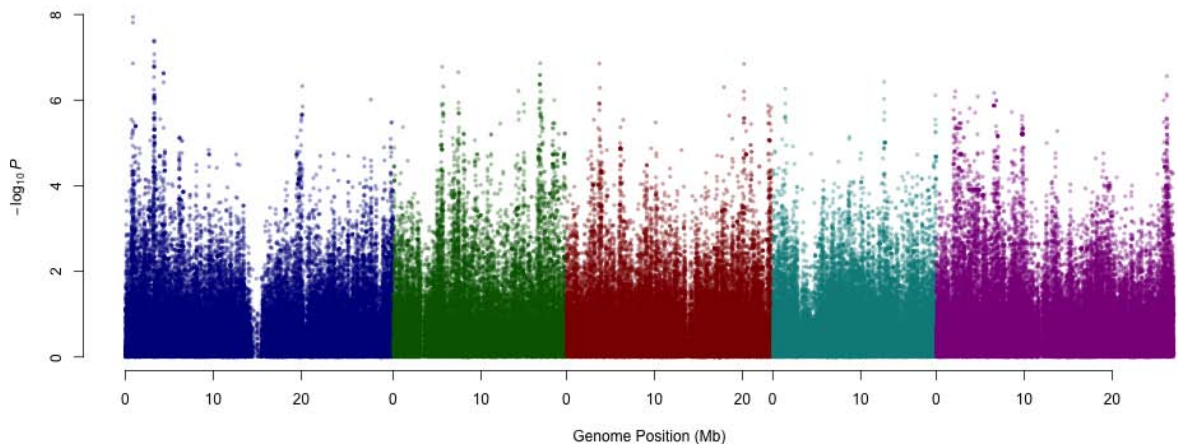

**Predictive ability assessed by cross validation**

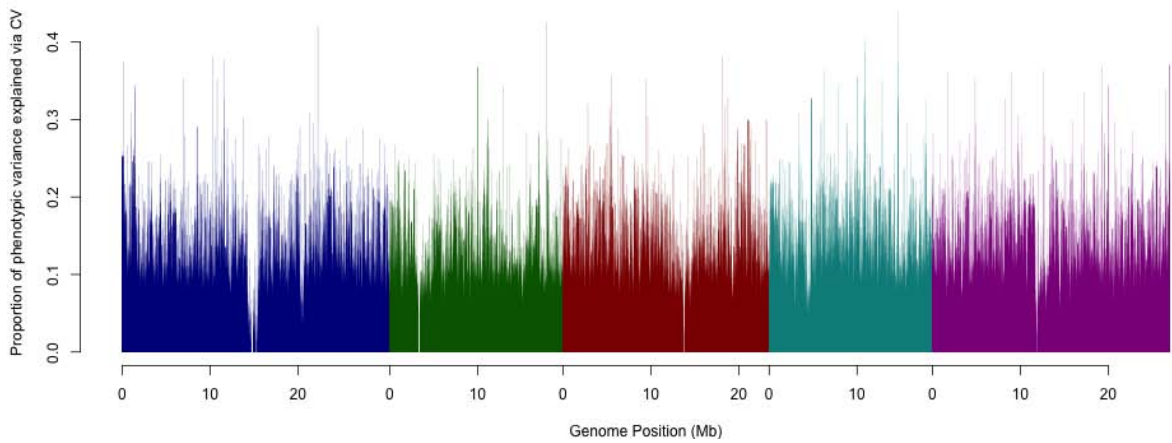

**Supplementary Figure 48** - Results of GWAS *p*-values and cross-validated predictive ability for DSDS50

**Comparison of *p*-values and predictive ability**

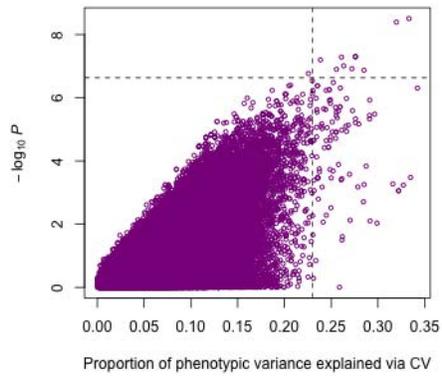

**Genome-wide association mapping via Wilcoxon test**

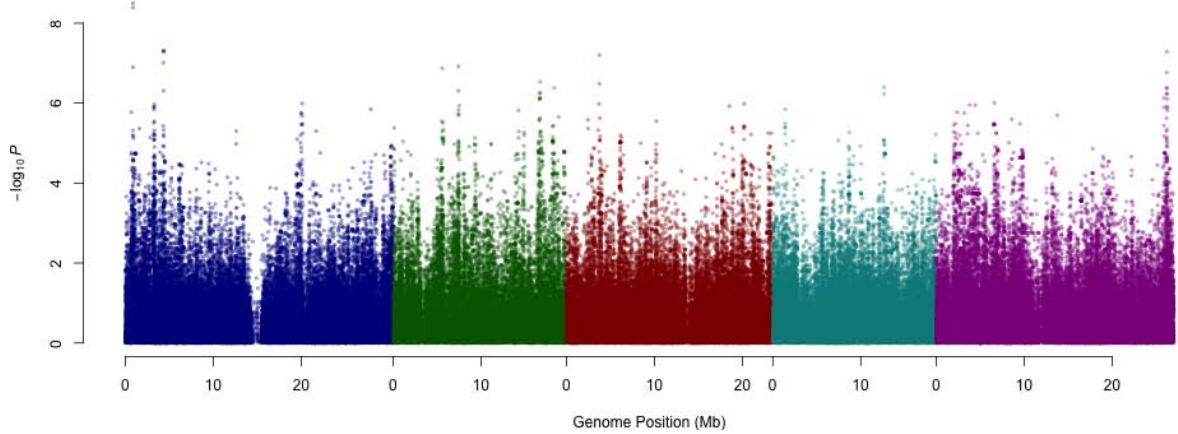

**Predictive ability assessed by cross validation**

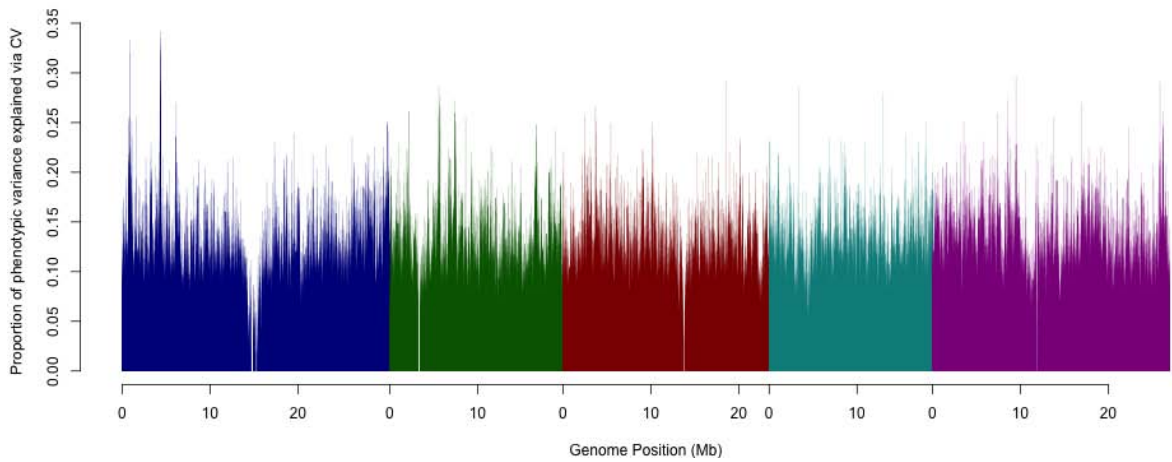

**Supplementary Figure 49** - Results of GWAS *p*-values and cross-validated predictive ability for Storage 56 days